\documentclass[a4paper,11pt]{article}
\pdfoutput=1
\usepackage{jheppub}
\usepackage{amsthm}
\usepackage{amsmath}
\usepackage{float}
\usepackage{graphicx}
\usepackage{subcaption}
\usepackage{amssymb}
\usepackage{bbold}
\usepackage[normalem]{ulem}
\usepackage{xspace}
\usepackage{slashed}
\usepackage[utf8]{inputenc}
\usepackage[T1]{fontenc}
\usepackage{subcaption}
\usepackage{scalerel,stackengine}
\allowdisplaybreaks

\newcommand*\diff{\mathop{}\!\mathrm{d}}

\newcommand{\nn}{\nonumber}

\newcommand{\be}{\begin{eqnarray}}
\newcommand{\ee}{\end{eqnarray}}

\newcommand{\ml}{\mathcal}
\newcommand{\bs}{\boldsymbol}

\begin{document}
\title{Quantum Simulation of Light-Front QCD for Jet Quenching in Nuclear Environments}

\author{Xiaojun Yao}
\affiliation{Center for Theoretical Physics, Massachusetts Institute of Technology, Cambridge, MA 02139 USA}
\affiliation{InQubator for Quantum Simulation, University of Washington, Seattle, WA 98195 USA}
\emailAdd{xjyao@mit.edu, xjyao@uw.edu}

\preprint{MIT-CTP 5435, IQuS@UW-21-037}
\abstract{We develop a framework to simulate jet quenching in nuclear environments on a quantum computer. The formulation is based on the light-front Hamiltonian dynamics of QCD. The Hamiltonian consists of three parts relevant for jet quenching studies: kinetic, diffusion and splitting terms. In the basis made up of $n$-particle states in momentum space, the kinetic Hamiltonian is diagonal. Matrices representing the diffusion and splitting parts are sparse. The diffusion part of the Hamiltonian depends on  classical background gauge fields, which need to be sampled classically before constructing quantum circuits for the time evolution. The cost of the sampling scales linearly with the time length of the evolution and the momentum grid volume. The framework automatically keeps track of quantum interference and thus it can be applied to study the Landau-Pomeranchuk-Migdal effect in cases with more than two coherent splittings, which is beyond the scope of state-of-the-art analyses, no matter whether the medium is static or expanding, thin or thick, hot or cold. We apply this framework to study a toy model and gluon in-medium radiation on a small lattice. The essence of the Landau-Pomeranchuk-Migdal effect is observed in the quantum simulation results of both the toy model and the gluon case, which is quantum decoherence caused by medium interactions that suppresses the total radiation probability.}
\maketitle

\section{Introduction}
In high energy collisions, partons of large virtuality are produced from hard scatterings, which then radiate and hadronize subsequently, forming collimated sprays of particles called jets. Studying jet production can deepen our understanding of both perturbative and nonperturbative aspects of Quantum Chromodynamics (QCD), which is the theory for strong interaction in the Standard Model. In recent years, jet and jet substructure observables in proton-proton collisions have been intensively investigated in both theory and experiment~\cite{Butterworth:2008iy,Ellis:2009su,Stewart:2010tn,Ellis:2010rwa,Abdesselam:2010pt,Altheimer:2012mn,Larkoski:2013eya,Altheimer:2013yza,Dasgupta:2013ihk,Larkoski:2014wba,Adams:2015hiv,Chien:2015cka,Larkoski:2015kga,Moult:2016cvt,Frye:2016okc,Frye:2016aiz,Kang:2016mcy,Kang:2016ehg,Kolodrubetz:2016dzb,Moult:2016fqy,Chien:2016led,Moult:2017jsg,Moult:2017okx,Larkoski:2017jix,Kang:2018jwa,Ebert:2018lzn,Moult:2018jjd,Chien:2018lmv,Kang:2018vgn,Dasgupta:2018nvj,Asquith:2018igt,Marzani:2019hun,Hoang:2019ceu,Kang:2019prh,Chien:2019gyf,Chien:2019osu,Stewart:2022ari}.

In heavy ion collisions, jets serve as useful probes of the quark-gluon plasma (QGP), a strongly coupled fluid produced shortly after the collision. High energy partons with large virtuality are produced even earlier, much before the formation of the QGP close to thermal equilibrium. The initial hard production of partons is followed by subsequent parton showers and when the produced partons traverse the QGP, further radiation induced by the medium can happen. Eventually partons hadronize into color neutral particles at freezeout. By comparing jets produced in proton-proton and heavy ion collisions, we are able to learn how the QGP modifies the parton shower. Jets can be thought of as external to the QGP, since the large energy scale involved in the jet production is much bigger than the typical temperature of the QGP fireball, which falls in the range $\sim[150,600]$ MeV. In this sense, a jet can also be treated as an open quantum system embedded in the QGP fireball~\cite{Vaidya:2020cyi,Vaidya:2020lih}. Nevertheless, the soft ingredients of jets cannot be fully distinguished from the QGP fireball in general.

To understand and interpret experimentally measured jet and jet substructure observables in heavy ion collisions, at least three aspects of jet-medium dynamics need theoretical studies: jet energy loss, medium response and selection bias. First, when high energy partons traverse the QGP, they interact with the soft medium and as a result lose energy and momentum. This is the original idea of jet quenching in heavy ion collisions. Furthermore, the lost energy and momentum evolve in the QGP fireball, which may or may not thermalize completely to become part of the QGP, and eventually turn into particles that still have some correlation with the original high energy partons losing energy and momentum. Due to the remaining correlation, some of the particles produced in this way are reconstructed as part of the final jets. Finally, since jets of wider opening angles lose more energy than those with narrower opening angles,  when experimentalists reconstruct jets of a given energy or transverse momentum, more narrower jets are selected due to the power-law decrease in jet spectra. Jet energy loss has been studied widely for a long time, while in recent years, more studies focused on understanding medium response~\cite{CasalderreySolana:2004qm,Ruppert:2005uz,Chaudhuri:2005vc,CasalderreySolana:2006sq,Chesler:2007an,Gubser:2007ga,Chesler:2007sv,Chesler:2008wd,Chesler:2008uy,Neufeld:2008fi,Neufeld:2008dx,Qin:2009uh,Neufeld:2009ep,Gubser:2009sn,Chesler:2011nc,Betz:2010qh,Ayala:2012bv,Ayala:2014sua,Floerchinger:2014yqa,Tachibana:2014lja,Yan:2017rku,Chen:2017zte,Tachibana:2020mtb,Casalderrey-Solana:2020rsj} and selection bias~\cite{Brewer:2021hmh}.

Jet energy loss has been studied in both the strong coupling~\cite{Chesler:2014jva,Chesler:2015nqz,Casalderrey-Solana:2014bpa,Casalderrey-Solana:2015vaa,Casalderrey-Solana:2016jvj,Hulcher:2017cpt,Casalderrey-Solana:2018wrw,Casalderrey-Solana:2019ubu} and weak coupling limits. In the weak coupling (perturbative) approach, an important quantum interference effect that needs consideration is called the Landau-Pomeranchuk-Migdal (LPM) effect. The LPM effect suppresses in-medium radiation because of quantum decoherence, caused by soft momentum exchange with the medium that modifies the phase in the time evolution in a random way. Early perturbative studies of the LPM effect focused on the case with a static medium and just one splitting, i.e., with one incoming parton and two outgoing partons for an initial quark state~\cite{Gyulassy:1993hr,Wang:1994fx,Baier:1994bd,Baier:1996kr,Zakharov:1996fv,Baier:1996sk,Gyulassy:1999zd,Gyulassy:2000fs,Wiedemann:2000za,Arnold:2002ja}, and were later generalized for an incoming gluon~\cite{CasalderreySolana:2011rz,MehtarTani:2011tz,Ovanesyan:2011xy,MehtarTani:2011gf,MehtarTani:2012cy,Blaizot:2012fh,Blaizot:2013hx,Blaizot:2013vha,Ghiglieri:2015ala} and expanding media~\cite{Salgado:2003gb,Adhya:2019qse}. The difficulty of analyzing the LPM effect lies in that the soft momentum transfer from the medium and the parton splitting do not commute, which requires one to keep track of both in a time-ordered way. The soft momentum exchange process in the time evolution can be analyzed by studying a time evolution equation for a two-point correlation function, which describes the propagation of a single parton in the medium, undergoing transverse momentum broadening due to diffusion. The soft momentum exchange is encoded in terms of a ``potential'' term in the equation, which can be calculated in the opacity expansion or modeled. The description of the soft momentum exchange can be improved by expanding the ``potential'' term perturbatively at high frequency on top of a harmonic oscillator form~\cite{Mehtar-Tani:2019ygg,Barata:2021wuf}. Recent studies have attempted to investigate cases with two coherent splittings~\cite{Arnold:2020uzm,Arnold:2021pin,Arnold:2022epx}, but the analysis becomes extremely complicated due to multiple interfering diagrams where the daughter partons have overlapped formation times. Therefore, it is extremely challenging to analyze the LPM effect for cases with more than two coherent splittings, especially when the medium is time dependent.

In this paper, we propose a framework for quantum simulation of jet quenching in hot and/or dense nuclear environments, which can help us to study multiple coherent splittings in a generic medium. Quantum simulation of quantum dynamics has been proposed long time ago~\cite{feynman1986quantum} and is developing rapidly in recent years~\cite{Devoret2013,annurev-conmatphys-031119-050605,doi:10.1063/1.5088164,google_supremacy,Lamm:2018siq,Bauer:2019qxa,Mueller:2019qqj,Wei:2019rqy,Smith2019,Barata:2020jtq,Liu:2020eoa,Liu:2020wtr,Buser:2020cvn,Kan:2021nyu,Martyn:2021eaf,Klco:2021lap,Bauer:2021gup,Czajka:2021yll,Ciavarella:2022zhe,Bauer:2022hpo}. For applications in quantum field theory, it has been shown that scalar field theory with the $\phi^4(x)$ interaction can be efficiently simulated on a quantum computer~\cite{Jordan:2011ci,Jordan:2012xnu,Jordan:2017lea,Klco:2018zqz}. Later studies investigated fermionic fields~\cite{Jordan:2014tma} and gauge theories in low dimensions~\cite{hauke2013quantum,kuhn2014quantum,Klco:2018kyo,Zache:2018cqq,Klco:2019evd,Chakraborty:2020uhf,Nguyen:2021hyk,deJong:2021wsd,Ciavarella:2021nmj,Gonzalez-Cuadra:2022hxt}. Quantum simulation has been explored to study open quantum systems in heavy ion collisions such as heavy quarks and jets~\cite{DeJong:2020riy,Barata:2021yri}. Furthermore, hadron structure can also be studied on a quantum computer by using basis light-front quantization approach~\cite{Qian:2021jxp}. In the noisy intermediate-scale quantum (NISQ) era~\cite{preskill2018quantum}, error mitigation techniques~\cite{He:2020udd,Pascuzzi:2021mhw} are crucial for useful applications of quantum computers.

To simulate jet quenching on a quantum computer, we will apply the light-front Hamiltonian formulation of QCD~\cite{Brodsky:1997de,Bakker:2013cea} to describe the in-medium time evolution of high energy partons. The light-front Hamiltonian approach has been used to study the time evolution of a high energy quark inside a heavy nucleus, where the time evolution equation is solved classically~\cite{Li:2020uhl,Li:2021zaw}. The Hamiltonian relevant for jet quenching can be decomposed into three parts: a kinetic term for the phase change in the time evolution, a diffusion term accounting for the transverse momentum broadening due to the soft kicks from the medium, and a splitting term that governs radiation of partons and their recombination. The random transverse momentum exchange between partons and the medium can be described by an external classical background gauge field that satisfies certain correlations. These correlation functions depend on the medium properties such as its temperature. The classical background field results in a random change of the kinetic energy, which leads to a random phase in the time evolution and is the crucial part for the quantum decoherence in the LPM effect. The classical background field needs to be sampled classically before constructing quantum circuits and the cost of the sampling scales linearly with the time length of the evolution and the momentum grid volume. We note that Ref.~\cite{Barata:2021yri} used a similar approach
to study the jet quenching parameter, which only involves the kinetic and diffusive parts of the Hamiltonian and does not contain radiation in the quantum evolution. To study jet quenching phenomenon using Ref.~\cite{Barata:2021yri}, one still needs to use some perturbative treatment of radiation with the jet quenching parameter as an input, which suffers from extreme complications to analyze multiple coherent radiations, as explained above. Here we include the radiation Hamiltonian in the quantum evolution and thus being able to treat radiation beyond perturbation and deal with multiple coherent radiations where the daughter partons have overlapped formation times.

Furthermore, Ref.~\cite{Barata:2021yri} focused on a 1-body quantum mechanical system. The algorithm used therein is efficient and the efficiency originates in making both the kinetic and diffusive parts of the Hamiltonian diagonal in two different bases that can be swapped efficiently via quantum Fourier transform. This algorithm has also been used in showing that scalar field theory can be quantum simulated efficiently~\cite{Jordan:2012xnu,Jordan:2017lea}, where quantum Fourier transform swaps the field and its canonical momentum at each spatial point efficiently.
Although efficient, this algorithm does not apply to gauge theories and thus jet quenching studies in general. In this work, instead of using field values at each spatial point as a basis, we will use $n$-particle states in momentum space as the basis of the Hilbert space and write down matrix elements for the three parts of the Hamiltonian. It will turn out that in this basis the kinetic term is diagonal and thus can be efficiently simulated. Furthermore, the matrices of the diffusion and splitting Hamiltonians are sparse, indicating that we are very likely able to efficiently simulate them on a quantum computer. After discretizing momenta and encoding all the basis states in the qubit register, we can construct quantum gates for the Hamiltonian evolution. We will also discuss how to construct quantum circuits for multi-parton cases, by using the circuits for the single-parton case as building blocks, which is important if we want to scale up the quantum simulation and crucially replies on the diagonality and sparsity of the Hamiltonian matrix elements.

The initial state of the time evolution for jet quenching is given by one or many partons (quarks and gluons) with definite momenta, colors and spins, properly (anti)symmetrized, which can be easily constructed in the qubit register since it is a linear combination of the basis states with known coefficients. 
The standard Trotterization method will then be applied to simulate the Hamiltonian evolution. At the end of the time evolution, we perform measurements by projecting the final state onto a state with certain number of partons with specific momenta, colors and spins, that is properly (anti)symmetrized. Radiation spectra can then be estimated from the measurement results by repeating the time evolution and the projective measurement multiple times. Our approach automatically keeps track of quantum interference, since it is based on the quantum evolution of a wavefunction, i.e., it evolves on the amplitude level. Therefore, our framework can be easily used to coherently study the LPM effect for more than two splittings with overlapped formation times, no matter whether the medium is time independent or time dependent, thin or thick, hot or cold, which has never been done. In the future, with fault-tolerant quantum computers that have a few hundred logical qubits, we will be able to use this framework to study QCD jet quenching in nuclear environments and learn new physical insights into the LPM effect.

We will first apply the formalism to study a toy model that can be encoded by five qubits, to demonstrate how to construct a quantum circuit from a Hamiltonian. The toy model consists of scalar particles, which means we neglect the spin and color degrees of freedom that are present in QCD. To reduce the size of the Hilbert space, we simply consider a $2+1$ dimensional system with only one transverse direction. Both the longitudinal and transverse momenta have two levels. We include both $1$-particle and $2$-particle states in the Hilbert space, which allows us to study the quantum decoherence effect in one splitting. Classical background fields are also used to describe the random transverse momentum exchanges in the toy model, which are sampled classically. By explicitly constructing a quantum circuit for the time evolution of the toy model and running simulations on the IBM Qiskit simulator, we compare the total radiation probabilities in vacuum and in the medium for an initially virtual $1$-particle state. We find that the probability of having two particles in the final state is smaller in the medium, which means the quantum decoherence effect that suppresses radiation is observed in the quantum simulation results of the toy model, which is the essence of the LPM effect.

We will then apply the formalism to study the LPM effect in gluon radiation. By using a small momentum lattice in 3 dimension, we can encode both 1-gluon and 2-gluon states on a 15-qubit system. By using discretized light-front Hamiltonian of QCD, we are able to simulate the time evolution of an initially virtual 1-gluon state in both vacuum and the medium. The essence of the LPM effect is also observed in the simulation results of the time evolution.

This paper is organized as follows: in section~\ref{sect:formalism} we will give an overview of the framework, which includes state initialization, Hamiltonian time evolution and final measurements. We will introduce the light-front Hamiltonian of QCD to describe the in-medium dynamics of high energy partons and explain the $n$-particle basis of the Hilbert space. The matrix elements of the three parts of the Hamiltonian: the kinetic, diffusion and splitting terms will be given explicitly in the following section~\ref{sect:matrix}, together with a discussion on the sampling of classical background fields. Furthermore, quantum simulation of the toy model for studying the quantum decoherence effect cased by medium interactions will be discussed in section~\ref{sect:toy}, with an explicit construction of the quantum circuit for the time evolution. Simulation results that are based on the IBM Qiskit quantum simulator will also be shown. Then we will study the quantum simulation of gluon radiation on a small momentum grid in section~\ref{sect:gluon}. Finally, we will conclude and give an outlook in section~\ref{sect:conclusion}.

\section{Formalism}
\label{sect:formalism}

\begin{figure}[t]
\centering
\includegraphics[height=2.2in]{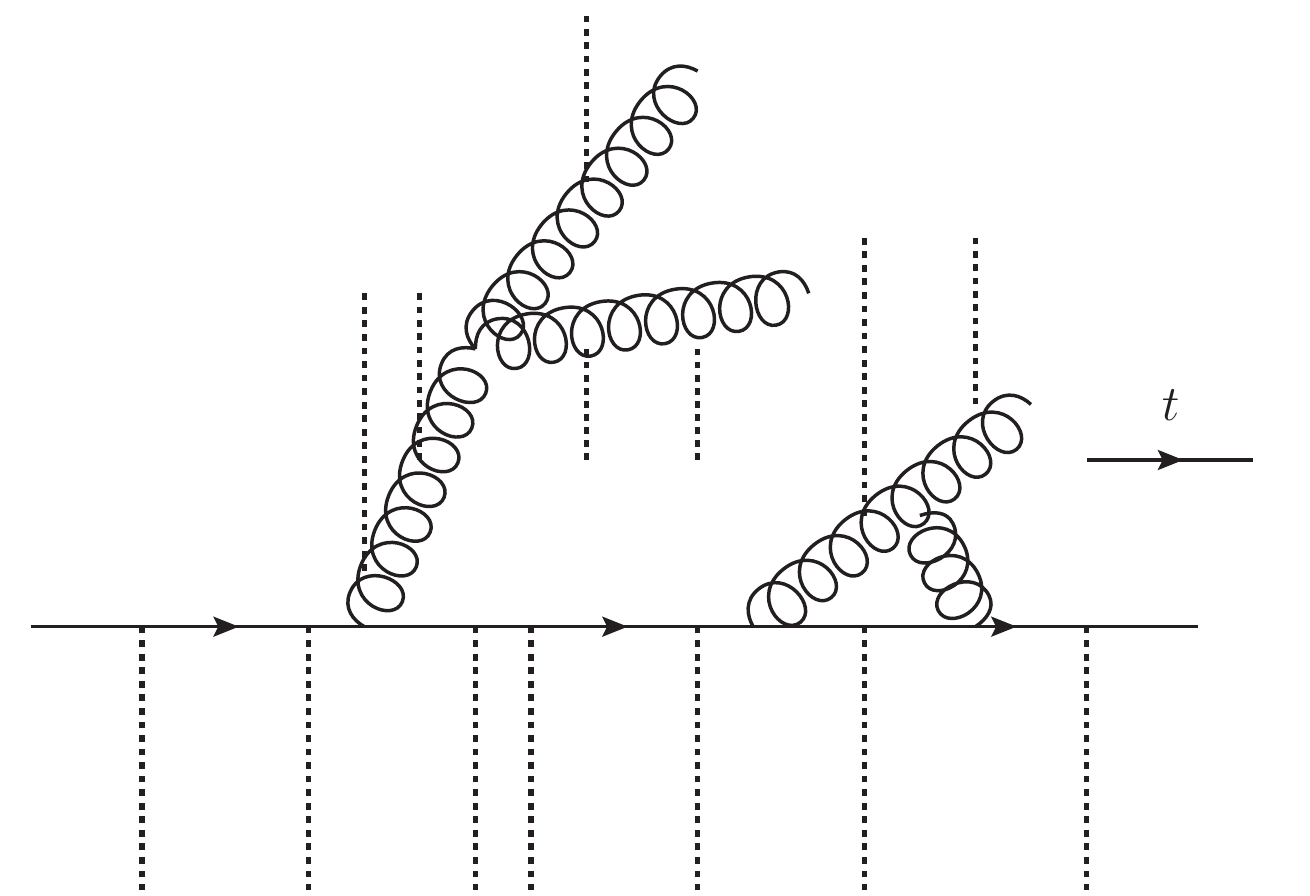}
\caption{Typical diagram describing the LPM effect in jet quenching on the amplitude level, which includes free propagation, soft momentum exchange (labeled by dashed lines) and splitting/recombination. The solid lines with arrows indicate the propagation of quarks in time, while the curly lines are for the propagation of gluons.}
\label{fig:lpm}
\end{figure}

A typical diagram to understand the LPM effect in jet quenching is depicted in Fig.~\ref{fig:lpm}, which describes the time evolution of a quantum state initiated by an incoming parton that undergoes subsequent soft momentum exchanges, splitting and recombination. The diagram is on the amplitude level. To calculate physical observables, one needs to sum over the amplitudes of all diagrams with the same final state. In general, the number of diagrams grow exponentially with the number of splittings and their quantum interference is extremely difficult to account for in an approach based on perturbation theory.  

\begin{figure}[t]
\centering
\includegraphics[height=2.2in]{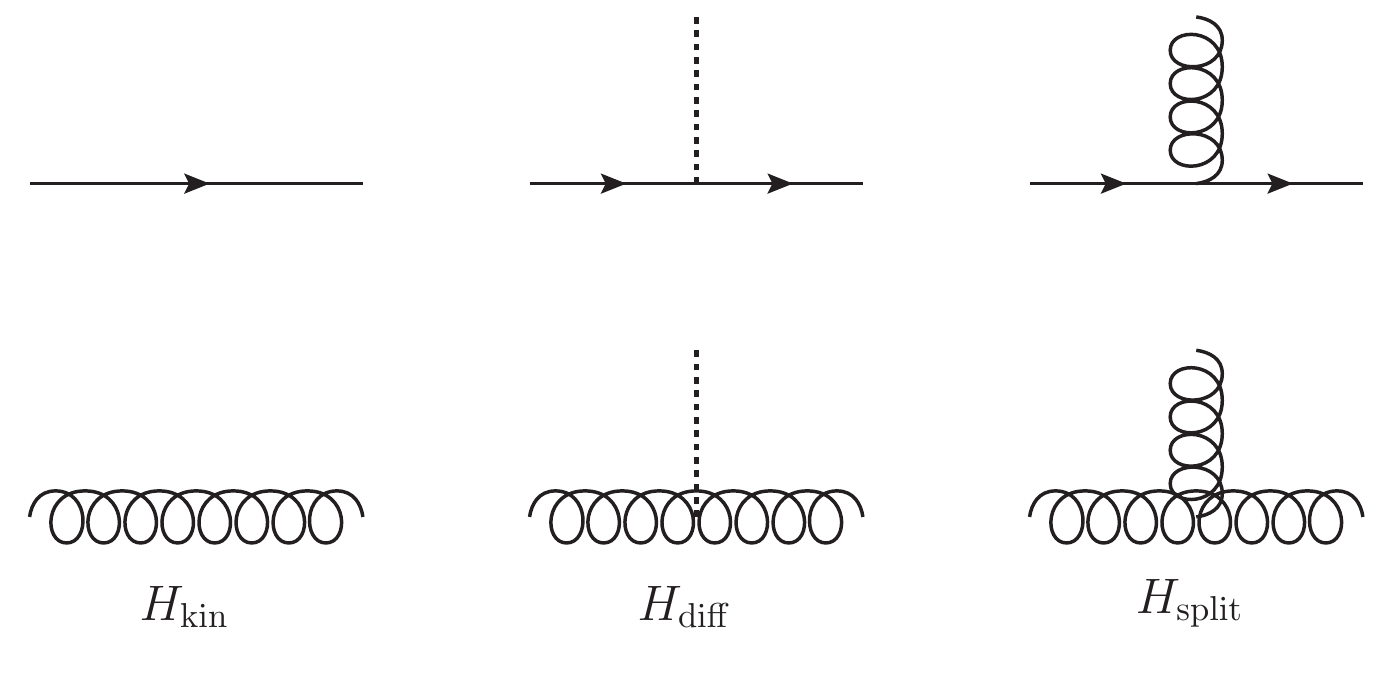}
\caption{Three parts of the Hamiltonian for studying the LPM effect in jet quenching: the kinetic, diffusion and splitting/recombination terms. Matrix elements of these Hamiltonians will be explicitly given in section~\ref{sect:matrix}.}
\label{fig:lpm_H}
\end{figure}

To simulate the time evolution of jets and study the LPM effect on a quantum computer, we need a Hamiltonian description of the evolution, which includes the kinetic term, diffusion caused by soft momentum transfer from the medium and splitting/recombination, as depicted in Fig.~\ref{fig:lpm_H}. In this work, we will use the light-front Hamiltonian of QCD~\cite{Brodsky:1997de,Bakker:2013cea} to describe the dynamics of high energy partons and their interactions with nuclear media. A brief introduction to the light-front Hamiltonian of QCD can be found in appendix~\ref{app:lfqcd}. We will first discuss the light-front Hamiltonian dynamics for studying the LPM effect in jet quenching in section~\ref{sect:h}. Then  in section~\ref{sect:hilbert} we will introduce the computational basis of the Hilbert space for the quantum simulation. 

\subsection{Light-Front Hamiltonian Dynamics}
\label{sect:h}
The light-front Hamiltonian dynamics is determined by
\begin{align}
2i\frac{\partial}{\partial x^+} |\Psi\rangle = H |\Psi\rangle \,,
\end{align}
where $x^+=x^0+x^3$ is the light-cone time.\footnote{The factor of $2$ on the left-hand side is just a convention. When defining the light-front Hamiltonian, we integrate the Hamiltonian density with the integral measure
\be
\int \diff x^- \diff^2 x_\perp \ml{H} \,.
\ee
On the other hand, we know the Lorentz invariant measure in spacetime is
\be
\int \diff^4 x = \frac{1}{2}\int \diff x^+ \diff x^- \diff^2 x_\perp \,.
\ee
Therefore, for consistency we need to treat $\frac{1}{2}x^+$ as the ``time'' conjugated to the Hamiltonian.

Another way of seeing this factor of $2$ is to note that $\frac{\partial}{\partial x^+}$ is associated with $P_+$ and defining $P_+$ involves
\be
\int \varepsilon_{+12-} \diff x^- \diff x^1 \diff x^2\,,
\ee
where the Levi-Civita tensor is normalized by
\be
\varepsilon_{+12-} = \frac{1}{2} \,,
\ee
when one uses the convention $x^+=x^0+x^3$. See e.g., Ref.~\cite{Brodsky:1997de}.} Our convention of the light-cone coordinates and the construction of the light-front Hamiltonian of QCD can be found in appendix~\ref{app:lfqcd}. The light-front Hamiltonian of QCD can be written as
\begin{align}
\label{eqn:H}
H & = \int \diff x^- \diff^2 x_\perp \bigg(
i \psi_+^\dagger \big(- \slashed{D}_\perp +im \big) \frac{1}{\partial^+} \big( \slashed{D}_\perp +im \big) \psi_+
-g \psi^\dagger_+ A^{-a}T^a \psi_+ \\
&\qquad\qquad\qquad\quad +\frac{1}{4}F_\perp^{ija} F_{\perp ij}^a -\frac{1}{8} ( \partial^+ A^{-a} )^2 + \frac{1}{2}( \partial^+ A^{i a}_\perp ) ( -\partial_i A^{-a} + g f^{abc} A^{-b} A_{\perp i}^c ) \bigg) \,, \nn 
\end{align}
where $i=1,2$ and $j=1,2$ denote the transverse components and are implicitly summed over. The $-$ component of the gauge field is not dynamical and is related to the dynamical components via
\be
\label{eqn:A-a}
A^{-a} = \frac{2}{\partial^+} \partial^i A^{ia}_\perp - \frac{2g}{\partial^{+2}} \Big(
f^{abc} ( \partial^+ A^{ib}_\perp ) A^{i c}_\perp - 2\psi_+^\dagger T^a \psi_+
\Big) \,,
\ee
where $\partial^{+2}=(\partial^+)^2$. The light-front Hamiltonian~\eqref{eqn:H} is time independent. The dynamical fields $\psi_+^i$ and $A_\perp^{ia}$ at zero time $x^+=0$ can be expanded in terms of creation and annihilation operators in momentum space 
\begin{align}
\label{eqn:mode_expansion}
\psi_+^i(x^+=0,x_\perp, x^-) &= \sum_{\sigma=\pm\frac{1}{2}} \int_{k^+>0} \frac{\diff k^+ \diff^2 k_\perp}{2(2\pi)^3 k^+} 
\Big( b^i(k,\sigma) u_+(k, \sigma) e^{-ik\cdot x} + d^{i\dagger}(k,\sigma) v_+(k,\sigma) e^{ik\cdot x} \Big)\,, \nn\\
A^{ib}_\perp(x^+=0, x_\perp, x^-) &= \sum_{\lambda=\pm} \int_{k^+>0}\frac{\diff k^+ \diff^2 k_\perp}{2(2\pi)^3 k^+} \Big(
a^b(k,\lambda) \varepsilon_\perp^i(\lambda) e^{-ik\cdot x} + a^{b\dagger}(k,\lambda) \varepsilon_\perp^{i*}(\lambda) e^{ik\cdot x}
\Big) \,,
\end{align}
where $a,b,d$ ($a^\dagger,b^\dagger,d^\dagger$) are annihilation (creation) operators for gluons, quarks and antiquarks respectively. Here $\sigma$ denotes quark spins, $\lambda$ represents gluon polarizations and $\epsilon_\perp(\lambda)$ is the corresponding polarization tensor in the transverse plane. The modes are constrained to have positive $k^+$ here, because we want to study collinear radiation processes in which all the daughter partons have large $+$ momenta, as their mother parton. Soft radiation processes can also happen in reality and involve the zero mode. The zero mode is known to be important for vacuum properties~\cite{Bender:1992yd,Ji:2020baz}. Since the LPM effect is mainly studied for collinear radiation, we will not discuss soft radiation and the effect of the zero mode here, which are left to future studies.

The Hamiltonian can be quantized by imposing the following (anti-)commutation relations:
\begin{align}
\big\{ b^i(k,\sigma), b^{j\dagger}(k',\sigma') \big\} = \big\{ d^i(k,\sigma), d^{j\dagger}(k',\sigma') \big\} = 2(2\pi)^3 k^+  \delta^{ij} \delta_{\sigma \sigma'} \delta^3(k-k')\,,\\
\big[ a^b(k, \lambda), a^{c\dagger}(k',\lambda') \big] = 2(2\pi)^3 k^+ \delta_{\lambda\lambda'} \delta^{bc} \delta^3(k-k') \,,\nn
\end{align}
where $\delta^3(k-k') = \delta(k^+ - k'^{+}) \delta^2(k_\perp-k'_\perp)$. In our notations, $\delta(x)$ represents a Dirac delta function while $\delta_x$ denotes a Kronecker delta function.

In the following, when we describe the soft momentum exchange between the QGP and high energy partons, which results in diffusion of the partons in the transverse plane, we will use a description based on a background gauge field $\bar{A}^{-a}$~\cite{Blaizot:2012fh}.\footnote{In the rest frame of a high energy parton, the nuclear medium is moving fast. It has been shown that the only non-zero component of the gauge field generated by the nuclear medium that affects the parton is the $-$ component~\cite{Gelis:2005pt}. Boosting back to the frame where the parton is moving fast only rescales the $-$ component and does not make the other vanishing components nonvanishing.} The $\bar{A}^{-a}$ field is classical and will be discussed in detail in section~\ref{sect:diffuse}. To incorporate the classical background field into the Hamiltonian, we simply apply the replacement
\be
\label{eqn:A+Abar}
A^{-a} \to A^{-a}+ \bar{A}^{-a} \,,
\ee
of which the right hand side is the new $-$ component of the gauge field appearing in the Hamiltonian, with $A^{-a}$ given by Eq.~\eqref{eqn:A-a} and $\bar{A}^{-a}$ the classical background field. In general, the classical background field depends on the light-cone time $x^+$, so under the replacement~\eqref{eqn:A+Abar} the light-front Hamiltonian becomes time dependent through $\bar{A}^{-a}$
\be
H \to H(x^+) = H[\bar{A}^{-a}(x^+)] \,.
\ee

The Hamiltonian can be split into three parts for studying the LPM effect in jet quenching:
\be
H(x^+) = H_{\rm kin} + H_{\rm diff}(x^+) + H_{\rm split} \,.
\ee
Here $H_{\rm kin}$ describes the free theory of quarks and gluons on the light front and induces a phase change for each parton in the time evolution. $H_{\rm diff}$ represents the interaction between the QGP medium and quarks/gluons in the system, which originates from Glauber exchanges induced by the background fields and results in the transverse momentum broadening of partons. $H_{\rm split}$ gives the interaction between quarks and gluons, describing the splitting process of $n$ high energy partons going into $n+1$ partons and the inverse process, i.e., recombination of partons. It is necessary to include recombination of partons in $H_{\rm split}$ for it to be Hermitian and for the time evolution to be unitary. 

To simulate the in-medium jet evolution on a digital quantum computer from $x^+=0$ to a time $x^+ \equiv 2t$,\footnote{Here $t$ is just a short hand notation for $x^+/2$ and should be distinguished from $x^0$ used in the definition of $x^+ = x^0+x^3$.} we decompose the total time length into $N_t$ small pieces with a step size $\Delta t = t/N_t$ and apply the standard Trotterization method:
\begin{align}
\Big( e^{ -i( H_{\rm kin} + H_{\rm diff} + H_{\rm split} ) \Delta t } \Big)^{N_t} |\Psi\rangle = \Big( \prod_{j} e^{-i H_j \Delta t} e^{\ml{O}((\Delta t)^2)} \Big)^{N_t} |\Psi\rangle \,,
\end{align}
where each $H_j$ is chosen such that we know how to construct the quantum circuit for it and $\sum_j H_j = H_{\rm kin} + H_{\rm diff} + H_{\rm split}$. The error $\ml{O}((\Delta t)^2)$ on the right hand side comes from nonzero commutators $[H_j,H_k]\neq0$ ($j\neq k$).
When $N_t$ is large, the correction term $\ml{O}((\Delta t)^2)$ can be neglected. Then simulating the in-medium jet evolution can be realized by constructing quantum gates implementing the Hamiltonian dynamics determined by each $H_j$. The convergence rate of the Trotterization can be further improved by including higher-order corrections. 

To write out matrix elements for each part of the Hamiltonian, we need to choose a basis of the Hilbert space to project the Hamiltonian. In the next subsection, we will explain the basis constructed from $n$-particle states in momentum space.

\subsection{Hilbert Space}
\label{sect:hilbert}
To formulate the Hamiltonian dynamics on a digital quantum computer, we need to first construct a basis of the physical Hilbert space and discretize it so that we can encode quantum states in terms of qubits and represent the Hamiltonian as quantum gates. We also want the Hamiltonian matrix to be diagonal or sparse, so we may be able to efficiently simulate them on a quantum computer. To this end, we use $n$-particle states in the light-front momentum space to construct the basis of the Hilbert space. A $1$-particle state can be labeled as
\be
\label{eqn:1-state}
\big| q/g,\, k^+>0,\, k_x,\, k_y,\, {\rm color},\, {\rm spin} \big\rangle : \quad \frac{b^{i\dagger}(k,\sigma)|0\rangle}{\sqrt{2(2\pi)^3k^+}} \,,~
\frac{d^{i\dagger}(k,\sigma)|0\rangle}{\sqrt{2(2\pi)^3k^+}}\,,~ \frac{a^{b\dagger}(k,\lambda)|0\rangle}{\sqrt{2(2\pi)^3k^+}}\,, 
\ee
which is obtained by applying a creation operator ($a^{b\dagger}$, $b^{i\dagger}$ or $d^{i\dagger}$) on the vacuum state. The normalization factor $1/\sqrt{2(2\pi)^3k^+}$ is chosen for later convenience. Here $q/g$ indicates whether the state is a quark or a gluon. There is no ghost state since in the light-front Hamiltonian formulation of QCD, the light-cone gauge $A^+=0$ is chosen and ghosts are decoupled from gluons. The momentum of a state is specified by the $+$ and transverse components: $(k^+,k_x,k_y)$. In the light-front approach, the $+$ component is always non-negative. Furthermore, we have constrained the Hilbert space to contain only states with positive $k^+$ since we focus on the LPM effect in collinear radiation (see the discussion below Eq.~\eqref{eqn:mode_expansion}). We leave the inclusion of the zero mode to future studies. A quark or an antiquark state has three degrees of freedom in color. We will label both states as $q$, i.e., a quark state and then differentiate them by the color degrees of freedom. In other words, a quark state has six degrees of freedom in color in our notation, which requires three qubits to encode.
A gluon state has eight degrees of freedom in color, which also requires three qubits to encode. The spin degree of freedom has two possibilities for both quark and gluon states, which needs one qubit to store. (For gluon states, by spin we mean the polarization.) The quark and gluon states are normalized as
\begin{align}
\label{eqn:1particle_norm}
\big\langle q,\, k_1^+,\, k_{1\perp},\, i_1,\, \sigma_1 \big| q,\, k_2^+,\, k_{2\perp},\, i_2,\, \sigma_2 \big\rangle
& = \delta(k_1^+-k_2^+) \delta^2(k_{1\perp} - k_{2\perp}) \delta_{i_1i_2} \delta_{\sigma_1 \sigma_2} \,, \\
\big\langle g,\, k_1^+,\, k_{1\perp},\, a_1,\, \lambda_1 \big| g,\, k_2^+,\, k_{2\perp},\, a_2,\, \lambda_2 \big\rangle 
& = \delta(k_1^+-k_2^+) \delta^2(k_{1\perp} - k_{2\perp}) \delta_{a_1a_2} \delta_{\lambda_1 \lambda_2} \,,\nn
\end{align}
where $i$ and $a$ denote the fundamental and adjoint color indices and $\sigma$ and $\lambda$ represent the spin degrees of freedom.

A general $n$-particle basis state can be written as
\be
\label{eqn:n-state}
\bigotimes_{i=1}^n \big| q/g,\, k^+>0,\, k_x,\, k_y,\, {\rm color},\, {\rm spin} \big\rangle_i
\ee
where the $i$-th and $j$-th states ($i\neq j$) generally differ in momenta and/or quantum numbers. A single $n$-particle basis state cannot be physical, since physical states of multiple particles need to be properly (anti)symmetrized. For studying the LPM effect, if we start with a $1$-particle state, the Hamiltonian evolution will guarantee the final state is properly (anti)symmetrized, since the (anti)symmetric properties of the boson (fermion) creation and annihilation operators are already included in the construction of the Hamiltonian. To simulate the time evolution of a more general initial state for jet quenching, the initial state needs proper (anti)symmetrization. Then the Hamiltonian evolution will lead to a properly (anti)symmetrized final state.

The basis of the Hilbert space consists of $n$-particle states for all integers $n$.
To simulate the LPM effect in processes with $N$ particles in total (which can happen in cases with one initial parton having $N-1$ splittings or two initial partons having $N-2$ splittings, etc), we need to include all the $1$-particle states, $2$-particle states and all the way to $N$-particle states in the basis, in order to describe the system. In principle, states with more than $N$ particles can also affect the time evolution through loop effects, i.e., they only exist as intermediate states and are absent in the final states measured. To reduce the loop effects, one may truncate the states with a much higher particle number such as $2N$.

Before moving on to the detailed discussion of the Hamiltonian, we give an estimate of the qubit cost. For each $1$-particle state, to distinguish a quark state from a gluon one, two degrees of freedom are required. We also need eight color degrees of freedom (a quark state only has six degrees of freedom in color but the more demanding case in terms of the register resource is given by a gluon state) and two spin degrees of freedom. To encode the basis states on a digital quantum computer, we need to truncate and discretize momenta. We assume the ranges of momenta are given by
\begin{align}
k^+\in (0, K_{\rm max}^+ ]\,,\quad\qquad k_x\in [-K_{\rm max}^\perp , K_{\rm max}^\perp ]\,,\quad\qquad k_y\in [-K_{\rm max}^\perp , K_{\rm max}^\perp ] \,.
\end{align}
With step sizes set by $\Delta k^+,\Delta k^\perp,\Delta k^\perp$ for the $+,x,y$ components respectively, the number of degrees of freedom in momenta is given by $N^+N_\perp^2$ where
\be
N^+ = \frac{K_{\rm max}^+}{\Delta k^+} \,,\quad\qquad N_\perp = \frac{2K_{\rm max}^\perp}{\Delta k^\perp} +1 \,.
\ee
Therefore, the number of qubits needed to represent all the $1$-particle states is estimated as
\be
\log_2( 2^5N^+N_\perp^2 ) \,.
\ee
Encoding all the $n$-particle states (fixed $n$) requires a number of qubits given by
\be
\log_2 \Big( \big( 2^5N^+N_\perp^2 \big)^n \Big) \,.
\ee
If we want to study the LPM effect in processes with $N$ particles in total with loop effects from states of more than $N$ particles neglected, we have to include all the $n$-particle states where $n=1,2,\cdots,N$. The total number of qubits needed in the register then is
\be
\label{eqn:Nqubits}
\log_2\Big( \sum_{n=1}^N \big(2^5N^+N_\perp^2\big)^n \Big) \,.
\ee
One can reduce the qubit cost in the register for special cases. For example, if we study the LPM effect in a process initiated by one parton, the qubit cost is given by
\be
\log_2 \!\Big( \sum_{n=1}^N  \frac{1}{n!}\big(2^5N^+N_\perp^2\big)^n \Big) \,,
\ee
where the $1/{n!}$ factor originates from the constraint that $k^+ > 0$ and the total $+$ component of the momentum is conserved in each splitting. In general, the qubit cost is estimated by Eq.~\eqref{eqn:Nqubits}. If we choose $N^+=N_\perp=100$, the cost of qubit numbers is about $50$ for one initial parton having one splitting and about $75$ for two splittings, according to Eq.~\eqref{eqn:Nqubits}. To go beyond the scope of current studies of the LPM effect, we will simulate the case with one initial parton and three splittings, which needs about $100$ qubits. In the NISQ era, a quantum simulation using $100$ qubits is possible, but error mitigation techniques are necessary for physical applications. Fault-tolerant quantum computers with $100$ qubits may become available in the near future. 

\subsection{State Initialization and Measurement}
For studies of the LPM effect, the initial state contains a number of partons with specific momenta, colors and spins and each of them can be either a quark or a gluon. 
Therefore the initial state is just a linear combination of $n$-particle basis states, properly (anti)symmetrized, and it can be easily initialized in the qubit register. The initialization is much simpler than cases where the initial states involve hadrons such as protons, which are nontrivial linear superposition of all $n$-particle states, with coefficients that are a priori unknown. Adiabatic state preparation has been proposed to prepare such complicated initial states by starting with free particles and then slowly turning on the interaction~\cite{farhi2000quantum}. Here we only focus on quantum simulation of the LPM effect for one or a number of initial partons that are off-shell. Quantum simulation of the whole heavy ion collision where the initial state consists of two heavy nuclei, which are complicated nuclear bound states, is beyond the scope of our current study.

The final state contains multi-particle states due to splitting in the time evolution. To extract the radiation spectrum from the final state, we project the final state onto a specific $n$-particle state with given momenta, colors and spins, which corresponds to a specific state in the computational basis or a linear combination of the basis states with known coefficients. So the measurement is simply projective. The time evolution and the projective measurement need repeating multiple times since the quantum state collapses after the measurement. After collecting enough statistics, one will be able to calculate the radiation spectrum. Color and spin degrees of freedom may be averaged, depending on the radiation spectrum of interest.

\section{Matrix Elements of the Light-Front Hamiltonian of QCD}
\label{sect:matrix}
In this section, we will write down matrix elements of the light-front Hamiltonian of QCD in the computational basis introduced in the previous section, for the kinetic $H_{\rm kin}$, diffusion $H_{\rm diff}$ and splitting $H_{\rm split}$ terms. As we will see, they are either diagonal or sparse in this basis, which is important for potentially efficient quantum simulation.

\subsection{Kinetic Term}
\label{sect:kinetic}
The kinetic energy parts of the Hamiltonian are given by
\begin{align}
H_{f,\,{\rm kin}} &= \sum_i \sum_{\sigma=\pm\frac{1}{2}} \int_{k^+>0} \frac{\diff k^+ \diff^2 k_\perp}{2(2\pi)^3 k^+} \frac{{\bs k}_\perp^2}{k^+} \Big( b^{i\dagger}(k,\sigma) b^{i}(k,\sigma)
+d^{i\dagger}(k,\sigma) d^{i}(k,\sigma) \Big) \,,\\
H_{g,\,{\rm kin}} &= \sum_b \sum_{\lambda=\pm} \int_{k^+>0} \frac{\diff k^+ \diff^2k_\perp }{2(2\pi)^3k^+} \frac{{\bs k}_\perp^2}{k^+} a^{b\dagger}(k,\lambda) a^b(k,\lambda) \,, \nn
\end{align}
for quarks and gluons respectively. The derivation of these terms can be found in appendix~\ref{app:lfqcd}. Matrix elements of the kinetic terms in the basis of Eq.~\eqref{eqn:1-state} are given by
\begin{align}
\label{eqn:1particle_Hkin}
\big\langle q,\, k_1^+,\, k_{1\perp},\, i_1,\, \sigma_1 \big| H_{q,\,{\rm kin}} \big| q,\, k_2^+,\, k_{2\perp},\, i_2,\, \sigma_2 \big\rangle
& =  \frac{{\bs k}_{1\perp}^2}{k_1^+} \delta(k_1^+-k_2^+) \delta^2(k_{1\perp} - k_{2\perp}) \delta_{i_1i_2} \delta_{\sigma_1 \sigma_2} \,, \nn\\
\big\langle g,\, k_1^+,\, k_{1\perp},\, a_1,\, \lambda_1 \big| H_{g,\,{\rm kin}} \big| g,\, k_2^+,\, k_{2\perp},\, a_2,\, \lambda_2 \big\rangle 
& = \frac{{\bs k}_{1\perp}^2}{k_1^+} \delta(k_1^+-k_2^+) \delta^2(k_{1\perp} - k_{2\perp}) \delta_{a_1a_2} \delta_{\lambda_1 \lambda_2} \,. 
\end{align}
These matrix elements can be easily generalized to the case with $n-$particle states that are symbolically represented as $\bigotimes_{i=1}^n |i\rangle \equiv |123\cdots n\rangle $ where $|i\rangle$ labels the $i$-th particle state $ |q/g,\, k^+,\, k_\perp,\, {\rm color},\, {\rm spin}\rangle_i$:
\begin{align}
&\big\langle 1'2'3'\cdots n' \big| H_{\rm kin} \big| 123\cdots n \big\rangle \\
& \quad = \sum_{i=1}^n \big\langle i' \big| H_{\rm kin} \big| i \big\rangle \big\langle 1'2'3'\cdots (i'-1)(i'+1)\cdots n' \big| 123\cdots (i-1)(i+1)\cdots n \big\rangle \nn \\
& \quad = \sum_{i=1}^n \frac{{\bs k}_{i\perp}^2}{k_i^+} \delta_{1'1} \delta_{2'2} \cdots \delta_{n'n}\,, \nn
\end{align}
where $\langle i'| H_{\rm kin} | i \rangle$ is given by Eq.~\eqref{eqn:1particle_Hkin} and $\delta_{i'i}$ is a short hand notation for the production of the Dirac delta functions of momenta and the Kronecher delta functions of colors and spins for parton $i'$ and parton $i$. No cross terms of the form $\langle i' | H_{\rm kin} | j \rangle$ ($i\neq j$) appear in the matrix elements involving two $n$-particle states. We want to emphasize this is just a result of our choice of the computational basis. Such cross terms $\langle i' | H_{\rm kin} | j \rangle$ ($i\neq j$) are physical and can be accounted for when the quantum state is properly (anti)symmetrized, i.e., such cross terms will show up in the matrix elements of $H_{\rm kin}$ involving two physical states.

With our choice of the computational basis, the kinetic term $H_{{\rm kin}}$ is diagonal and the diagonal element is given by the light-cone energy of the corresponding $n$-particle state:
\be
\sum_{i=1}^n \frac{{\bs k}_{i\perp}^2}{k_i^+} \,,
\ee
where the summation is over all the constitutes in the $n$-particle state.
The time evolution induced by the kinetic Hamiltonian is just a phase, which can be efficiently simulated on a quantum computer.
We will give an explicit construction of the quantum circuit for the kinetic evolution in sections~\ref{sect:toy} and~\ref{sect:gluon}.

\subsection{Diffusion Term}
\label{sect:diffuse}
To describe the diffusion process in the transverse plane caused by the soft momentum transfer from the medium, we replace the $A^{-a}$ field in Eq.~\eqref{eqn:H} with $A^{-a}+\bar{A}^{-a}$ where $A^{-a}$ is determined by the dynamical field degrees of freedom as shown in  Eq.~\eqref{eqn:A-a} and $\bar{A}^{-a}$ denotes a classical background field. 
We follow Ref.~\cite{Blaizot:2012fh} to describe the medium as a source of the background gauge field $\bar{A}^{-a}$, which can be time dependent.
We assume the background field is $x^-$ independent
\be
\bar{A}^{-a}(x^+, x^-, x_\perp) = \bar{A}^{-a}(x^+, x^-=0, x_\perp) \,,
\ee
since a high energy parton has a large $+$ component of momentum $k^+$, thus only probing the medium at a small $x^- \sim {1}/{k^+}$. From now on, we will omit the dependence of the background gauge field on the $x^-$ coordinate. 

We further assume the random background field satisfies the two-point correlation
\be
\label{eqn:Abar_corr_space}
\big\langle \bar{A}^{-a}(x^+, x_\perp)
\bar{A}^{-b}(y^+, y_\perp) \big\rangle = \delta^{ab} \delta(x^+ - y^+) \gamma({\bs x}_\perp - {\bs y}_\perp) \,.
\ee
The random background fields at different light-cone times are assumed independent. One can replace the $\delta(x^+ - y^+)$ function with some other functions in $x^+ - y^+$ to describe some correlation between the random background fields at different times. The $\gamma({\bs x}_\perp - {\bs y}_\perp)$ function accounts for nontrivial correlation between background fields at the same light-cone time but different transverse positions. The model used in Ref.~\cite{Blaizot:2012fh} for a hot nuclear environment is motivated from the hard-thermal-loop calculation of the Landau damping phenomenon
\be
\label{eqn:gamma_HTL}
\gamma({\bs x}_\perp - {\bs y}_\perp) = g^2 \int\frac{\diff^2 q_\perp}{(2\pi)^2} e^{i{\bs q}_\perp \cdot ({\bs x}_\perp - {\bs y}_\perp)} \frac{\pi T m_D^2}{({\bs q}_\perp^2+m_D^2)^2} \,,
\ee
where $T$ denotes the temperature of the plasma and $m_D$ is the Debye mass. Our framework of the quantum simulation for jet quenching is general and the construction does not depend on any specific form of the correlation function. For cold nuclear environments, one can replace Eq.~\eqref{eqn:gamma_HTL} with corresponding correlation functions.

To transform to momentum space, we use the Fourier transform defined by
\be
\bar{A}^{-a}(x^+, x_\perp) = \int \frac{\diff q^- \diff^2q_\perp}{2(2\pi)^3} e^{-iq^-x^+/2 - iq_\perp \cdot x_\perp} \bar{A}^{-a}(q^-, q_\perp) \,.
\ee
Applying $\int\diff R^+ \diff^2R_\perp \int \diff r^+ \diff^2 r_\perp\, e^{ik^-r^+/2 + ik_\perp\cdot r_\perp}$ to Eq.~\eqref{eqn:Abar_corr_space}, where $R^+ = (x^++y^+)/2$, $r^+ = x^+-y^+$ and similarly for the transverse components, we find the correlation function of the background gauge field in momentum space is given by
\be
\big\langle \bar{A}^{-a}(k^-, k_\perp)
\bar{A}^{-b}(-k^-, -k_\perp) \big\rangle = \delta^{ab} \gamma({\bs k}_\perp) \int\diff R^+ \diff^2R_\perp \,.
\ee
It turns out to be easier to use the mixed space representation
\be
\label{eqn:AAmixed}
\big\langle \bar{A}^{-a}(x^+, k_\perp)
\bar{A}^{-b}(y^+, -k_\perp) \big\rangle = \delta^{ab} \delta(x^+-y^+) \gamma({\bs k}_\perp) \int \diff^2R_\perp \,,
\ee
where $\int \diff^2R_\perp$ gives the area of the transverse plane.

The quark diffusion term in the Hamiltonian can be obtained from terms of the form $\psi_+^\dagger \bar{A}^- \psi_+$. Since the background field $\bar{A}^{-a}$ is $x^-$ independent, we find
\be
\partial^+ \bar{A}^{-a}(x^+,x_\perp) = \frac{\partial}{\partial x_+} \bar{A}^{-a}(x^+,x_\perp) = 2 \frac{\partial}{\partial x^-} \bar{A}^{-a}(x^+,x_\perp) = 0\,.
\ee
Therefore the term $(\partial^+ A^{-a} + \partial^+ \bar{A}^{-a})^2$ in the Hamiltonian~\eqref{eqn:H} is irrelevant to the quark diffusion process, which is not obvious from the beginning, since $A^{-a}$ contains $\psi_+^\dagger T^a \psi_+$. With this simplification, the quark diffusion Hamiltonian can be written as
\begin{align}
H_{q,\,{\rm diff}} &= -g\int\diff x^- \diff^2 x_\perp \psi_+^\dagger(x) \bar{A}^{-a}(x)T^a \psi_+(x) \\
&= -g\int\diff x^- \diff^2 x_\perp \sum_{\sigma_1,\sigma_2} \int_{k_1^+>0} \frac{\diff k_1^+ \diff^2k_{1\perp}}{2(2\pi)^3k^+_1}
\int_{k_2^+>0} \frac{\diff k_2^+ \diff^2k_{2\perp}}{2(2\pi)^3k^+_2} \nn\\
& \qquad\quad \Big( b^{i\dagger}(k_1,\sigma_1) u_+^\dagger(k_1, \sigma_1) e^{ik_1\cdot x} + d^{i}(k_1,\sigma_1) v_+^\dagger(k_1,\sigma_1) e^{-ik_1\cdot x} \Big) \bar{A}^{-a}(x)T^a_{ij} \nn\\
& \qquad\quad 
\Big( b^j(k_2,\sigma_2) u_+(k_2, \sigma_2) e^{-ik_2\cdot x} + d^{j\dagger}(k_2,\sigma_2) v_+(k_2,\sigma_2) e^{ik_2\cdot x} \Big) \,.\nn
\end{align}
Since $\bar{A}^a(x)$ is $x^-$ independent, the integration over $x^-$ can be carried out to give a delta function in the $+$ component of the momenta:
\begin{align}
\label{eqn:integral_x^-}
\int \diff x^- e^{i(k_1^+ x^- \pm k_2^+ x^-)/2} = 2(2\pi) \delta(k_1^+ \pm k_2^+) \,.
\end{align}
Since both $k_1^+>0$ and $k_2^+>0$, the delta function with the plus sign vanishes. Then we have
\begin{align}
H_{q,\,{\rm diff}}=& -g \sum_{\sigma_1,\sigma_2} \int_{k_1^+>0} \frac{\diff k^+_1}{2(2\pi)(k_1^+)^2} \int \frac{\diff^2 k_{1\perp}}{(2\pi)^2}
\int \frac{\diff^2 k_{2\perp}}{(2\pi)^2} \\
& \qquad\quad \Big( b^{i\dagger}(k_1,\sigma_1) T_{ij}^a b^j(k_2,\sigma_2) u_+^\dagger(k_1,\sigma_1) u_+(k_2,\sigma_2) \bar{A}^{-a}(x^+, {\bs k}_{1\perp} - {\bs k}_{2\perp})\nn\\
& \quad\quad  +  d^{i}(k_1,\sigma_1) T_{ij}^a d^{j\dagger}(k_2,\sigma_2) v_+^\dagger(k_1,\sigma_1) v_+(k_2,\sigma_2) \bar{A}^{-a}(x^+, -{\bs k}_{1\perp} + {\bs k}_{2\perp}) \Big)\bigg|_{k_2^+=k_1^+}\nn \,.
\end{align}
When $k_1^+ \gg k_{1\perp}, k_{2\perp}, m$, we have
\begin{align}
u_+^\dagger(k_1,\sigma_1) u_+(k_2,\sigma_2) \big|_{k_1^+=k_2^+} &= k_1^+\delta_{\sigma_1\sigma_2} + \ml{O}\Big(\frac{k_{1\perp}}{k_1^+}, \frac{k_{2\perp}}{k_1^+}, \frac{m}{k_1^+} \Big) \,,\\
v_+^\dagger(k_1,\sigma_1) v_+(k_2,\sigma_2) \big|_{k_1^+=k_2^+} &= k_1^+\delta_{\sigma_1\sigma_2} + \ml{O}\Big(\frac{k_{1\perp}}{k_1^+}, \frac{k_{2\perp}}{k_1^+}, \frac{m}{k_1^+} \Big) \,, \nn
\end{align}
which means in the high energy limit, the spin of a quark does not change under a small transverse perturb. Under the high energy approximation, we take the leading terms and obtain
\begin{align}
\label{eqn:Hq_diff}
&H_{q,\,{\rm diff}}= \\
& -g \sum_{\sigma} \int_{k_1^+>0} \frac{\diff k^+_1}{2(2\pi)k_1^+} \int\frac{\diff^2 k_{1\perp}}{(2\pi)^2}
\int\frac{\diff^2 k_{2\perp}}{(2\pi)^2} \Big( b^{i\dagger}(k_1,\sigma) T_{ij}^a b^j(k_2,\sigma) \bar{A}^{-a}(x^+, {\bs k}_{1\perp} - {\bs k}_{2\perp}) \nn\\
& \qquad\qquad\qquad\qquad\qquad\qquad\qquad + d^{i}(k_1,\sigma) T_{ij}^a d^{j\dagger}(k_2,\sigma) \bar{A}^{-a}(x^+, -{\bs k}_{1\perp} + {\bs k}_{2\perp}) \Big)\bigg|_{k_2^+=k_1^+}\nn \,,
\end{align}
in which up to a constant, we can switch the order of $d^{i}(k_1,\sigma)$ and $d^{j\dagger}(k_2,\sigma)$ in the second term and obtain a negative sign due to the anticommutation relation.

The gluon diffusion Hamiltonian can be similarly worked out, which involves terms of the form $A_\perp  \bar{A}^- A_\perp$. First, the $F_\perp^2$ term in the Hamiltonian~\eqref{eqn:H} does not involve any $\bar{A}^-$ field, so it is irrelevant for the gluon diffusion process. Furthermore the term $(\partial^+ A^{-a} + \partial^+ \bar{A}^{-a})^2$ in Eq.~\eqref{eqn:H} is also irrelevant since the background gauge field $\bar{A}^{-a}$ is $x^-$ independent. The remaining part of the gluon Hamiltonian for consideration is
\begin{align}
\int \diff x^- \diff^2 x_\perp \,\frac{1}{2}( \partial^+ A^{i a}_\perp ) \big( -\partial_i (A^{-a}+\bar{A}^{-a}) + g f^{abc} (A^{-b}+\bar{A}^{-b}) A_{\perp i}^c \big) \,.
\end{align}
Integration by parts and using $\partial^+ \bar{A}^{-a} = 0$ lead to the following Hamiltonian describing the gluon diffusion process (we omit terms without any $\bar{A}^{-a}$)
\begin{align}
H_{g,\,{\rm diff}} &= -\frac{g}{2} f^{abc} \int \diff x^- \diff^2 x_\perp \, A^{i a}_\perp(x) \bar{A}^{-b}(x) \partial^+ A_{\perp i}^c(x) \\
& = -\frac{g}{2} f^{abc} \int \diff x^- \diff^2 x_\perp
\sum_{\lambda_1,\lambda_2} \int_{k^+_1>0}\frac{\diff k^+_1 \diff^2 k_{1\perp}}{2(2\pi)^3 k^+_1}
\int_{k^+_2>0}\frac{\diff k^+_2 \diff^2 k_{2\perp}}{2(2\pi)^3 k^+_2} \nn\\
& \qquad\qquad \Big(
a^a(k_1,\lambda_1) \varepsilon_\perp^i(\lambda_1) e^{-ik_1\cdot x} + a^{a\dagger}(k_1,\lambda_1) \varepsilon_\perp^{i*}(\lambda_1) e^{ik_1\cdot x} \Big) \bar{A}^{-b}(x) \nn\\
& \qquad\qquad \Big( -ik_2^+
a^c(k_2,\lambda_2) \varepsilon_{\perp i}(\lambda_2) e^{-ik_2\cdot x} + ik_2^+ a^{c\dagger}(k_2,\lambda_2) \varepsilon_{\perp i}^{*}(\lambda_2) e^{ik_2\cdot x} \Big) \,.\nn
\end{align}
Since the background gauge field $\bar{A}^{-a}$ is $x^-$ independent, we can use Eq.~\eqref{eqn:integral_x^-} to show 
\begin{align}
H_{g,\,{\rm diff}} &= -\frac{ig}{2} f^{abc} \sum_{\lambda_1,\lambda_2} \int_{k_1^+>0}\frac{\diff k_1^+}{2(2\pi)k_1^+} \int\frac{\diff^2 k_{1\perp}}{(2\pi)^2} \int\frac{\diff^2 k_{2\perp}}{(2\pi)^2} \\
& \qquad\qquad\qquad \Big( a^a(k_1,\lambda_1) \varepsilon_\perp^i(\lambda_1) a^{c\dagger}(k_2,\lambda_2) \varepsilon_{\perp i}^{*}(\lambda_2) \bar{A}^{-b}(x^+, -{\bs k}_{1\perp} + {\bs k}_{2\perp}) \nn\\
& \qquad\qquad\ \quad - a^{a\dagger}(k_1,\lambda_1) \varepsilon_\perp^{i*}(\lambda_1) a^c(k_2,\lambda_2) \varepsilon_{\perp i}(\lambda_2) \bar{A}^{-b}(x^+, {\bs k}_{1\perp} - {\bs k}_{2\perp}) \Big) \bigg|_{k_1^+ = k_2^+} \,. \nn
\end{align}
In the high energy limit $k_1^+ \gg k_{1\perp}, k_{2\perp}, m$, the polarizations $\lambda_1$ and $\lambda_2$ are defined with respect to the same axis along which $k_1^+$ is aligned. So we have the simplification
\be
\sum_{i=1,2}\epsilon_\perp^i (\lambda_1) \epsilon_{\perp i}^* (\lambda_2) = - \delta_{\lambda_1 \lambda_2} \,.
\ee
Then we have
\begin{align}
\label{app:Hg_diff}
& H_{g,\,{\rm diff}} \\
&= \frac{ig}{2} f^{abc} \sum_{\lambda} \int_{k_1^+>0}\frac{\diff k_1^+}{2(2\pi)k_1^+} \int\frac{\diff^2 k_{1\perp}}{(2\pi)^2} \int\frac{\diff^2 k_{2\perp}}{(2\pi)^2} \Big(
a^a(k_1,\lambda) a^{c\dagger}(k_2,\lambda)  \bar{A}^{-b}(x^+, -{\bs k}_{1\perp} + {\bs k}_{2\perp}) \nn\\
&\qquad\qquad\qquad\qquad\qquad\qquad\qquad\qquad - a^{a\dagger}(k_1,\lambda)  a^c(k_2,\lambda)  \bar{A}^{-b}(x^+, {\bs k}_{1\perp} - {\bs k}_{2\perp})
\Big)\bigg|_{k_1^+ = k_2^+} \,. \nn
\end{align}
We are allowed to switch the order of $a^a(k_1,\lambda)$ and $a^{c\dagger}(k_2,\lambda) $ in the first term of $H_{g,\,{\rm diff}}$ since the commutator is proportional to $\delta^{ac}$ which is symmetric and thus vanishing when contracted with $f^{abc}$.

With Eqs.~\eqref{eqn:Hq_diff} and~\eqref{app:Hg_diff} describing the transverse diffusion processes for quarks and gluons, we can write out the matrix elements of the diffusion Hamiltonian
\begin{align}
\label{eqn:1particle_Hdiff}
&\big\langle q,\, k_1^+,\, k_{1\perp},\, i_1,\, \sigma_1 \big| H_{q,\,{\rm diff}}(x^+) \big| q,\, k_2^+,\, k_{2\perp},\, i_2,\, \sigma_2 \big\rangle \\
=& \begin{cases}
-\frac{g}{(2\pi)^2}\delta(k_1^+-k_2^+) \delta_{\sigma_1\sigma_2} T^a_{i_1i_2} \bar{A}^{-a}(x^+, {\bs k}_{1\perp} - {\bs k}_{2\perp})  \quad {\rm for\ quark} \\
+\frac{g}{(2\pi)^2}\delta(k_1^+-k_2^+) \delta_{\sigma_1\sigma_2} T^a_{i_2i_1} \bar{A}^{-a}(x^+, {\bs k}_{1\perp} - {\bs k}_{2\perp})    \quad {\rm for\ antiquark}
\end{cases}
\,,\nn \\
&\big\langle g,\, k_1^+,\, k_{1\perp},\, a_1,\, \lambda_1 \big| H_{g,\,{\rm diff}}(x^+) \big| g,\, k_2^+,\, k_{2\perp},\, a_2,\, \lambda_2 \big\rangle \nn \\
=&\, \frac{ig}{ 2(2\pi)^2}\delta(k_1^+-k_2^+) \delta_{\lambda_1\lambda_2} \big(f^{a_2ba_1}-f^{a_1ba_2}\big) \bar{A}^{-b}(x^+, {\bs k}_{1\perp} - {\bs k}_{2\perp}) \,.\nn
\end{align}
These matrices are Hermitian if we have $\bar{A}^{-a}({\bs k}_\perp) = \bar{A}^{-a}(-{\bs k}_\perp)$. In these matrix elements, nontrivial color rotations occur in addition to the transverse momentum exchange.

It is easy to generalize the matrix elements for $n$-particle states $\bigotimes_{i=1}^n |i\rangle \equiv |123\cdots n\rangle $ where $|i\rangle$ labels the $i$-th particle state $ |q/g,\, k^+,\, k_\perp,\, {\rm color},\, {\rm spin}\rangle_i$. Since the diffusion process does not change the number of particles in the state and only changes the transverse momentum and color of the state, a matrix element involving two states with different particle numbers vanishes
\be
\big\langle 1'2'3'\cdots n'\big| H_{\rm diff} \big| 123\cdots m \big\rangle = 0\,,\qquad {\rm if\ } n\neq m\,.
\ee
When the two states have the same number of particles, we have
\begin{align}
\label{eqn:Hdiff_multi}
& \big\langle 1'2'3'\cdots n'\big| H_{\rm diff} \big| 123\cdots n \big\rangle \\
& \quad = \sum_{i=1}^n \big\langle i' \big| H_{\rm diff} \big| i \big\rangle \big\langle 1'2'3'\cdots (i'-1)(i'+1)\cdots n' \big| 123\cdots (i-1)(i+1)\cdots n \big\rangle \nn\\
& \quad = \sum_{i=1}^n \big\langle i' \big| H_{\rm diff} \big| i \big\rangle \delta_{1'1}\delta_{2'2}\cdots \delta_{(i'-1)(i-1)}\delta_{(i'+1)(i+1)}\cdots \delta_{n'n} \,. \nn
\end{align}
Cross terms of the form $\langle i' | H_{\rm diff} | j \rangle$ ($i\neq j$) are accounted for by properly (anti)symmetrized quantum states, as in the case of the kinetic term discussed above. 
The matrix elements between two states with different $k^+_i$s, spins or polarizations also vanish, no matter whether they have the same number of particles or not. Therefore, the matrix representing the diffusion Hamiltonian is sparse and thus we expect that encoding it on a quantum computer does not require an exponential number of gates.

The diffusion Hamiltonian that we have constructed is general and valid not only for background fields satisfying Eq.~\eqref{eqn:AAmixed}, but also for other background fields that satisfy certain higher-point correlation functions, which will only affect our sampling method when generating the background fields. Once the classical background fields are sampled at each time step, they can be plugged into the diffusion Hamiltonian constructed above. In section~\ref{sect:sample}, we will discuss how to sample the background fields according to Eq.~\eqref{eqn:AAmixed}.

\subsection{Splitting Term}
\label{sect:split}
Finally we work out the matrix elements of the  Hamiltonian describing the parton splitting process and its inverse. The full Hamiltonian~\eqref{eqn:H} contains both $1\to2$ and $1\to3$ splittings, as well as $2\to1$, $2\to2$ and $3\to1$ processes. For simplicity, we will focus on the $1\to2$ splitting and its inverse process in this paper. The Hamiltonian for the other processes is either one order higher in the coupling strength $g$ or at least one order higher in the inverse of the large longitudinal momentum $\frac{1}{\partial^+}$ than the $1\to2$ splitting. Therefore, these $1\to3$, $2\to2$ and $3\to1$ processes are suppressed in the high energy limit, either by the coupling strength or by the large longitudinal momentum $1/k^+$. For completeness, all the operators in the Hamiltonian~\eqref{eqn:H} describing splitting processes are listed in appendix~\ref{app:split}, organized by powers of $g$ and $\frac{1}{\partial^+}$.

The $1\to2$ splitting and its inverse process that involve quarks happen at the order $\ml{O}(\frac{g}{\partial^+})$. The relevant Hamiltonian is
\begin{align}
H_{q,\,{\rm split}} &= -g \int\diff x^- \diff x^2_\perp \\
& \qquad\qquad \bigg[ \psi_+^\dagger A_{\perp i} \gamma^i \gamma^j \Big(\frac{\partial_{\perp j}}{\partial^+} \psi_+\Big) + \Big( \frac{\partial_{\perp i}}{\partial^+} \psi_+^\dagger \Big) A_{\perp j} \gamma^i \gamma^j \psi_+ +2 \psi_+^\dagger T^a \psi_+ \Big( \frac{\partial^i}{\partial^+} A^{ia}_\perp \Big) \bigg] \,, \nn
\end{align}
where we have neglected terms proportional to the quark mass $m$. The $1\to2$ splitting and its inverse with three gluons involved start to occur at the order $\ml{O}(g)$. In other words, the $1\to2$ splitting with quarks involved is suppressed by one power in $\frac{1}{\partial^+}$ with respect to that with only gluons involved and thus suppressed in the high energy limit. Collecting relevant terms shown in appendix~\ref{app:split}, we find the splitting Hamiltonian with three gluons involved can be written as
\begin{align}
H_{g,\,{\rm split}} &= gf^{abc} \int\diff x^- \diff x^2_\perp \bigg[ \big( \partial^+ A_\perp^{ia} \big) \Big( \frac{\partial^j}{\partial^+} A_\perp^{jb} \Big) A_{\perp i}^{c} - \big( \partial^i A_\perp^{ja} \big) A_{\perp i}^b A_{\perp j}^c
\bigg] \,. 
\end{align}

The matrix elements of the $1\to2$ splitting for a quark or a gluon are given by
\begin{align}
\label{eqn:1to2_Hsplit}
&\big\langle q, k_2^+, k_{2\perp}, i_2, \sigma_2 ; g, q^+, q_\perp, a, \lambda \big| H_{q,\,{\rm split}} \big| q, k_1^+, k_{1\perp}, i_1, \sigma_1  \big\rangle  \\[4pt] 
&\quad = -\frac{g}{\sqrt{2(2\pi)^3q^+k_1^+k_2^+}}\delta(k_1^+ - k_2^+ - q^+) \delta^2(k_{1\perp} - k_{2\perp} - q_\perp)\nn \\
&\qquad \times \bar{u}(k_2,\sigma_2) \bigg( \epsilon_\perp^i \gamma^i\gamma^j \frac{k_{1\perp}^j}{k_1^+} T^a_{i_2i_1} + \frac{k_{2\perp}^i}{k_2^+}\gamma^i\gamma^j \epsilon_\perp^j T^a_{i_2i_1}
+ 2T^a_{i_2i_1} \frac{q_\perp^i}{q^+} \epsilon_\perp^i 
\bigg) u(k_1,\sigma_1) \,, \nn\\
&\big\langle g, -k_2^+, -k_{2\perp}, a_2, \lambda_2 ; g, -k_3^+, -k_{3\perp}, a_3, \lambda_3 \big| H_{g,\,{\rm split}} \big| g, k_1^+, k_{1\perp}, a_1, \lambda_1  \big\rangle \nn \\[4pt] 
& \quad = -\frac{ig}{\sqrt{2(2\pi)^3k_1^+k_2^+k_3^+}} f^{a_1a_2a_3} \delta(k_1^+ + k_2^+ + k_3^+) \delta^2(k_{1\perp} + k_{2\perp} + k_{3\perp})\nn \\
&\qquad \bigg( k_1^+\epsilon_\perp^i(\lambda_1) \Big[ \frac{k_{2\perp}^j }{k_2^+} \epsilon_\perp^j(\lambda_2) \epsilon_{\perp i}(\lambda_3)
- \frac{k_{3\perp}^j }{k_3^+} \epsilon_\perp^j(\lambda_3) \epsilon_{\perp i}(\lambda_2)
\Big] 
+ k_2^+\epsilon_\perp^i(\lambda_2) \Big[ \frac{k_{3\perp}^j }{k_3^+} \epsilon_\perp^j(\lambda_3) \epsilon_{\perp i}(\lambda_1) \nn\\
&\qquad - \frac{k_{1\perp}^j }{k_1^+} \epsilon_\perp^j(\lambda_1) \epsilon_{\perp i}(\lambda_3)
\Big] 
+ k_3^+\epsilon_\perp^i(\lambda_3) \Big[ \frac{k_{1\perp}^j }{k_1^+} \epsilon_\perp^j(\lambda_1) \epsilon_{\perp i}(\lambda_2)
- \frac{k_{2\perp}^j }{k_2^+} \epsilon_\perp^j(\lambda_2) \epsilon_{\perp i}(\lambda_1)
\Big] \nn\\
&\qquad - k_{1\perp}^i \epsilon_\perp^j(\lambda_1) \Big[ \epsilon_{\perp i}(\lambda_2) \epsilon_{\perp j}(\lambda_3) - \epsilon_{\perp i}(\lambda_3) \epsilon_{\perp j}(\lambda_2)\Big]
 - k_{2\perp}^i \epsilon_\perp^j(\lambda_2) \Big[ \epsilon_{\perp i}(\lambda_3) \epsilon_{\perp j}(\lambda_1) \nn\\
& \qquad - \epsilon_{\perp i}(\lambda_1) \epsilon_{\perp j}(\lambda_3) \Big] 
- k_{3\perp}^i \epsilon_\perp^j(\lambda_3) \Big[ \epsilon_{\perp i}(\lambda_1) \epsilon_{\perp j}(\lambda_2) - \epsilon_{\perp i}(\lambda_2) \epsilon_{\perp j}(\lambda_1)\Big]
\bigg) \,,\nn
\end{align}
where we used negative momenta to label the outgoing states in the splitting involving three gluons, which allows us to easily keep track of the signs. Physical states should have positive $+$ components of momenta (we omit the zero mode in the current study) and the matrix elements of the splitting Hamiltonian for physical outgoing states can be easily obtained by flipping the signs of the momenta for the outgoing particles. The matrix elements of the splitting Hamiltonian can be easily generalized for cases with $n$ initial partons, which describe $n\to n+1$ splitting processes:
\begin{align}
& \big\langle 1'2'3'\cdots n'(n'+1)\big| H_{\rm split} \big| 123\cdots n \big\rangle \\
& \quad = \sum_{i=1}^n \big\langle i' (n'+1) \big| H_{\rm split} \big| i \big\rangle \big\langle 1'2'3'\cdots (i'-1)(i'+1)\cdots n' \big| 123\cdots (i-1)(i+1)\cdots n \big\rangle \nn\\
& \quad = \sum_{i=1}^n \big\langle i'(n'+1) \big| H_{\rm split} \big| i \big\rangle \delta_{1'1}\delta_{2'2}\cdots \delta_{(i'-1)(i-1)}\delta_{(i'+1)(i+1)}\cdots \delta_{n'n} \,, \nn
\end{align}
where terms of the form $\langle i'(n'+1)|H_{\rm split} | j \rangle$ ($i\neq j$) do not contribute. They are properly accounted for by the (anti)symmetric property of a quantum state. As can be seen, the matrix for the splitting Hamiltonian is also sparse. 

The matrix representing the splitting Hamiltonian is not Hermitian. Its Hermitian conjugate gives the matrix for the inverse process, which describes parton recombination. It is essential to include parton recombination to reproduce the virtual correction diagrams in the usual Feynman diagram approach to study the LPM effect.

\subsection{Sampling Classical Background Field}
\label{sect:sample}
The diffusion part of the Hamiltonian is light-cone time dependent and the dependence is through the random classical background field $\bar{A}^{-a}$, which satisfies the correlation~\eqref{eqn:AAmixed}. To generate the matrix elements of the diffusion Hamiltonian, we need to generate the random classical background fields at each time step in the Trotterization, which can be done by sampling random variables according to the correlation. If the time is discretized, the correlation can be written as
\be
\big\langle \bar{A}^{-a}(x^+, k_\perp)
\bar{A}^{-b}(y^+, -k_\perp) \big\rangle = \frac{1}{\Delta x^+}\delta^{ab} \delta_{x^+y^+} \gamma({\bs k}_\perp) \int\diff^2 R_\perp \,,
\ee
where $\delta_{x^+y^+}$ is a Kronecker delta function for the discretized light-cone time and $\Delta x^+$ is the grid size in the direction of the light-cone time. The delta function in time means the classical background fields at different times are independent, and thus can be sampled independently. At a given time $x^+$, the correlation that governs the distribution of the background field is written as
\be
\label{eqn:AA_sametime}
\big\langle \bar{A}^{-a}(x^+, k_\perp)
\bar{A}^{-a}(x^+, -k_\perp) \big\rangle = \frac{1}{\Delta x^+} \gamma({\bs k}_\perp) \int\diff^2 R_\perp \equiv \tilde{\gamma}({\bs k}_\perp)\,,
\ee
which almost corresponds to the width of a Gaussian distribution for the random variable $\bar{A}^{-a}(x^+, k_\perp)$ (note that we assume the QGP is overall color neutral $\langle \bar{A}^{-a} \rangle=0$). In general the sign is different between the $k_\perp$ arguments of the two random fields. In other words, $\bar{A}^{-a}(x^+, k_\perp)$ and $\bar{A}^{-a}(x^+, -k_\perp)$ are two different random variables for $k_\perp\neq0$.\footnote{As a side remark, we discuss how to sample $\bar{A}^{-a}(x^+, k_\perp)$ and $\bar{A}^{-a}(x^+, -k_\perp)$ as two different random variables in a more general case: First, when $k_\perp=0$, $\bar{A}^{-a}(x^+, 0_\perp)$ can be generated by sampling a Gaussian random variable with the variance $\tilde{\gamma}(0_\perp)$ 
\be
\big\langle \bar{A}^{-a}(x^+, 0_\perp)
\bar{A}^{-a}(x^+, 0_\perp) \big\rangle = \tilde{\gamma}(0_\perp) \,.
\ee
Next for $k_\perp\neq0$, by using Eq.~\eqref{eqn:AA_sametime} we can show 
\begin{align}
\Big\langle  \big[\bar{A}^{-a}(x^+, k_\perp)+\bar{A}^{-a}(x^+, -k_\perp)\big] \big[\bar{A}^{-a}(x^+, k_\perp)+\bar{A}^{-a}(x^+, -k_\perp) \big]     \Big\rangle = 2\tilde{\gamma}(0_\perp)+2 \tilde{\gamma}({\bs k}_\perp) \,,\nn\\
\Big\langle  \big[\bar{A}^{-a}(x^+, k_\perp) - \bar{A}^{-a}(x^+, -k_\perp)\big] \big[ \bar{A}^{-a}(x^+, k_\perp) - \bar{A}^{-a}(x^+, -k_\perp) \big]     \Big\rangle = 2\tilde{\gamma}(0_\perp) - 2 \tilde{\gamma}({\bs k}_\perp) \,,
\end{align}
which means $\bar{A}^{-a}(x^+, k_\perp)+\bar{A}^{-a}(x^+, -k_\perp)$ and $\bar{A}^{-a}(x^+, k_\perp) - \bar{A}^{-a}(x^+, -k_\perp)$ are two Gaussian random variables with the variances $2\tilde{\gamma}(0_\perp)+2\tilde{\gamma}({\bs k}_\perp)$ and $2\tilde{\gamma}(0_\perp)-2\tilde{\gamma}({\bs k}_\perp)$ respectively. We can then independently sample two Gaussian random variables $X_1$ and $X_2$ from these two Gaussian distributions and finally obtain $(X_1+ X_2)/2$ and $(X_1- X_2)/2$ as the sampled classical background fields for $\bar{A}^{-a}(x^+, k_\perp)$ and $\bar{A}^{-a}(x^+, -k_\perp)$ respectively.} However, we note that for the diffusion Hamiltonian to be Hermitian, we must set $\bar{A}^{-a}(x^+, k_\perp) = \bar{A}^{-a}(x^+, -k_\perp)$. Therefore $\bar{A}^{-a}(x^+, k_\perp)$ and $\bar{A}^{-a}(x^+, -k_\perp)$ correspond to the same random variable, which can be sampled from a Gaussian distribution with the variance $\tilde{\gamma}({\bs k}_\perp) = \tilde{\gamma}(|{\bs k}_\perp|)$.

This method requires $\ml{O}(tV_k)$ classical samplings to generate the random background fields for the construction of the quantum circuits describing the diffusion Hamiltonian evolution, where $V_k$ denotes the volume of the momentum space, i.e., the number of lattice points $V_k = N^+N_\perp^2$. The quantum simulation with a given set of classical background fields corresponds to one particular trajectory for an initial state. In practice, one needs to repeat the classical sampling and the simulation of the diffusion process for multiple trajectories. Physical results are obtained by averaging over multiple trajectories. 
An interesting question is whether one can simulate the diffusion Hamiltonian evolution more efficiently by using some random quantum circuit~\cite{Alexandru:2019dmv} or modifying the Quantum Signal Processing algorithm~\cite{low2017optimal,martyn2021efficient}. This is left for future studies. 

\section{Quantum Simulation of Toy Model}
\label{sect:toy}
In this section we consider a simple toy model in which we neglect the color, spin and flavor (quark or gluon) degrees of freedom discussed in the previous sections and focus on the case with only one transverse direction. The purpose is to demonstrate how to construct quantum gates to describe the time evolution driven by the three pieces of the Hamiltonian, in order to study the LPM effect in jet quenching. Then we will show some simulation results of the toy model that are obtained from the IBM Qiskit simulator. The more complicated case in QCD will be discussed in the next section.

\subsection{Toy Model}
Here we construct a toy model to demonstrate the construction of quantum gates describing the time evolution driven by the three pieces of the Hamiltonian. The toy model we consider describes the dynamics of scalar particles in $2+1$ dimension with only $1\to2$ splitting and its inverse. Instead of deriving the Hamiltonian from the light-front quantization of scalar field theory in $2+1$ dimension, we use a ``bottom-up'' approach where  we write down phenomenological matrix elements that describe the kinetic, diffusion and splitting processes, which is enough for our purpose to demonstrate the construction of quantum gates relevant for the studies of the LPM effect.

With a limited number of qubits, we discretize the $+$ and $\perp$ components of the momenta as
\begin{align}
k^+ \in K^+_{\rm max} \{ 0.5, 1\}\,,\qquad k_\perp \in K^\perp_{\rm max} \{ 0, 1\}  \,,
\end{align}
where $k_\perp$ has only one component, rather than $x$ and $y$ components as in the previous sections. With more qubits available, one would add the second transverse component and further divide each momentum component into finer levels and eventually take the continuum limit. We will study the case with one initial particle and only one splitting, which means the Hilbert space consists of $1$-particle and $2$-particle states. According to our discussion in section~\ref{sect:hilbert}, totally five qubits are needed to encode all the quantum states in this case. For each particle, we need one qubit to encode the transverse momentum and another for the $+$ component of the momentum. The correspondence between the qubit representation and the momentum state of a particle is given by
\begin{align}
\label{eqn:qubit_for_k}
|00\rangle : & \quad k^+ = 0.5\,,\ k_\perp = 0\,,\\
|01\rangle : & \quad k^+ = 0.5\,,\ k_\perp = 1\,, \nn\\
|10\rangle : & \quad k^+ = 1\,,\ k_\perp = 0\,, \nn\\
|11\rangle : & \quad k^+ = 1\,,\ k_\perp = 1\,, \nn
\end{align}
where we have labeled the momenta by fractions of the maximum values. To encode both $1$-particle and $2$-particle states, we first need one qubit to distinguish them. Then we need another four qubits to represent the $2$-particle states (representing the $1$-particle states only requires two qubits).
We list the values of the five qubits from left to right to describe a quantum state as $|q_1q_2q_3q_4q_5\rangle$. We use the following rules when encoding the states:
\begin{align}
\label{eqn:state_qbit_repre}
| \underbrace{q_1}_\text{separate $1$- and $2$-particle states} \overbrace{q_2 q_3}^\text{describe momenta of the 2nd particle} \underbrace{q_4 q_5}_\text{describe momenta of the 1st particle}  \rangle \,,
\end{align}
where the momentum state of a particle is represented as in Eq.~\eqref{eqn:qubit_for_k}.
In this way, the $1$-particle state is represented as
\begin{align}
\label{eqn:1parton_qubit}
|000q_4q_5\rangle \,,
\end{align}
where the second and the third $0$s from the left have no physical meaning since this is a $1$-particle state. On the other hand, the $2$-particle state is labeled as
\begin{align}
\label{eqn:2parton_qubit}
|1q_2q_3q_4q_5\rangle \,.
\end{align}
The setup can be easily generalized for multiple particles and cases requiring more qubits to represent $1$-particle states such as those having more levels in the momentum discretization and degrees of freedom in color and spin: We will assign a certain number of qubits to label the number of particles in the state; Then for each particle, we will use a fixed number of qubits to represent its particle species, discretized momenta, color and spin degrees of freedom, as demonstrated in Eq.~\eqref{eqn:state_qbit_repre}. This setup may not be the most efficient encoding scheme in terms of the number of qubits needed. But in this setup the Pauli matrix representation of the Hamiltonian for multi-particle states can be easily obtained from that for 1-particle states, as will be discussed below and in appendix~\ref{app:alter}.

\subsubsection{Kinetic Term}
The kinetic term is diagonal in the $n$-particle basis we haven chosen. 
We first consider the $1$-particle kinetic term, which only involves two qubits and will serve as a building block of the full kinetic term. In the basis given by Eq.~\eqref{eqn:qubit_for_k}, which is listed in the order $|00\rangle,|01\rangle,|10\rangle,|11\rangle$, the kinetic term is given by
\be
\label{equ:kin_1par}
H_{\rm kin}^{(1)} = \frac{(K_{\rm max}^\perp)^2}{K_{\rm max}^+}{\rm diag}\big( 0,\, 2,\, 0,\, 1\big) \,,
\ee
which means $\langle 01 |H_{\rm kin}^{(1)} | 01 \rangle = 2(K_{\rm max}^\perp)^2/K_{\rm max}^+$, $\langle 11 |H_{\rm kin}^{(1)} | 11 \rangle = (K_{\rm max}^\perp)^2/K_{\rm max}^+$ and all the other matrix elements vanish.

We can easily generalize this to the case involving both $1$-particle and $2$-particle states (it is easier to write a code for the generalization than to write them out explicitly).
In the basis of the five qubits introduced in \eqref{eqn:state_qbit_repre}, the matrix elements of the kinetic Hamiltonian are given by
\begin{align}
\label{eqn:Hkin_toy}
&\langle 00000 | H_{\rm kin} | 00000 \rangle = 0\,, \qquad\qquad\qquad\quad~ \langle 00001 | H_{\rm kin} | 00001 \rangle = 2(K_{\rm max}^\perp)^2/K_{\rm max}^+\,, \\
&\langle 00010 | H_{\rm kin} | 00010 \rangle = 0\,, \qquad\qquad\qquad\quad~ \langle 00011 | H_{\rm kin} | 00011 \rangle = (K_{\rm max}^\perp)^2/K_{\rm max}^+\,, \nn\\
&\langle 10000 | H_{\rm kin} | 10000 \rangle = 0\,, \qquad\qquad\qquad\quad~ \langle 10001 | H_{\rm kin} | 10001 \rangle = 2(K_{\rm max}^\perp)^2/K_{\rm max}^+\,, \nn\\
&\langle 10010 | H_{\rm kin} | 10010 \rangle = 0\,, \qquad\qquad\qquad\quad~ \langle 10011 | H_{\rm kin} | 10011 \rangle = (K_{\rm max}^\perp)^2/K_{\rm max}^+\,, \nn\\
&\langle 10100 | H_{\rm kin} | 10100 \rangle = 2(K_{\rm max}^\perp)^2/K_{\rm max}^+\,, \quad \langle 10101 | H_{\rm kin} | 10101 \rangle = 4(K_{\rm max}^\perp)^2/K_{\rm max}^+\,, \nn\\
&\langle 10110 | H_{\rm kin} | 10110 \rangle = 2(K_{\rm max}^\perp)^2/K_{\rm max}^+\,, \quad \langle 10111 | H_{\rm kin} | 10111 \rangle = 3(K_{\rm max}^\perp)^2/K_{\rm max}^+\,, \nn\\
&\langle 11000 | H_{\rm kin} | 11000 \rangle = 0\,, \qquad\qquad\qquad\quad~ \langle 11001 | H_{\rm kin} | 11001 \rangle = 2(K_{\rm max}^\perp)^2/K_{\rm max}^+\,, \nn\\
&\langle 11010 | H_{\rm kin} | 11010 \rangle = 0\,, \qquad\qquad\qquad\quad~ \langle 11011 | H_{\rm kin} | 11011 \rangle = (K_{\rm max}^\perp)^2/K_{\rm max}^+\,, \nn\\
&\langle 11100 | H_{\rm kin} | 11100 \rangle = (K_{\rm max}^\perp)^2/K_{\rm max}^+\,, \quad \langle 11101 | H_{\rm kin} | 11101 \rangle = 3(K_{\rm max}^\perp)^2/K_{\rm max}^+\,, \nn\\
&\langle 11110 | H_{\rm kin} | 11110 \rangle = (K_{\rm max}^\perp)^2/K_{\rm max}^+\,, \quad \langle 11111 | H_{\rm kin} | 11111 \rangle = 2(K_{\rm max}^\perp)^2/K_{\rm max}^+\,, \quad \nn
\end{align}
and all the other matrix elements are vanishing.

\subsubsection{Diffusion Term}
The diffusion part of the Hamiltonian changes transverse momenta of particles and depends on an external classical background field, which is needed to construct the relevant Hamiltonian. Here we just assume the classical background fields at each momentum grid have been generated by using the sampling method described in section~\ref{sect:sample} for each time step in the time evolution. For notational consistency, we still use $\bar{A}^-$ to label the classical background fields here, even though our toy model has no gauge fields.
Since our toy model has only two levels in the transverse momentum, we only need the classical background fields $\bar{A}^-$ at two values of the transverse momenta $0$ and $K_{\rm max}^\perp$. In the case of only one particle, the diffusion term in the basis given by Eq.~\eqref{eqn:qubit_for_k} is given by
\begin{align}
\label{eqn:Hdiff_single}
&\langle 00 | H_{\rm diff}^{(1)} | 00 \rangle = g_d \bar{A}^-(0)\,, \quad \langle 01 | H_{\rm diff}^{(1)} | 01 \rangle = g_d \bar{A}^-(0)\,, \quad \langle 10 | H_{\rm diff}^{(1)} | 10 \rangle = g_d \bar{A}^-(0)\,, \\
& \langle 11 | H_{\rm diff}^{(1)} | 11 \rangle = g_d \bar{A}^-(0)\,, \quad \langle 01 | H_{\rm diff}^{(1)} | 00 \rangle = \langle 00 | H_{\rm diff}^{(1)} | 01 \rangle = g_d \bar{A}^-(K_{\rm max}^\perp) \,,\nn\\ 
& \langle 11 | H_{\rm diff}^{(1)} | 10 \rangle = \langle 10 | H_{\rm diff}^{(1)} | 11 \rangle = g_d \bar{A}^-(K_{\rm max}^\perp) \,, \nn
\end{align}
and all the others are zero, where $g_d$ is the coupling constant in the diffusion term and we have used $\bar{A}^-(K_{\rm max}^\perp) = \bar{A}^-(-K_{\rm max}^\perp)$.

The part of the diffusion Hamiltonian involving $\bar{A}^-(0)$ is proportional to an identity operator, which means its effect is to change the global phase of the state and thus does not change any physics. Therefore it is legitimate to ignore the $\bar{A}^-(0)$ term in the diffusion Hamiltonian. We will do so in the following.

Using Eq.~\eqref{eqn:Hdiff_multi}, we can generalize the diffusion Hamiltonian to the five qubit case introduced in Eq.~\eqref{eqn:state_qbit_repre} leads to
\begin{align}
\label{eqn:Hdiff_toy}
&\langle 00001 | H_{\rm diff} | 00000 \rangle = \langle 00000 | H_{\rm diff} | 00001 \rangle =
\langle 00011 | H_{\rm diff} | 00010 \rangle = \langle 00010 | H_{\rm diff} | 00011 \rangle \nn\\
=\,&\langle 10001 | H_{\rm diff} | 10000 \rangle = \langle 10000 | H_{\rm diff} | 10001 \rangle = \langle 10011 | H_{\rm diff} | 10010 \rangle = \langle 10010 | H_{\rm diff} | 10011 \rangle 
 \nn \\
=\,&\langle 10101 | H_{\rm diff} | 10100 \rangle = \langle 10100 | H_{\rm diff} | 10101 \rangle = \langle 10111 | H_{\rm diff} | 10110 \rangle = \langle 10110 | H_{\rm diff} | 10111 \rangle 
 \nn \\
=\,&\langle 11001 | H_{\rm diff} | 11000 \rangle = \langle 11000 | H_{\rm diff} | 11001 \rangle = \langle 11011 | H_{\rm diff} | 11010 \rangle = \langle 11010 | H_{\rm diff} | 11011 \rangle 
 \nn \\
=\,&\langle 11101 | H_{\rm diff} | 11100 \rangle = \langle 11100 | H_{\rm diff} | 11101 \rangle = \langle 11111 | H_{\rm diff} | 11110 \rangle = \langle 11110 | H_{\rm diff} | 11111 \rangle 
 \nn \\
=\,&\langle 10100 | H_{\rm diff} | 10000 \rangle = \langle 10000 | H_{\rm diff} | 10100 \rangle = \langle 11100 | H_{\rm diff} | 11000 \rangle = \langle 11000 | H_{\rm diff} | 11100 \rangle 
 \nn \\
=\,&\langle 10101 | H_{\rm diff} | 10001 \rangle = \langle 10001 | H_{\rm diff} | 10101 \rangle = \langle 11101 | H_{\rm diff} | 11001 \rangle = \langle 11001 | H_{\rm diff} | 11101 \rangle 
 \nn \\
=\,&\langle 10110 | H_{\rm diff} | 10010 \rangle = \langle 10010 | H_{\rm diff} | 10110 \rangle = \langle 11110 | H_{\rm diff} | 11010 \rangle = \langle 11010 | H_{\rm diff} | 11110 \rangle 
 \nn \\
=\,&\langle 10111 | H_{\rm diff} | 10011 \rangle = \langle 10011 | H_{\rm diff} | 10111 \rangle = \langle 11111 | H_{\rm diff} | 11011 \rangle = \langle 11011 | H_{\rm diff} | 11111 \rangle 
 \nn \\
=\,& g_d \bar{A}^-(K_{\rm max}^\perp)\,,
\end{align}
and all the other matrix elements are zero,
where we have neglected the global phase change caused by the $\bar{A}^-(0)$ term.\footnote{Rigorously speaking, the phases for 1-particle and 2-particle states differ by a factor of two, i.e., they are $g_d\bar{A}^-(0)$ and $2g_d\bar{A}^-(0)$ respectively. However, this difference does not affect the radiation probability that we want to study here.} In practice, we only need to sample one Gaussian random variable $\bar{A}^-(K_{\rm max}^\perp)$ at each time step.

\subsubsection{Splitting Term}
Finally we discuss the construction of the splitting part of the Hamiltonian. Due to the momentum conservation in $k^+$ and $k_\perp$, only the following $1\to2$ splitting process can happen in our toy model:
\begin{align}
\label{eqn:toy_split}
|00010\rangle & \to |10000\rangle \\
|00011\rangle & \to |10001\rangle + |10100\rangle \,. \nn
\end{align}
In the first process, the initial particle with $k^+=K_{\rm max}^+$ and $k_\perp=0$ splits into two particles that both have $k^+=0.5K_{\rm max}^+$ and $k_\perp=0$. In the second process, the initial particle with $k^+=K_{\rm max}^+$ and $k_\perp=K_{\rm max}^\perp$ splits into two particles, one with $k^+=0.5K_{\rm max}^+$ and $k_\perp=0$ and the other with $k^+=0.5K_{\rm max}^+$ and $k_\perp=K_{\rm max}^\perp$. The splitting process described in Eq.~\eqref{eqn:toy_split} symmetrizes the final state, up to a normalization. The matrix elements of the splitting Hamiltonian are given by
\begin{align}
\label{eqn:Hsplit_toy}
& \langle 10000 | H_{\rm split} | 00010 \rangle = \langle 00010 | H_{\rm split} | 10000 \rangle = \langle 10001 | H_{\rm split} | 00011 \rangle  \\
=\,& \langle 00011 | H_{\rm split} | 10001 \rangle 
= \langle 10100 | H_{\rm split} | 00011 \rangle = \langle 00011 | H_{\rm split} | 10100 \rangle = g_s \,,\nn
\end{align}
and all the other matrix elements vanish, where $g_s$ is the coupling constant in the splitting Hamiltonian. Here we choose the coupling constants in the diffusion and splitting Hamiltonians to be independent, which is just a feature of the toy model we constructed here. In the QCD case, these two couplings are related.

We have written out explicitly the matrix elements of the Hamiltonian in the toy model. In the next subsection, we will show how to construct quantum gates to describe the relevant Hamiltonian evolution. 

\subsection{Construction of Quantum Circuit}
\label{sect:decomposition}
We use a general method to construct the quantum circuit~\cite{DBLP:books/daglib/0046438}. In general, when we have a matrix $(H_{ij})$ representing a given Hamiltonian $H$, we can construct the corresponding quantum gates by first projecting the matrix onto the basis made up of tensor products of Pauli matrices:
\begin{align}
H = \sum_{\mu_1,\mu_2,\cdots \mu_n} a_{\mu_1\mu_2\cdots \mu_n} \sigma_1^{\mu_1} \otimes \sigma_2^{\mu_2} \otimes \cdots \otimes \sigma_n^{\mu_n}\,,
\end{align}
where we have assumed the matrix can be encoded by $n$ qubits. Here $\sigma_i^{\mu_i}$ indicates the Pauli matrices for the $i$-th qubit and $\sigma^\mu = (\mathbb{1}, \sigma^x, \sigma^y, \sigma^z)$. The linear combination coefficients can be obtained by
\begin{align}
\label{eqn:coeff_decom}
a_{\mu_1\mu_2\cdots \mu_n} = \frac{1}{2^n}{\rm Tr}\Big[ H \big( \sigma_1^{\mu_1} \otimes \sigma_2^{\mu_2} \otimes \cdots \otimes \sigma_n^{\mu_n} \big) \Big] \,,
\end{align}
where we have a matrix multiplication between $H$ and $\sigma_1^{\mu_1} \otimes \sigma_2^{\mu_2} \otimes \cdots \otimes \sigma_n^{\mu_n}$ inside the trace.

After obtaining the linear combination coefficients $a_{\mu_1\mu_2\cdots \mu_n}$, we can construct the quantum gates for the time evolution $e^{-i H \Delta t}$. Using the Trotterization method, we can write
\begin{align}
e^{-i H \Delta t} = e^{\ml{O}((\Delta t)^2)}\prod_{\mu_1,\mu_2,\cdots \mu_n} e^{ -i \Delta t\, a_{\mu_1\mu_2\cdots \mu_n} \sigma_1^{\mu_1} \otimes \sigma_2^{\mu_2} \otimes \cdots \otimes \sigma_n^{\mu_n} } \,.
\end{align}
Therefore, once we know how to construct quantum gates for the time evolution determined by one of the tensor products of Pauli matrices, we can construct a circuit for the full time evolution determined by $H$. Without loss of generality, we discuss how to construct the quantum gates for
\be
e^{-i\theta\,\sigma_1^{\mu_1} \otimes \sigma_2^{\mu_2} \otimes \cdots \otimes \sigma_n^{\mu_n}} \,.
\ee
The strategy is to change the basis of each single qubit such that all the Pauli matrices $\sigma_i^{\mu_i}$ become either $\mathbb{1}_i$ or $\sigma_i^z$. If the original Pauli matrix $\sigma_i^{\mu_i}=\mathbb{1}_i$ or $\sigma_i^z$, nothing needs to be done for the $i$-th qubit. If the original Pauli matrix $\sigma_i^{\mu_i}$ is $\sigma_i^x$, then we apply the Hadamard gate
\be
h = \frac{1}{\sqrt{2}}\begin{pmatrix}
1 & 1 \\
1 & -1 
\end{pmatrix} \,,
\ee
in the beginning and apply its inverse (which turns out to be itself) in the end of the circuit segment such that
\be
h_i\, e^{-i\theta \sigma_i^x} h_i = e^{-i\theta \sigma_i^z} \,,
\ee
where the subscript $i$ indicates the Hadamard gate acts on the $i$-th qubit.
Similarly, if the original Pauli matrix is $\sigma_i^{\mu_i}=\sigma_i^y$, we apply 
\be
R_x = \frac{1}{\sqrt{2}}\begin{pmatrix}
1 & -i \\
-i & 1 
\end{pmatrix}\,,
\ee
and its inverse in the beginning and the end of the circuit segment respectively such that
\be
\label{eqn:ytoz}
(R_x)_i\, e^{-i\theta \sigma_i^y} (R_x^\dagger)_i = e^{-i\theta \sigma_i^z} \,,
\ee
where the subscript $i$ indicates the $R_x$ rotation gate acts on the $i$-th qubit.
The $R_x$ rotation gate can be decomposed as 
\be
R_x = S^\dagger h\, S^\dagger \,, \qquad S = \begin{pmatrix}
1 & 0 \\
0 & i 
\end{pmatrix}\,,
\ee
which can be useful in the construction of the quantum circuit.

\begin{figure}[t]
\centering
\includegraphics[width=0.7\textwidth]{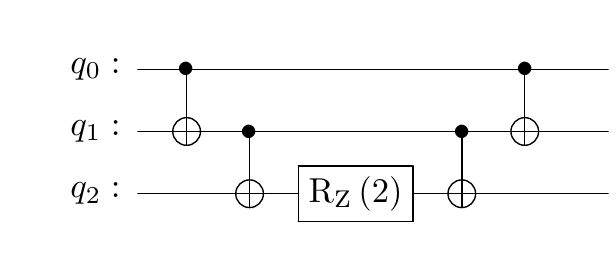}
\caption{Quantum circuit for the unitary evolution $e^{-i\theta \sigma_1^z \otimes \sigma_2^z \otimes \sigma_3^z }$. Every two-qubit gate in the circuit is a CNOT gate with the black dot indicating the control qubit. The argument of the $z$-rotation represents the index of the qubit on which the rotation acts. The $z$-rotation gate is given by $R_z(\theta,j) = e^{-i\theta\sigma^z_j/2} = {\rm diag}(e^{-i\theta/2}, e^{+i\theta/2})$ where $\theta$ is the rotation angle (note the factor of $1/2$ in the rotation gate definition) and $j$ denotes the $j$-th qubit.}
\label{fig:tensorZ}
\end{figure}

In a nutshell, we only need to focus on constructing quantum gates for
\be
e^{-i\theta \sigma_1^z \otimes \cdots \otimes \sigma_m^z} \,,
\ee
where we have omitted the identity matrices and relabeled the indexes in the subscripts. Standard circuits exist to realize such unitary transformations. For example, the quantum circuit for $e^{-i\theta \sigma_1^z \otimes \sigma_2^z \otimes \sigma_3^z }$ is shown in Fig.~\ref{fig:tensorZ}, which can be easily generalized for more $\sigma^z$s.

Now we are ready to construct the quantum circuit for the time evolution of the toy model. We will show the quantum gates for the kinetic, diffusion and splitting terms in the Hamiltonian. 

\subsubsection{Kinetic Term}
\begin{figure}[t]
\raggedright
\includegraphics[height=0.16\textwidth]{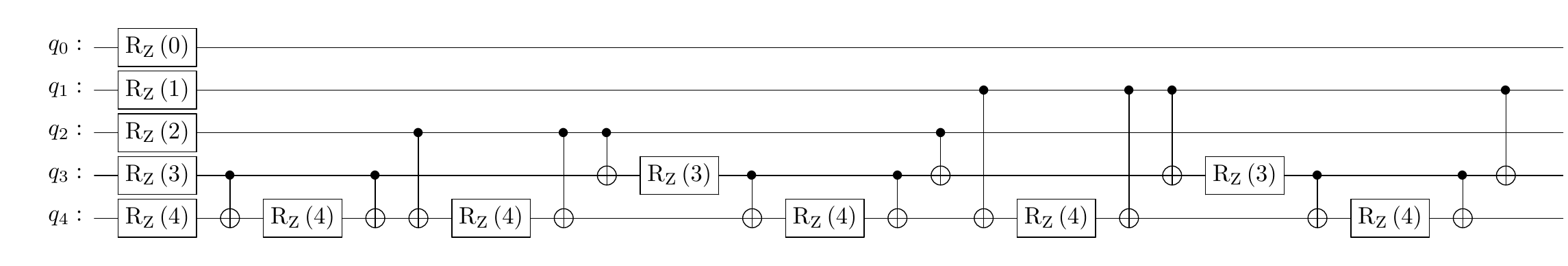}
\includegraphics[height=0.16\textwidth]{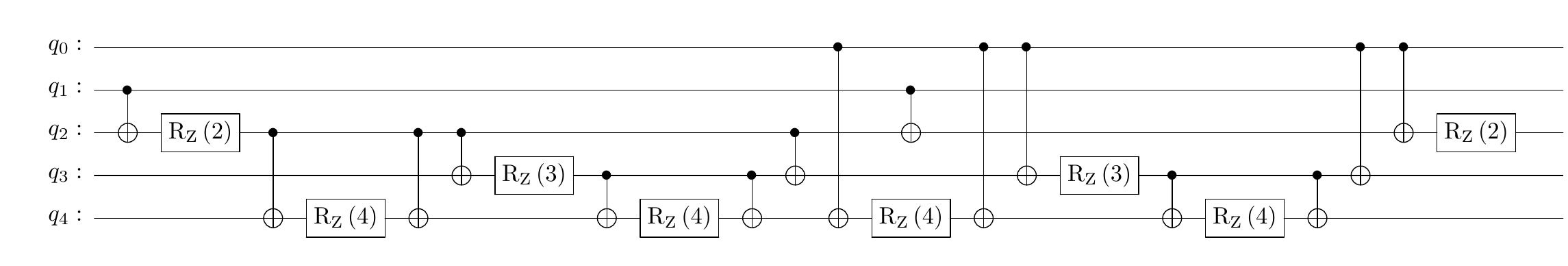}
\includegraphics[height=0.16\textwidth]{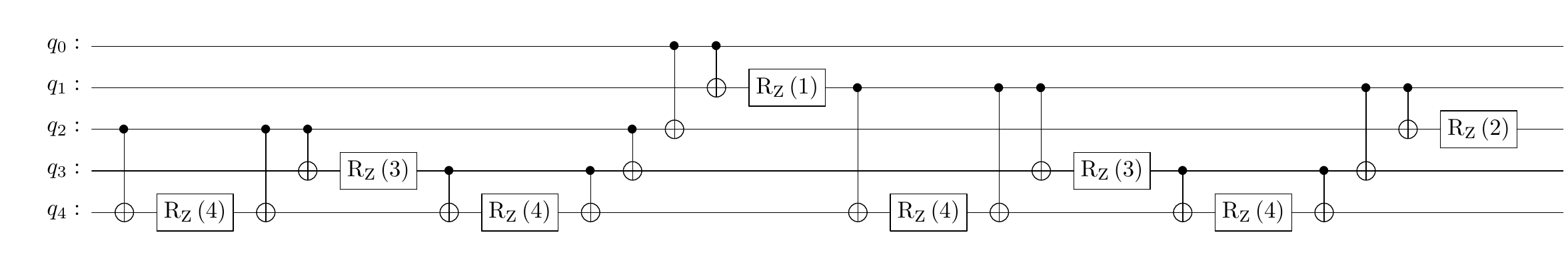}
\includegraphics[height=0.16\textwidth]{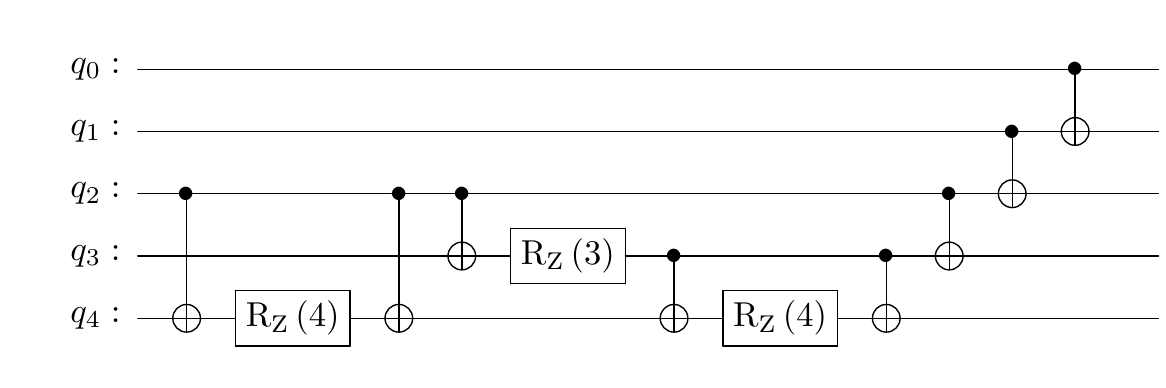}
\caption{Quantum circuit for the time evolution determined by the kinetic part of the Hamiltonian. Every two-qubit gate in the circuit is a CNOT gate with the black dot indicating the control qubit. The argument of the $z$-rotation represents the index of the qubit on which the rotation acts.  The $z$-rotation gate is given by $R_z(i) = e^{-iC\Delta t \, \sigma^z_i (K_{\rm max}^\perp)^2/K_{\rm max}^+}$ with the constants $C$ given in Eq.~\eqref{eqn:Hkin_5q}.}
\label{fig:kin}
\end{figure}

Since the kinetic Hamiltonian is diagonal, its decomposition into Pauli matrices only involves $\mathbb{1}_i$ and $\sigma_i^z$.
Using the procedure described above, the kinetic Hamiltonian in Eq.~\eqref{eqn:Hkin_toy} can be decomposed into
\begin{align}
\label{eqn:Hkin_5q}
H_{\rm kin} = & \frac{(K_{\rm max}^\perp)^2}{K_{\rm max}^+} \Big( \frac{27}{32} - \frac{15}{32} \sigma_5^z + \frac{5}{32} \sigma^z_4 - \frac{5}{32} \sigma^z_4 \otimes \sigma^z_5 - \frac{9}{32} \sigma^z_3 -\frac{3}{32} \sigma^z_3\otimes\sigma^z_5 + \frac{1}{32} \sigma^z_3\otimes\sigma^z_4 \nn\\
& - \frac{1}{32} \sigma^z_3\otimes\sigma^z_4\otimes\sigma^z_5 + \frac{7}{32}\sigma^z_2 - \frac{3}{32} \sigma^z_2 \otimes \sigma^z_5 + \frac{1}{32} \sigma^z_2 \otimes \sigma^z_4 - \frac{1}{32} \sigma^z_2 \otimes \sigma^z_4 \otimes \sigma^z_5 - \frac{1}{32} \sigma^z_2 \otimes \sigma^z_3 \nn\\
& - \frac{3}{32} \sigma^z_2 \otimes \sigma^z_3 \otimes \sigma^z_5 + \frac{1}{32} \sigma^z_2 \otimes \sigma^z_3 \otimes \sigma^z_4 - \frac{1}{32} \sigma^z_2 \otimes \sigma^z_3 \otimes \sigma^z_4 \otimes \sigma^z_5 - \frac{21}{32} \sigma^z_1
+ \frac{9}{32} \sigma^z_1\otimes\sigma_5^z \nn\\
&  - \frac{3}{32} \sigma^z_1\otimes\sigma^z_4 + \frac{3}{32} \sigma^z_1\otimes\sigma^z_4 \otimes \sigma^z_5 + \frac{15}{32} \sigma^z_1\otimes\sigma^z_3 - \frac{3}{32} \sigma^z_1\otimes\sigma^z_3\otimes\sigma^z_5 + \frac{1}{32} \sigma^z_1\otimes\sigma^z_3\otimes\sigma^z_4 \nn\\ 
& -\frac{1}{32} \sigma^z_1\otimes\sigma^z_3\otimes\sigma^z_4\otimes\sigma^z_5  - \frac{1}{32}\sigma^z_1\otimes\sigma^z_2 - \frac{3}{32} \sigma^z_1\otimes\sigma^z_2 \otimes \sigma^z_5 + \frac{1}{32} \sigma^z_1\otimes\sigma^z_2 \otimes \sigma^z_4 \nn\\
& - \frac{1}{32} \sigma^z_1\otimes\sigma^z_2 \otimes \sigma^z_4 \otimes \sigma^z_5 + \frac{7}{32} \sigma^z_1\otimes\sigma^z_2 \otimes \sigma^z_3 - \frac{3}{32} \sigma^z_1\otimes\sigma^z_2 \otimes \sigma^z_3 \otimes \sigma^z_5 \nn\\
& + \frac{1}{32} \sigma^z_1\otimes\sigma^z_2 \otimes \sigma^z_3 \otimes \sigma^z_4  - \frac{1}{32} \sigma^z_1\otimes\sigma^z_2 \otimes \sigma^z_3 \otimes \sigma^z_4 \otimes \sigma^z_5
\Big) \,,
\end{align}
where we have omitted identity operators for notational simplicity. For example, $\sigma^z_4$ shown above corresponds to $\mathbb{1}_1 \otimes \mathbb{1}_2 \otimes \mathbb{1}_3 \otimes \sigma^z_4 \otimes \mathbb{1}_5$ in the complete five qubit representation. The first term with the coefficient $27/32$ is an identity operator and only results in a global phase change, which will be neglected when we construct the quantum gates. The quantum circuit for the kinetic time evolution $e^{-i H_{\rm kin}\Delta t}$ is shown in Fig.~\ref{fig:kin}.

\begin{figure}[t]
\raggedright
\includegraphics[height=0.16\textwidth]{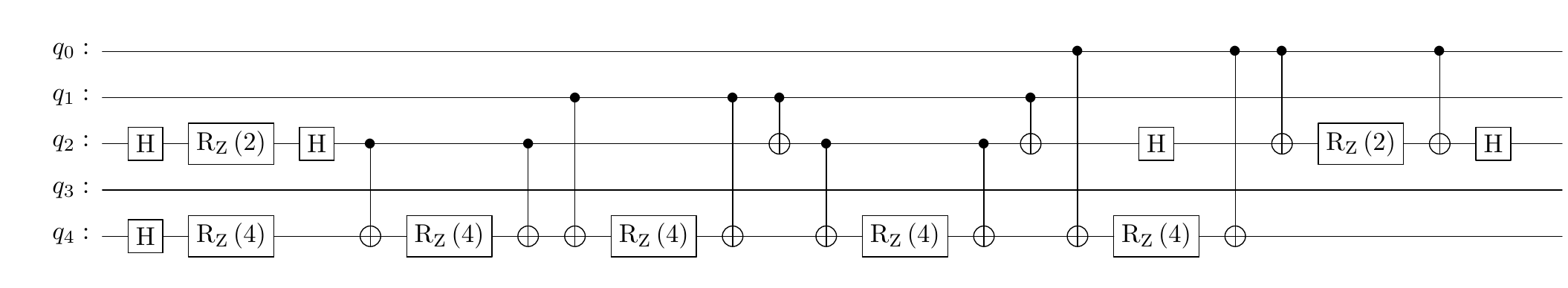}
\raggedright
\includegraphics[height=0.16\textwidth]{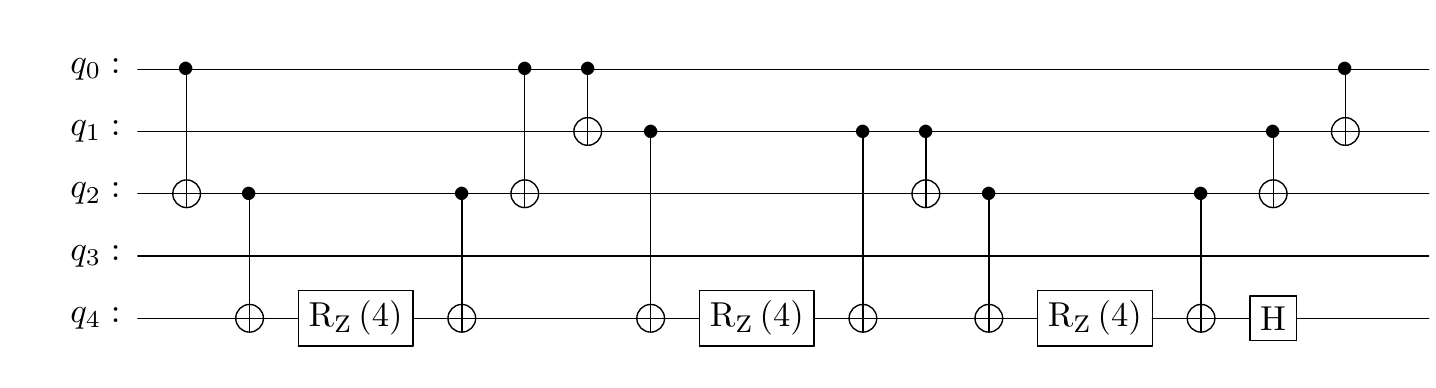}
\caption{Quantum circuit for the time evolution driven by the diffusion part of the Hamiltonian. Every two-qubit gate in the circuit is a CNOT gate with the black dot indicating the control qubit. The argument of the $z$-rotation represents the index of the qubit on which the rotation acts. The $z$-rotation gate is given by $R_z(i) = e^{- iC\Delta t \, g_d \bar{A}^-(K_{\rm max}^\perp) \sigma^z_i} $ with the coefficients $C$ given in Eq.~\eqref{eqn:Hdiff_5q} and $\bar{A}^-(K_{\rm max}^\perp)$ is the time dependent classical background field.}
\label{fig:diff}
\end{figure}

\subsubsection{Diffusion Term}
Similarly, the diffusion part of the Hamiltonian Eq.~\eqref{eqn:Hdiff_toy} can be decomposed into tensor products of Pauli matrices as
\begin{align}
\label{eqn:Hdiff_5q}
H_{\rm diff} =& g_d \bar{A}^-(K_{\rm max}^\perp) \Big( \frac{5}{8}\sigma_5^x + \frac{1}{2}\sigma_3^x + \frac{1}{8}\sigma_3^z \otimes \sigma_5^x + \frac{1}{8}\sigma_2^z \otimes \sigma_5^x + \frac{1}{8}\sigma_2^z \otimes \sigma_3^z \otimes \sigma_5^x - \frac{3}{8}\sigma_1^z \otimes \sigma_5^x \nn\\
& - \frac{1}{2} \sigma_1^z \otimes \sigma_3^x + \frac{1}{8} \sigma_1^z \otimes \sigma_3^z \otimes \sigma_5^x + \frac{1}{8} \sigma_1^z \otimes \sigma_2^z \otimes \sigma_5^x  + \frac{1}{8} \sigma_1^z \otimes \sigma_2^z \otimes \sigma_3^z \otimes \sigma_5^x
\Big) \,.
\end{align}
The quantum circuit for the diffusion time evolution $e^{-i H_{\rm diff} \Delta t }$ is shown in Fig.~\ref{fig:diff}.

\subsubsection{Splitting Term}
\begin{figure}[t]
\raggedright
\includegraphics[height=0.134\textwidth]{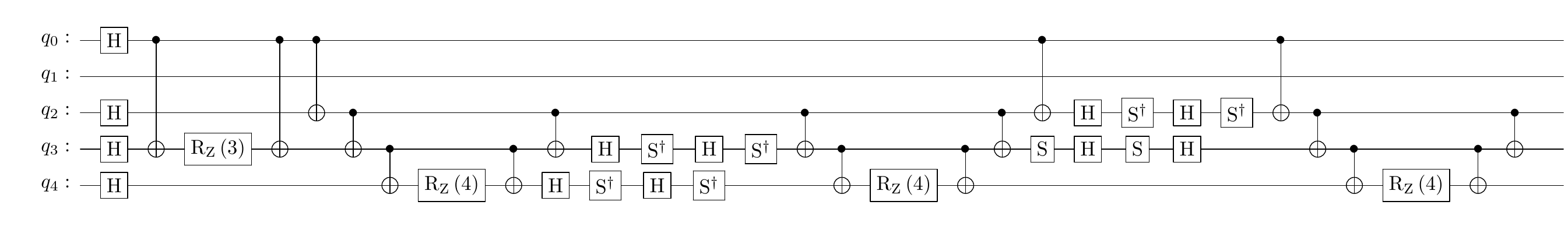}
\raggedright
\includegraphics[height=0.134\textwidth]{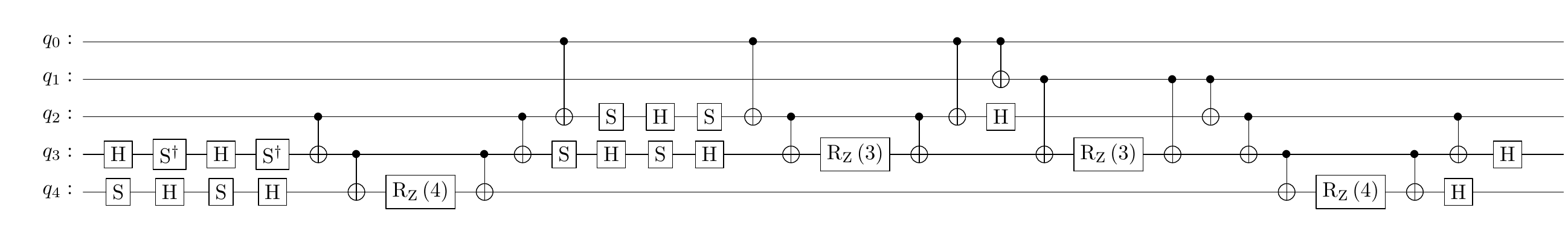}
\raggedright
\includegraphics[height=0.134\textwidth]{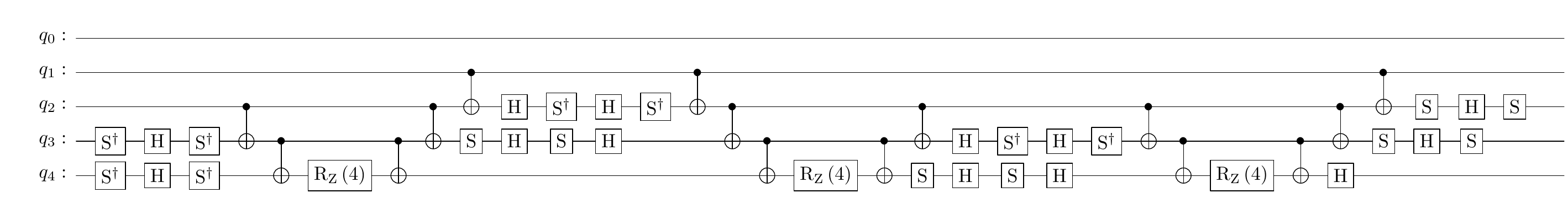}
\raggedright
\includegraphics[height=0.134\textwidth]{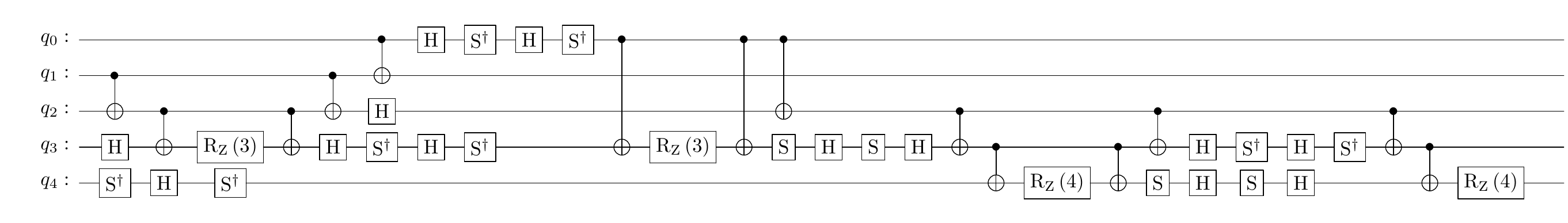}
\raggedright
\includegraphics[height=0.134\textwidth]{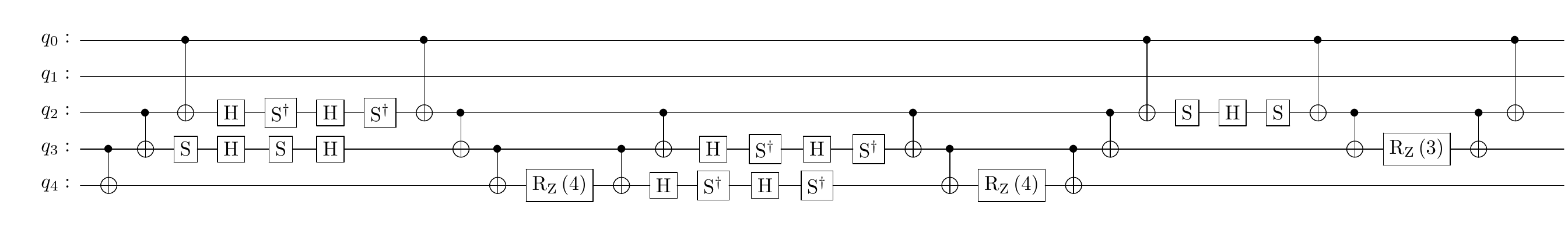}
\raggedright
\includegraphics[height=0.134\textwidth]{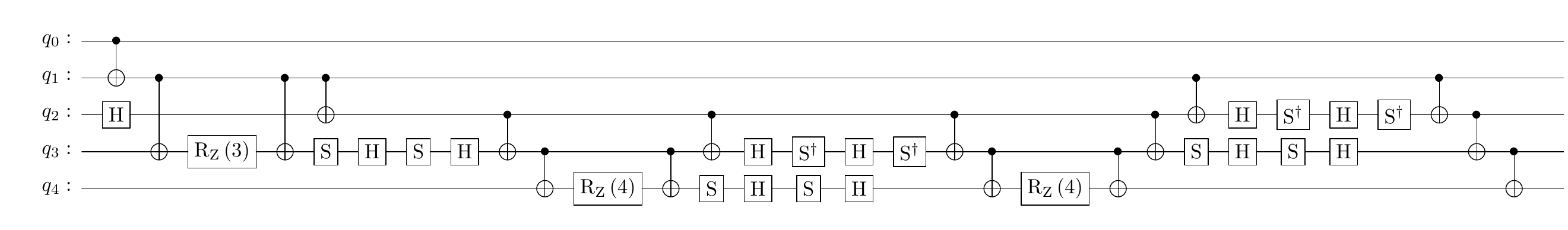}
\raggedright
\includegraphics[height=0.134\textwidth]{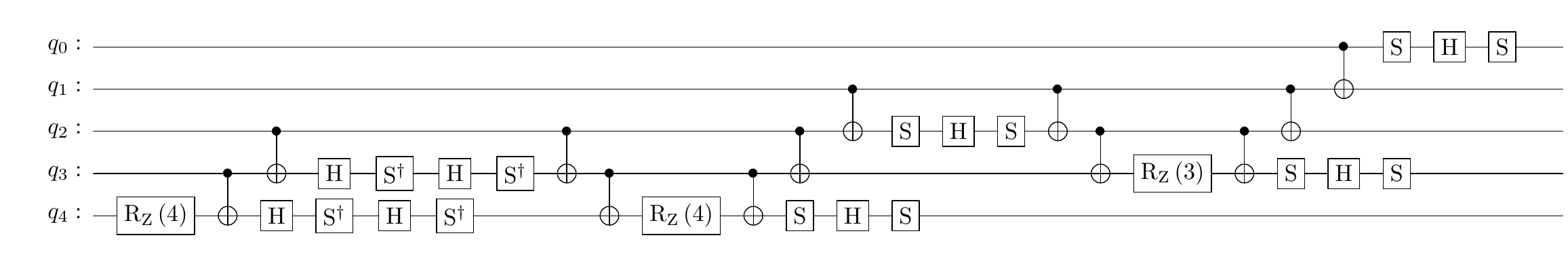}
\caption{Quantum circuit for the time evolution driven by the splitting part of the Hamiltonian. Every two-qubit gate in the circuit is a CNOT gate with the black dot indicating the control qubit. The argument of the $z$-rotation represents the index of the qubit on which the rotation acts. The $z$-rotation gate is given by $R_z(i) = e^{- iC\Delta t \, g_s \sigma^z_i}$ with the coefficients $C$ given in Eq.~\eqref{eqn:Hsplit_5q}.}
\label{fig:split}
\end{figure}

The part of the Hamiltonian for splitting is given by Eq.~\eqref{eqn:Hsplit_toy} and can be decomposed as
\begin{align}
\label{eqn:Hsplit_5q}
H_{\rm split} =& g_s \Big( \frac{1}{8}\sigma_1^x \otimes \sigma_4^x + \frac{1}{16}\sigma_1^x \otimes \sigma_3^x \otimes \sigma_4^x \otimes \sigma_5^x - \frac{1}{16}\sigma_1^x \otimes \sigma_3^x \otimes \sigma_4^y \otimes \sigma_5^y \\
&+ \frac{1}{16}\sigma_1^x \otimes \sigma_3^y \otimes \sigma_4^x \otimes \sigma_5^y + \frac{1}{16}\sigma_1^x \otimes \sigma_3^y \otimes \sigma_4^y \otimes \sigma_5^x + \frac{1}{8}\sigma_1^x \otimes \sigma_3^z \otimes \sigma_4^x + \frac{1}{8}\sigma_1^x \otimes \sigma_2^z \otimes \sigma_4^x \nn\\
&+ \frac{1}{16}\sigma_1^x \otimes \sigma_2^z \otimes \sigma_3^x \otimes \sigma_4^x \otimes \sigma_5^x - \frac{1}{16}\sigma_1^x \otimes \sigma_2^z \otimes \sigma_3^x \otimes \sigma_4^y \otimes \sigma_5^y \nn\\
&+ \frac{1}{16}\sigma_1^x \otimes \sigma_2^z \otimes \sigma_3^y \otimes \sigma_4^x \otimes \sigma_5^y + \frac{1}{16}\sigma_1^x \otimes \sigma_2^z \otimes \sigma_3^y \otimes \sigma_4^y \otimes \sigma_5^x + \frac{1}{8}\sigma_1^x \otimes \sigma_2^z \otimes \sigma_3^z \otimes \sigma_4^x \nn\\
&+\frac{1}{8}\sigma_1^y \otimes \sigma_4^y + \frac{1}{16}\sigma_1^y \otimes \sigma_3^x \otimes \sigma_4^x \otimes \sigma_5^y + \frac{1}{16}\sigma_1^y \otimes \sigma_3^x \otimes \sigma_4^y \otimes \sigma_5^x \nn\\
&- \frac{1}{16}\sigma_1^y \otimes \sigma_3^y \otimes \sigma_4^x \otimes \sigma_5^x + \frac{1}{16}\sigma_1^y \otimes \sigma_3^y \otimes \sigma_4^y \otimes \sigma_5^y + \frac{1}{8}\sigma_1^y \otimes \sigma_3^z \otimes \sigma_4^y + \frac{1}{8}\sigma_1^y \otimes \sigma_2^z \otimes \sigma_4^y \nn\\
&+ \frac{1}{16}\sigma_1^y \otimes \sigma_2^z \otimes \sigma_3^x \otimes \sigma_4^x \otimes \sigma_5^y + \frac{1}{16}\sigma_1^y \otimes \sigma_2^z \otimes \sigma_3^x \otimes \sigma_4^y \otimes \sigma_5^x \nn\\
&- \frac{1}{16}\sigma_1^y \otimes \sigma_2^z \otimes \sigma_3^y \otimes \sigma_4^x \otimes \sigma_5^x + \frac{1}{16}\sigma_1^y \otimes \sigma_2^z \otimes \sigma_3^y \otimes \sigma_4^y \otimes \sigma_5^y + \frac{1}{8}\sigma_1^y \otimes \sigma_2^z \otimes \sigma_3^z \otimes \sigma_4^y
\Big) \,.\nn
\end{align}
The quantum circuit for the time evolution driven by the splitting Hamiltonian $e^{-i H_{\rm split}\Delta t}$ is given in Fig.~\ref{fig:split}.

In appendix~\ref{app:alter}, we discuss an alternative way of decomposing the matrix elements of each Hamiltonian into tensor products of Pauli matrices, which illuminates an easy way to generalize the decomposition for states with more than two particles. This is important if we want to study a system consisting of many particles, since the generic way of decomposing into tensor products of Pauli matrices involves calculating an exponential number of coefficients and does not employ any property or symmetry of the system to simplify the decomposition.

This completes our construction of the quantum gates to describe the time evolution of the toy model.

\subsection{Simulation Results}
Using the quantum circuits constructed above, we can now simulate the time evolution of the toy model in both vacuum and the medium. We will perform the quantum simulation by using the Qiskit simulator package provided by IBM.

We will initialize the state as the $1$-particle state with $k^+=K_{\rm max}^+$ and $k_\perp=0$, which is represented as $|00010\rangle$ in the quantum register. The initial particle is off mass shell, which is caused by hard scattering or interaction with the medium. In the latter case where the radiation is medium-induced, what we call vacuum evolution should be thought of as in-medium evolution without the LPM effect. Since the quantum circuit constructed by the Qiskit package of IBM always initializes all the qubits to be in the $0$ states, we still need to apply the $\sigma_4^x$ gate to obtain the initial state we want. After the state initialization, we evolve the state in time by using the quantum circuits constructed. At the end of the time evolution, we measure the first qubit. The result ``$0$'' in the measurement corresponds to a $1$-particle state while the result ``$1$'' corresponds to a $2$-particle state. The simulation and the measurement need repeating multiple times, since each measurement returns either the result ``$0$'' or ``$1$'' and the wavefunction then collapses. Each repeating is called a shot. 

The parameters are chosen as follows for the results we are going to show: $K_{\rm max}^+=10$, $K_{\rm max}^\perp=1$, $g_d = 0.3$ and $g_s = 0.1$. In the toy model, everything is unitless. The time evolution starts at $t=0$. We choose $\Delta t = 0.01$ for the Trotterization step. To study the LPM effect in the medium, we will compare the total radiation probability in vacuum with that in the medium. In the former case, the dynamics is described by the kinetic and splitting terms of the Hamiltonian $H_{\rm kin}+H_{\rm split}$, while in the latter, all three parts of the Hamiltonian $H_{\rm kin}+H_{\rm diff}+H_{\rm split}$ are used in the description of the time evolution. The vacuum evolution can also be thought of describing medium-induced radiation without the LPM effect. The off-shell-ness of the parton in the case of medium-induced radiation is caused by $H_{\rm diff}$ during in-medium evolution before $t=0$. Then $H_{\rm diff}$ is turned off at $t=0$ so the time evolution after $t=0$ describes medium-induced radiation without the LPM effect. For the in-medium simulation, we also need to average the results over multiple trajectories. For each trajectory, the classical background fields need regenerating. At each time step of a trajectory, we sample the classical background field $\bar{A}^-(K_{\rm max}^+)$ by assuming it is described by a Gaussian distribution. The mean and the standard deviation of the Gaussian distribution are assumed to be $0$ and $3$ respectively.

\begin{figure}[t]
    \begin{subfigure}[t!]{0.5\textwidth}
    \centering
    \includegraphics[width=1.0\textwidth]{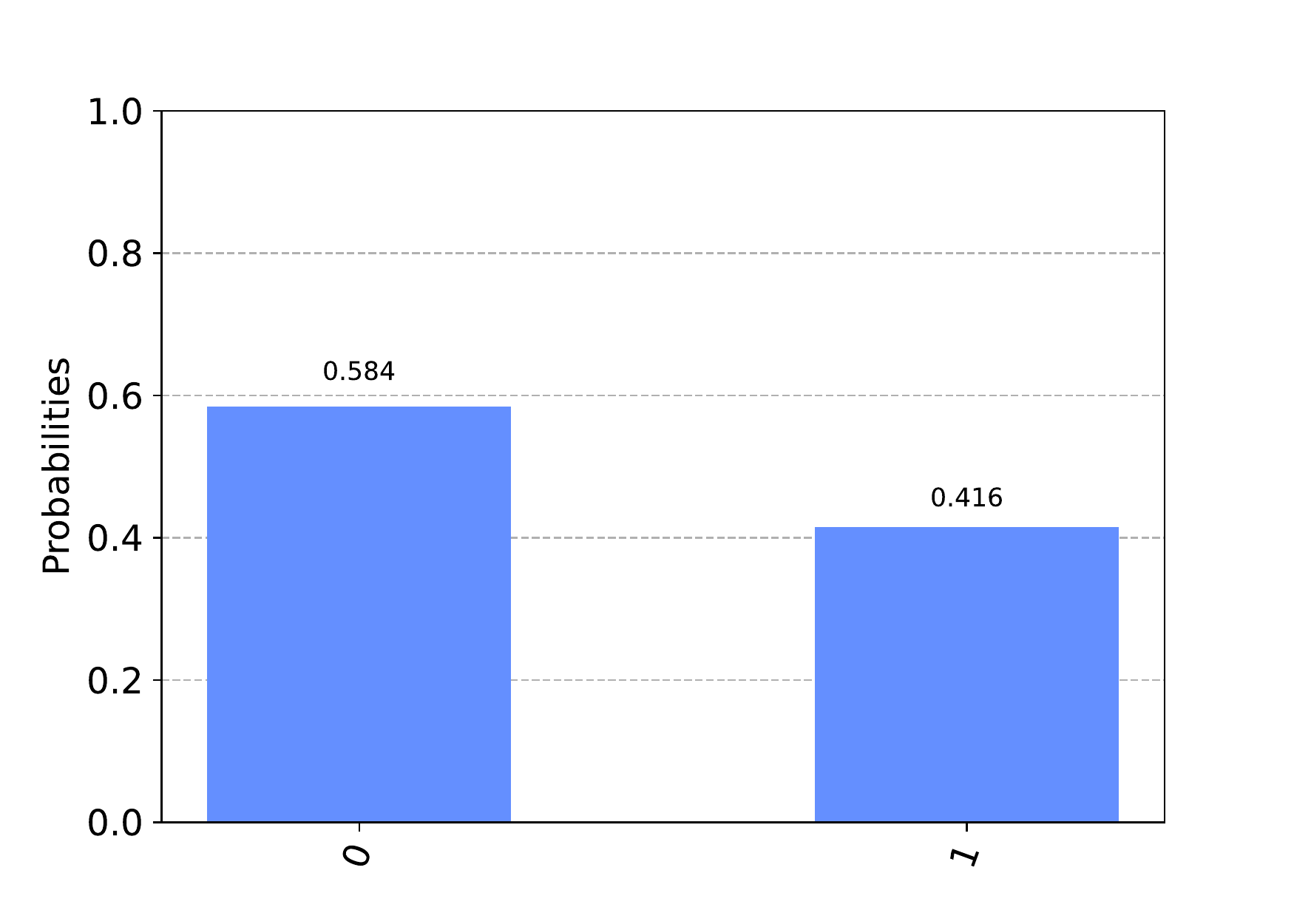}
    \caption{Vacuum case.}
    \end{subfigure}%
    ~
    \begin{subfigure}[t!]{0.5\textwidth}
    \centering
    \includegraphics[width=1.0\textwidth]{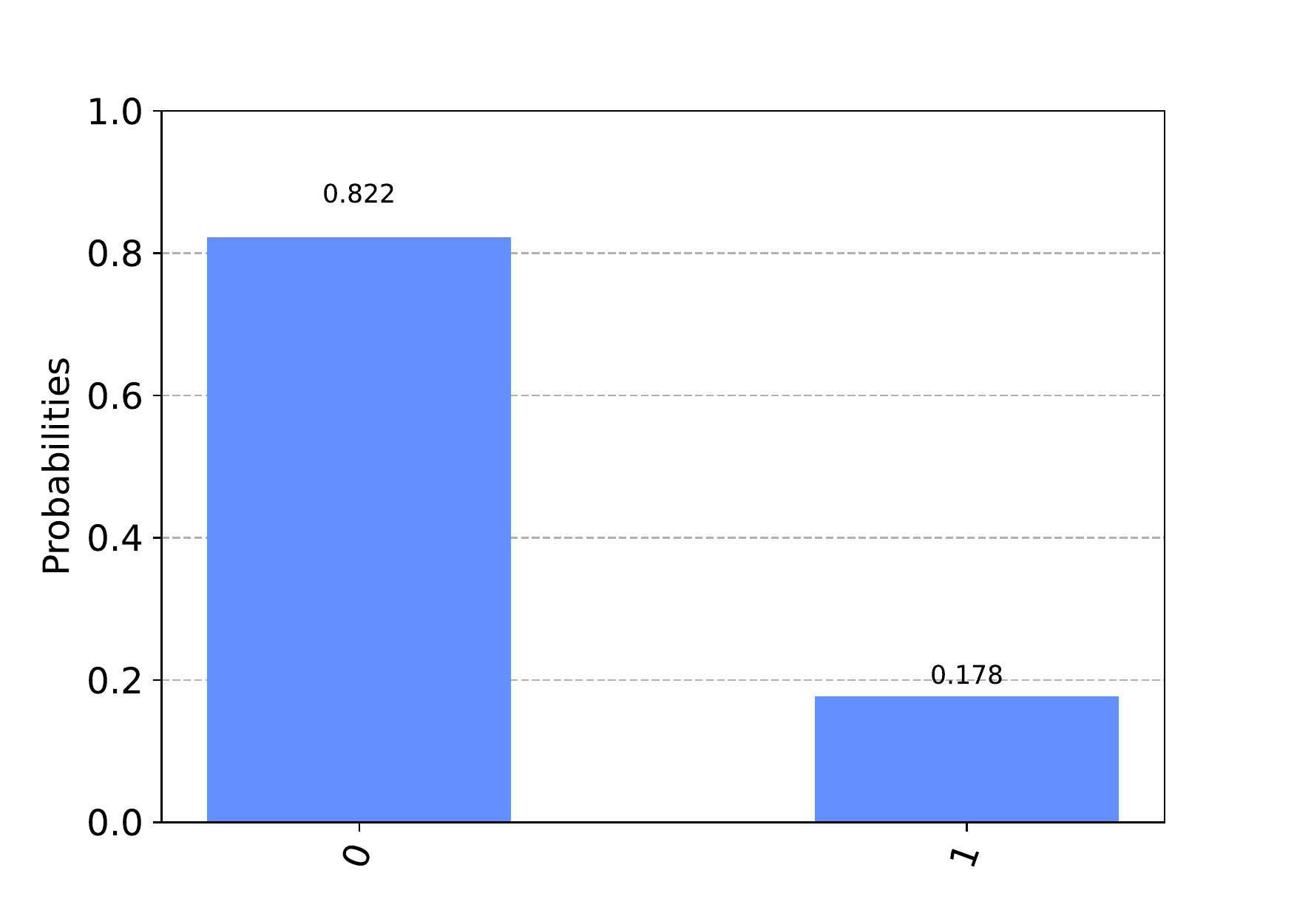}
    \caption{Medium case.}
    \end{subfigure}%
\caption{Quantum simulation results of the total radiation probabilities at time $t=7$ for the vacuum (left) and medium (right) cases. The measurement result ``$0$'' corresponds to no radiation while ``$1$'' indicates the occurrence of $1\to2$ splitting. The vacuum result also represents the case of medium-induced radiation without the LPM effect. The total radiation probability is suppressed in the medium case due to the quantum decoherence effect.}
\label{fig:results}
\end{figure}

The quantum simulation results of the total radiation probabilities at time $t=7$ are shown in Fig.~\ref{fig:results} for the vacuum and medium cases, where the result ``$0$'' indicates that no radiation happens and the final state is still a $1$-particle state while ``$1$'' represents that the $1\to2$ splitting occurs and the final state contains two particles. The vacuum result is obtained from $2^{20}$ shots while the medium result is obtained from averaging $500$ trajectories. The result for each trajectory is estimated from $2^{20}$ shots and every shot uses the same set of classical background fields sampled for the trajectory. It can be seen that once we turn on the diffusion Hamiltonian which originates from the transverse momentum exchange between partons and the medium, the radiation probability is suppressed.

\begin{figure}[t]
\centering
\includegraphics[width=0.7\textwidth]{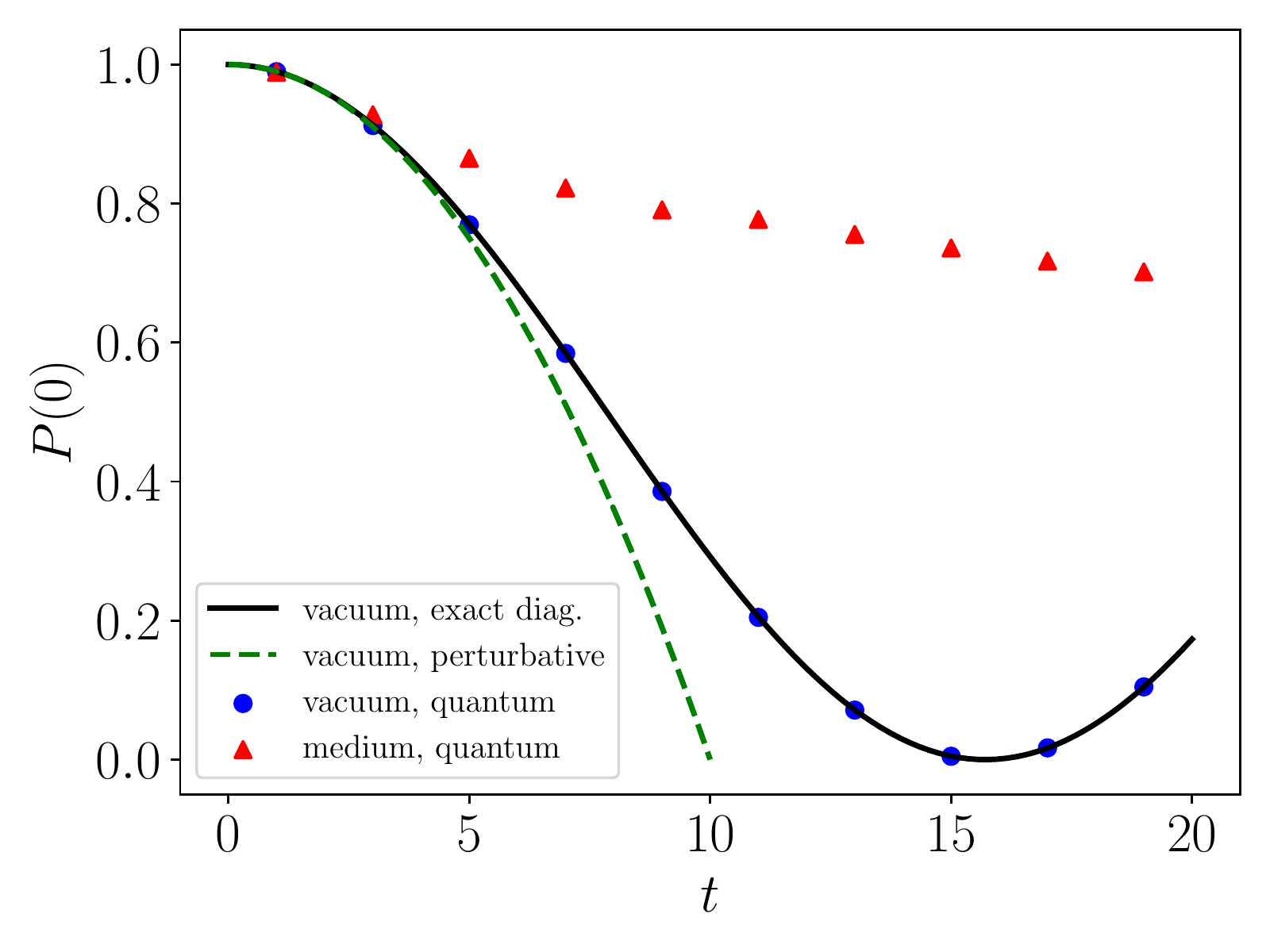}
\caption{Probabilities of no radiation in vacuum and in the medium as functions of time. The black solid line is obtained from exact diagonalization of the Hamiltonian in vacuum. The green dashed line is obtained from a first order perturbative calculation in vacuum. The blue and red points are obtained from quantum simulations. The vacuum result also represents the case of medium-induced radiation without the LPM effect.}
\label{fig:time_evol}
\end{figure}

To better understand the result, we calculate the probabilities of no radiation in vacuum and in the medium as functions of time, shown in Fig.~\ref{fig:time_evol}. The black solid line is obtained by exactly diagonalizing the vacuum Hamiltonian $H_{\rm vac} = H_{\rm kin} + H_{\rm split}$. The blue dots are obtained from a quantum simulation performed for the vacuum Hamiltonian, as described above. We carry out measurements at specific times corresponding to the horizontal locations of the blue dots. We see the quantum simulation results agree well with those obtained from exact diagonalization and phase rotation done classically, which indicates the Trotterization errors here are tiny with $\Delta t = 0.01$. The green dashed line is given by a first order perturbative calculation in the Schr\"odinger picture. Analytically, the quantum circuit with the measurement result ``1'' corresponds to
\begin{align}
\langle 1 | \rho(t) | 1 \rangle = \langle 1 | U(t,0) \rho(0) U(0,t) | 1 \rangle \,,
\end{align}
where $| 1 \rangle$ represents a 2-particle state and $U(t,0)$ is the unitary time evolution operator in the Schr\"odinger picture
$U(t,0) = \exp(-i H_{\rm vac}t)$ ($H_{\rm vac} = H_{\rm kin} + H_{\rm split}$). If we set the initial state to be a 1-particle state $\rho(0) = |0\rangle \langle0|$ and expand the unitary operator to first order in $H_{\rm vac}$, we obtain
\begin{align}
\langle 1 | \rho(t) | 1 \rangle = t^2 \langle 1 | H_{\rm vac} |0\rangle \langle0| H_{\rm vac} | 1 \rangle = g_s^2 t^2\,,
\end{align}
where we have used $\langle 1 | H_{\rm kin} |0\rangle = 0$ and $\langle 1 | H_{\rm split} |0\rangle = g_s$. This gives the total radiation probability and the probability of no radiation is $1-g_s^2t^2$.
We see that when $t$ is small, the first order perturbative result agrees well with the exact result in vacuum. At later times, the perturbative result deviates and it can be improved by expanding to second order in perturbation. We also note the vacuum result is oscillating in time, which is caused by higher order interactions. Finally, the quantum simulation results of the in-medium radiation process are marked as red upper triangles. The radiation probability in the medium is smaller than that in vacuum, for the time period that is studied here. (The first blue and red points from the left almost overlap with each other. But we have checked that indeed the in-medium radiation probability is smaller at that time point.) The suppression is caused by quantum decoherence, which is the essence of the LPM effect. One may worry that the vacuum result is oscillating in time so at late times the vacuum radiation probability will be smaller than the medium case. We will not discuss this issue here since the main motivation of the toy model is to show the construction of quantum circuits and study the quantum decoherence effect in the quantum simulation. We will come back to this issue in section~\ref{sect:gluon_simulation}.

\section{Quantum Simulation of Gluon Radiation in Medium}
\label{sect:gluon}
In this section, we discuss a more complicated case: gluon radiation in a quark-gluon plasma at thermal equilibrium. We focus on the gluon splitting $g\to gg$ process in the hot medium, since the quark splitting process $q\to qg$ is suppressed in the high energy limit, as explained in section~\ref{sect:split}. We will first construct the relevant discretized Hamiltonian.

\subsection{Discretized Hamiltonian}
The states in the computational basis~\eqref{eqn:n-state} have continuous momenta. To encode them on a quantum computer, we need to discretize the momenta. With discretized momenta, we want the 1-particle states to be normalized as 
\begin{align}
\label{eqn:1particle_norm_discrete}
\big\langle q,\, k_1^+,\, k_{1\perp},\, i_1,\, \sigma_1 \big| q,\, k_2^+,\, k_{2\perp},\, i_2,\, \sigma_2 \big\rangle & = \delta_{k_1^+ k_2^+} \delta_{k_{1x} k_{2x}} \delta_{k_{1y} k_{2y}} \delta_{i_1i_2} \delta_{\sigma_1 \sigma_2} \,, \\
\big\langle g,\, k_1^+,\, k_{1\perp},\, a_1,\, \lambda_1 \big| g,\, k_2^+,\, k_{2\perp},\, a_2,\, \lambda_2 \big\rangle & = \delta_{k_1^+k_2^+} \delta_{k_{1x} k_{2x}} \delta_{k_{1y} k_{2y}} \delta_{a_1a_2} \delta_{\lambda_1 \lambda_2} \,, \nn
\end{align}
where the Dirac delta functions of momenta in Eq.~\eqref{eqn:1particle_norm} become Kronecker delta functions of discretized momenta. When replacing Dirac delta functions with Kronecker ones, we also changed the dimensions of states: the mass dimension of a 1-particle state in Eq.~\eqref{eqn:1particle_norm} is $-1.5$ while in Eq.~\eqref{eqn:1particle_norm_discrete} the mass dimension is $0$. This can be seen from the discretized version of a Dirac delta function:
\be
\delta(k_{1\mu}-k_{2\mu}) \to \frac{1}{\Delta k_\mu} \delta_{k_{1\mu} k_{2\mu}}\,,
\ee
where $\Delta k_\mu$ is the lattice size of the momentum lattice along the $\mu$ direction. When writing down Eq.~\eqref{eqn:1particle_norm_discrete}, we implicitly multiplied Eq.~\eqref{eqn:1particle_norm} by $\Delta k^+ \Delta k_x \Delta k_y$.

To write down the discretized version of the Hamiltonian with the correct mass dimension, we need to take this multiplicative factor $\Delta k^+ \Delta k_x \Delta k_y$ into account. The general rule is that for each $n$-particle state involved in the matrix element of the Hamiltonian, we multiply the continuous version by a factor of $(\Delta k^+ \Delta k_x \Delta k_y)^{n/2}$. Applying this rule to Eqs.~(\ref{eqn:1particle_Hkin},~\ref{eqn:1particle_Hdiff},~\ref{eqn:1to2_Hsplit}) leads to
\begin{align}
\label{eqn:Hkin_dis1}
\big\langle q,\, k_1^+,\, k_{1\perp},\, i_1,\, \sigma_1 \big| H_{q,\,{\rm kin}} \big| q,\, k_2^+,\, k_{2\perp},\, i_2,\, \sigma_2 \big\rangle
& =  \frac{{\bs k}_{1\perp}^2}{k_1^+} \delta_{k_1^+k_2^+} \delta_{k_{1x} k_{2x}} \delta_{k_{1y} k_{2y}} \delta_{i_1i_2} \delta_{\sigma_1 \sigma_2} \,, \\
\big\langle g,\, k_1^+,\, k_{1\perp},\, a_1,\, \lambda_1 \big| H_{g,\,{\rm kin}} \big| g,\, k_2^+,\, k_{2\perp},\, a_2,\, \lambda_2 \big\rangle 
& = \frac{{\bs k}_{1\perp}^2}{k_1^+} \delta_{k_1^+k_2^+} \delta_{k_{1x} k_{2x}} \delta_{k_{1y} k_{2y}} \delta_{a_1a_2} \delta_{\lambda_1 \lambda_2} \,, \nn
\end{align}
for the kinetic Hamiltonian,
\begin{align}
\label{eqn:Hdiff_dis1}
&\big\langle q,\, k_1^+,\, k_{1\perp},\, i_1,\, \sigma_1 \big| H_{q,\,{\rm diff}}(x^+) \big| q,\, k_2^+,\, k_{2\perp},\, i_2,\, \sigma_2 \big\rangle \\
=& \begin{cases}
-\frac{g}{(2\pi)^2} \Delta k_{x} \Delta k_{y} 
\delta_{k_1^+k_2^+} \delta_{\sigma_1\sigma_2} T^a_{i_1i_2} \bar{A}^{-a}(x^+, {\bs k}_{1\perp} - {\bs k}_{2\perp})  \quad {\rm for\ quark} \\
+\frac{g}{(2\pi)^2} \Delta k_{x} \Delta k_{y}
\delta_{k_1^+k_2^+} \delta_{\sigma_1\sigma_2} T^a_{i_2i_1} \bar{A}^{-a}(x^+, {\bs k}_{1\perp} - {\bs k}_{2\perp})    \quad {\rm for\ antiquark}
\end{cases}
\,,\nn \\
&\big\langle g,\, k_1^+,\, k_{1\perp},\, a_1,\, \lambda_1 \big| H_{g,\,{\rm diff}}(x^+) \big| g,\, k_2^+,\, k_{2\perp},\, a_2,\, \lambda_2 \big\rangle \nn \\
=&\, \frac{ig}{ 2(2\pi)^2}\Delta k_{x} \Delta k_{y}
\delta_{k_1^+k_2^+} \delta_{\lambda_1\lambda_2} \big(f^{a_2ba_1}-f^{a_1ba_2}\big) \bar{A}^{-b}(x^+, {\bs k}_{1\perp} - {\bs k}_{2\perp}) \,,\nn
\end{align}
for the diffusion Hamiltonian and
\begin{align}
\label{eqn:Hsplit_dis1}
&\big\langle q, k_2^+, k_{2\perp}, i_2, \sigma_2 ; g, q^+, q_\perp, a, \lambda \big| H_{q,\,{\rm split}} \big| q, k_1^+, k_{1\perp}, i_1, \sigma_1  \big\rangle  \\[4pt]
&\quad = - g\sqrt{\frac{\Delta k^+ \Delta k_x \Delta k_y}{2(2\pi)^3q^+k_1^+k_2^+}}\delta_{k_1^+,\, k_2^++ q^+} \delta_{k_{1x},\, k_{2x} + q_x} \delta_{k_{1y},\, k_{2y} + q_y} \nn \\
&\qquad \times \bar{u}(k_2,\sigma_2) \bigg( \epsilon_\perp^i \gamma^i\gamma^j \frac{k_{1\perp}^j}{k_1^+} T^a_{i_2i_1} + \frac{k_{2\perp}^i}{k_2^+}\gamma^i\gamma^j \epsilon_\perp^j T^a_{i_2i_1}
+ 2T^a_{i_2i_1} \frac{q_\perp^i}{q^+} \epsilon_\perp^i 
\bigg) u(k_1,\sigma_1) \,, \nn\\
&\big\langle g, k_2^+, k_{2\perp}, a_2, \lambda_2 ; g, k_3^+, k_{3\perp}, a_3, \lambda_3 \big| H_{g,\,{\rm split}} \big| g, k_1^+, k_{1\perp}, a_1, \lambda_1  \big\rangle \nn \\[4pt] 
& \quad = -ig \sqrt{\frac{\Delta k^+ \Delta k_x \Delta k_y}{2(2\pi)^3k_1^+k_2^+k_3^+}} f^{a_1a_2a_3} \delta_{k_1^+,\, k_2^+ + q^+} \delta_{k_{1x},\, k_{2x} + q_x} \delta_{k_{1y},\, k_{2y} + q_y} \nn \\
&\qquad \bigg( k_1^+\epsilon_\perp^i(\lambda_1) \Big[ \frac{k_{2\perp}^j }{k_2^+} \epsilon_\perp^j(\lambda_2) \epsilon_{\perp i}(\lambda_3)
- \frac{k_{3\perp}^j }{k_3^+} \epsilon_\perp^j(\lambda_3) \epsilon_{\perp i}(\lambda_2)
\Big] 
- k_2^+\epsilon_\perp^i(\lambda_2) \Big[ \frac{k_{3\perp}^j }{k_3^+} \epsilon_\perp^j(\lambda_3) \epsilon_{\perp i}(\lambda_1) \nn\\
&\qquad - \frac{k_{1\perp}^j }{k_1^+} \epsilon_\perp^j(\lambda_1) \epsilon_{\perp i}(\lambda_3)
\Big] 
- k_3^+\epsilon_\perp^i(\lambda_3) \Big[ \frac{k_{1\perp}^j }{k_1^+} \epsilon_\perp^j(\lambda_1) \epsilon_{\perp i}(\lambda_2)
- \frac{k_{2\perp}^j }{k_2^+} \epsilon_\perp^j(\lambda_2) \epsilon_{\perp i}(\lambda_1)
\Big] \nn\\
&\qquad - k_{1\perp}^i \epsilon_\perp^j(\lambda_1) \Big[ \epsilon_{\perp i}(\lambda_2) \epsilon_{\perp j}(\lambda_3) - \epsilon_{\perp i}(\lambda_3) \epsilon_{\perp j}(\lambda_2)\Big]
+ k_{2\perp}^i \epsilon_\perp^j(\lambda_2) \Big[ \epsilon_{\perp i}(\lambda_3) \epsilon_{\perp j}(\lambda_1) \nn\\
& \qquad - \epsilon_{\perp i}(\lambda_1) \epsilon_{\perp j}(\lambda_3) \Big] 
+ k_{3\perp}^i \epsilon_\perp^j(\lambda_3) \Big[ \epsilon_{\perp i}(\lambda_1) \epsilon_{\perp j}(\lambda_2) - \epsilon_{\perp i}(\lambda_2) \epsilon_{\perp j}(\lambda_1)\Big]
\bigg) \,,\nn
\end{align}
for the splitting Hamiltonian. With discretized momenta, the correlation function of two classical background fields~\eqref{eqn:AA_sametime} can be written as
\be
\label{eqn:AA_discrete}
\big\langle \bar{A}^{-a}(x^+, k_\perp)
\bar{A}^{-a}(x^+, -k_\perp) \big\rangle = \frac{(2\pi)^2 \gamma({\bs k}_\perp)}{\Delta x^+ \Delta k_x \Delta k_y}\,.
\ee
As we discussed earlier, to make the diffusion Hamiltonian Hermitian, we set 
$\bar{A}^{-a}(x^+, k_\perp) = 
\bar{A}^{-a}(x^+, -k_\perp)$ as the same random variable, which can be sampled from a Gaussian distribution with the variance $\frac{(2\pi)^2 \gamma({\bs k}_\perp)}{\Delta x^+ \Delta k_x \Delta k_y}$.

\subsection{Hilbert Space}
We can neglect quark degrees of freedom since we focus on the $g\to gg$ process. With a limited number of qubits, we discretize the $+$, $x$ and $y$ components of momenta as
\begin{align}
\label{eqn:1gluon_k}
k^+ \in K^+_{\rm max} \{ 0.5, 1\}\,,\qquad k_x \in K^\perp_{\rm max} \{ 0, 1\}  \,, \qquad k_y \in K^\perp_{\rm max} \{ 0, 1\}\,.
\end{align}
As a result, we need 3 qubits to describe the momentum of a gluon, 1 for each component. Then we need another 3 qubits to describe the color of a gluon and 1 qubit for the polarization (spin). Totally we need 7 qubits to represent a gluon state:
\begin{align}
\label{eqn:1gluon}
| \underbrace{q_1q_2q_3}_\text{describe momentum} \overbrace{q_4 q_5q_6}^\text{describe color} \underbrace{q_7}_\text{describe polarization}  \rangle \,.
\end{align}
Since we have both 1-gluon and 2-gluon states in the process, we need 15 qubits to represent a state: 7 qubits for each gluon and 1 qubit to distinguish between the 1-gluon and 2-gluon states:
\begin{align}
\label{eqn:gluons}
| \underbrace{q_1}_\text{separate 1-gluon and 2-gluon states} \overbrace{q_2 q_3\dots q_8}^\text{describe 2nd gluon} \underbrace{q_9q_{10}\dots q_{15}}_\text{describe 1st gluon}  \rangle \,.
\end{align}
When $q_1=0$, the state is a 1-gluon state and the qubits $q_2q_3\dots q_8$ are redundant so we just set them to be all zeros:
\be
|0\ 0000000\ q_9q_{10}\dots q_{15}\rangle \,.
\ee
When $q_1=1$, the state is a 2-gluon state 
\be
|1\ q_2q_3\dots q_8\ q_9q_{10}\dots q_{15}\rangle \,.
\ee

Once we fix the computational basis, the matrix elements of each part of the Hamiltonian can be written down according to Eqs.~(\ref{eqn:Hkin_dis1},~\ref{eqn:Hdiff_dis1},~\ref{eqn:Hsplit_dis1}). Here we will not write these matrix elements out explicitly, neither their decomposition into tensor products of Pauli matrices, which becomes very lengthy but can be done. We have 15 qubits here and the generic method of decomposing the Hamiltonian into tensor products of Pauli matrices discussed in section~\ref{sect:decomposition} requires evaluating $4^{15}$ coefficients by using Eq.~\eqref{eqn:coeff_decom}. However, we know many of the coefficients are zeros since the Hamiltonian is sparse. The generic method of decomposition discussed in section~\ref{sect:decomposition} does not employ any property or structure of the system's Hamiltonian. A more efficient decomposition method that employs the structure of the system is illustrated in appendix~\ref{app:alter}, where we first construct $H_{\rm kin}$ and $H_{\rm diff}$ for 1-particle states and $H_{\rm split}$ for transitions between 1-particle and 2-particle states, and then use them as building blocks for states consisting of more particles. The strategy is to first construct Pauli matrix representations for smaller pieces of a Hamiltonian and then put all pieces together by tensor products. For the gluon radiation case in QCD, the new ingredient is the color and spin changes. Since the color part factorizes in the QCD Hamiltonian matrix elements, we can construct the Pauli matrix representations for the color change and the change of momentum and spin separately as smaller qubit systems and then take their tensor product. We discuss some useful decomposition formulas for these smaller qubit systems in appendix~\ref{app:qcd_decom}. 

\subsection{Simulation Results}
\label{sect:gluon_simulation}
Here we perform the simulation via keeping track of the statevector, i.e., the wavefunction, rather than using a quantum circuit consisting of 15 qubits.\footnote{The decomposition of each part of the Hamiltonian (kinetic, diffusion and splitting) into tensor products of Pauli matrices is straightforward, as explained in appendix~\ref{app:qcd_decom}. But constructing the corresponding quantum circuit in the IBM Qiskit simulator by appending single-qubit and CNOT gates to the circuit one by one in the code becomes extremely tedious.} The initial state is set as a 1-gluon state
\be
|\psi(t=0)\rangle = |0\ 0000000\ 1110000\rangle \,,
\ee
where $t=x^+/2$. One should think of the initial parton as being off mass shell, which can be caused by hard scattering or interaction with the medium. In the latter case, what we mean by the vacuum process is really medium-induced radiation without the LPM effect. What happens in the time evolution of medium-induced radiation without the LPM effect is that $H_{\rm diff}$ is turned on before $t=0$ which generates partons that are off-shell and thus radiating. Then $H_{\rm diff}$ is turned off at $t=0$, after which the time evolution describes medium-induced radiation without the LPM effect.
We time evolve the wavefunction according to
\begin{align}
\Big( e^{-iH_{\rm split}\Delta t} e^{-iH_{\rm kin}\Delta t} \Big)^{N_t} |\psi(t=0)\rangle \,,
\end{align}
and 
\begin{align}
\Big( e^{-iH_{\rm split}\Delta t} e^{-iH_{\rm diff}\Delta t} e^{-iH_{\rm kin}\Delta t} \Big)^{N_t} |\psi(t=0)\rangle \,,
\end{align}
for the vacuum and medium cases respectively. We choose $K_{\rm max}^+=100$ or $50$ GeV, $K_{\rm max}^\perp = 1$ GeV and the strong coupling $g=2$ at the scale 1 GeV, since the transverse momentum transferred is 1 GeV. The Trotterization time step is fixed to be $\Delta t=0.01$ fm/c. For the medium case, we need to sample classical background gauge fields $\bar{A}^{-a}$ at each time step from Gaussian distributions with the variances given in Eq.~\eqref{eqn:AA_discrete}. Classical background gauge fields with different momenta have different variances, but those with only different colors have the same variance. The variance depends on the function $\gamma({\bs k}_\perp)$. Here we use the model shown in Eq.~\eqref{eqn:gamma_HTL} for $\gamma({\bs k}_\perp)$. The temperature of the QGP is fixed to be $T=300$ MeV and can be easily made time dependent in our framework. The Debye mass is related to the temperature via
\be
m_D^2 = \frac{1}{3} \Big(N_c+\frac{N_f}{2} \Big)g^2 T^2\,,
\ee
where we take $N_c=3$ and $N_f = 3$. After updating the classical background gauge fields at each time step, we need to reconstruct the diffusion part of the Hamiltonian, which can be computationally expensive. Therefore, in practice we sample 3000 sets of $\bar{A}^{-a}$ and construct the corresponding $H_{\rm diff}$ that are saved in storage. At each time step, we just take one $H_{\rm diff}$ randomly from the 3000 ensemble.

\begin{figure}[t]
    \begin{subfigure}[t!]{0.49\textwidth}
        \centering
        \includegraphics[height=2.1in]{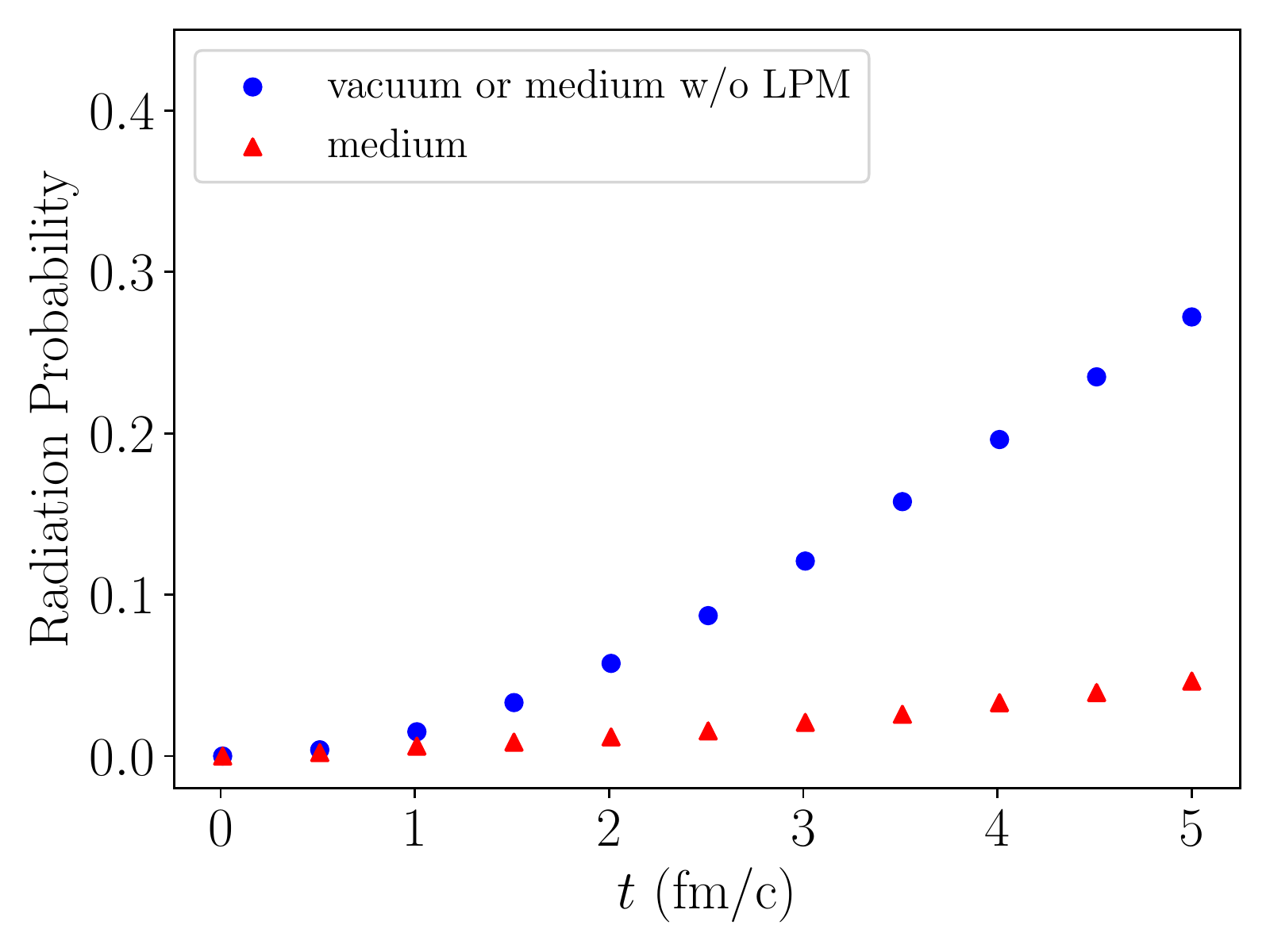}
        \caption{Initial $k^+=100$ GeV.}
        \label{fig:1}
    \end{subfigure}%
    ~
    \begin{subfigure}[t!]{0.49\textwidth}
        \centering
        \includegraphics[height=2.1in]{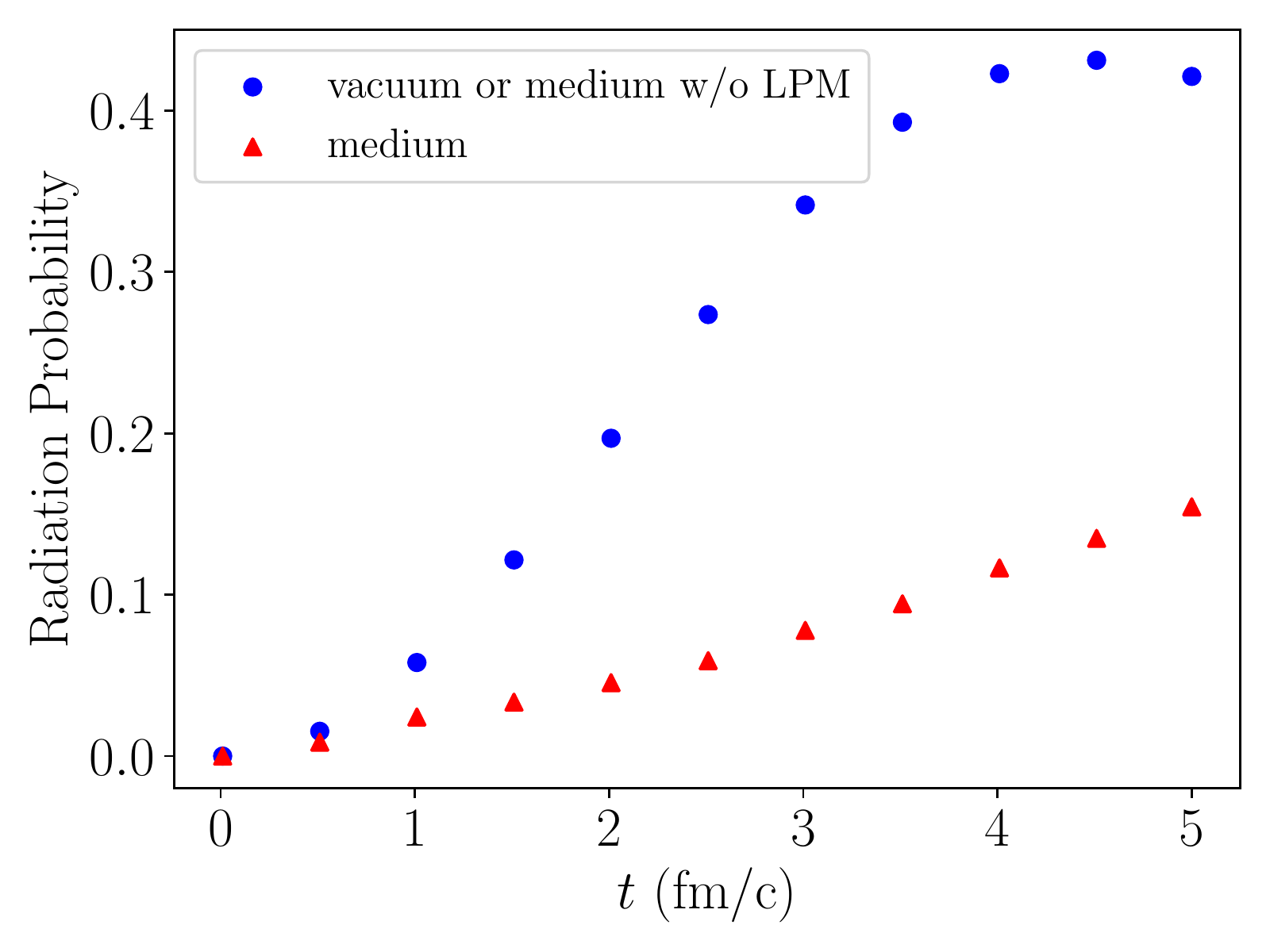}
        \caption{Initial $k^+=50$ GeV.}
        \label{fig:2}
    \end{subfigure}%
\caption{Gluon radiation probabilities as functions of time for both the vacuum (blue dots) and medium cases (red upper triangles) with two different initial $+$ momenta: $100$ GeV (left) and $50$ GeV (right). The vacuum result can also represent the case of medium-induced radiation without the LPM effect. The gluon radiation probability is suppressed in the medium.}
\label{fig:time_evol_qcd}
\end{figure}

The gluon radiation probabilities as functions of time are plotted in Fig.~\ref{fig:time_evol_qcd} for both the vacuum (or medium without the LPM effect) and medium cases with two different initial $+$ momenta. We see that the gluon radiation probability in the time period studied here is largely suppressed in the medium where random transverse momentum exchanges occur frequently and cause decoherence. Furthermore, we note that the radiation probability in the initial $k^+=50$ GeV case is larger than that in the initial $k^+=100$ GeV case, which means that more energetic partons lose less energy in the medium (we think of the vacuum case as medium-induced radiation without the LPM effect).

At late times in the case with an initial $k^+=50$ GeV, we observe the vacuum radiation probability starts to drop, which indicates the radiation probability in vacuum is oscillating in time, as already seen in Fig.~\ref{fig:time_evol} for the toy model. There are two potential reasons for the time oscillation. The first one is higher order correction. To see how higher order terms can result in oscillation, we consider a simple example of a two-level system with an interaction given by $g_x\sigma^x$. The transition amplitude between the ground state $|0\rangle$ and the excited state $|1\rangle$ is
\begin{align}
\langle 1 | e^{-ig_x\sigma^x t} | 0\rangle = -i\sin(g_xt) \,.
\end{align}
The oscillating behavior becomes manifest when $t\gtrsim 1/g_x$. Considering the prefactor in the splitting Hamiltonian~\eqref{eqn:Hsplit_dis1}, we conclude that for the time period studied here, higher order correction is not the reason behind the oscillating behavior seen in Fig.~\ref{fig:time_evol_qcd}.

The second potential reason of the oscillating behavior is the phase oscillation caused by an energy mismatch in the initial and final states. To see this more clearly, we use time-ordered perturbation theory in the interaction picture to calculate the transition amplitude between a $1$-parton state $|0\rangle$ and a $2$-parton state $|1\rangle$ (we assume these two states are eigenstates of the free Hamiltonian $H_0$, i.e., $H_0|i\rangle=E_i|i\rangle$ for $i=0,1$)
\begin{align}
\langle1| \ml{T} e^{-i\int_0^t\diff t_1 H_I(t_1)} |0\rangle = -i \int_0^t\diff t_1 e^{i(E_1-E_0)t_1} \langle1| H_I |0\rangle +\ml{O}(H_I^2)\,,
\end{align}
where $\ml{T}$ denotes the time-ordering operator and the interaction Hamiltonian in the interaction picture is given by
$H_I(t)= e^{iH_0t} H_I e^{-iH_0t}$. The time integration gives
\begin{align}
\int_0^t\diff t_1 e^{i(E_1-E_0)t_1} = 2e^{i(E_1-E_0)t/2} \frac{\sin({\frac{E_1-E_0}{2}t)}}{E_1-E_0} \,.
\end{align}
In the $t\to+\infty$ limit, the above time integral corresponds to a delta function in $E_1-E_0$:
\begin{align}
\lim_{t\to+\infty}\frac{2\sin{\frac{E_1-E_0}{2}t}}{E_1-E_0} = \pi \delta\Big(\frac{E_1-E_0}{2} \Big)\,,
\end{align}
which means the transition can only happen if the initial and final states have the same energy. The transition probability when $t$ is large is given by
\begin{align}
|\langle1| \ml{T} e^{-i\int_0^t\diff t_1 H_I(t_1)} |0\rangle|^2 = 2\pi\delta(E_1-E_0) | \langle 1|H_I|0\rangle |^2 t \,,
\end{align}
and the transition rate can be well defined in the $t\to+\infty$ limit. This is the case when we derive the Fermi's golden rule to calculate scattering cross sections for asymptotic states and the decay rate of an initial particle. Final states with mismatched energies will not contribute to cross sections or decay rates.

At any finite $t$, we see the transition probability $|\langle 1| \ml{T} e^{-i\int_0^t\diff t_1 H_I(t_1)} |0\rangle|^2 $ is oscillating in time if $E_1\neq E_0$. Since the momentum grid is very coarse here, an exact equality between the energies of the initial 1-gluon state and the final 2-gluon state is not possible. The typical energy gap in the gluon radiation process studied here is on the order of $2{\bs k}_{\perp}^2/k^+ \sim 0.02$ GeV for the initial $k^+=100$ GeV case and $0.04$ GeV for the initial $k^+=50$ GeV case. It will take about $10$ fm/c and $5$ fm/c to see the oscillation in radiation probability caused by the oscillating phase in the two cases respectively, which is consistent with the observation here. 



In short, the oscillating behavior of the vacuum radiation probability is caused by not having fine enough grids in the momenta, which results in a mismatch in the energies of the initial and final states. Despite this caveat, we still see the quantum decoherence effect in the time evolution shown in  Fig.~\ref{fig:time_evol_qcd}, which is the essence of the LPM effect. For future physical applications, one needs to perform quantum simulations with larger and finer grids in momenta. One also needs to properly take the continuum and infinite volume limits. As a sanity check, one should verify the well-known result of the LPM effect in one splitting from quantum simulation. Then one can move on to use quantum simulation to study the LPM effects in multiple splittings, which is beyond the scope of current analyses. These are left for future studies.

\section{Conclusions}
\label{sect:conclusion}
In this paper, we developed a framework to perform quantum simulation of jet quenching in nuclear environments. The quantum simulation automatically keeps track of quantum interference that is crucial in the studies of the LPM effect for multiple coherent radiations, since it simulates the time evolution of a state wavefunction. We used the light-front Hamiltonian of QCD to describe the time evolution of high energy partons in nuclear media. The light-front Hamiltonian relevant for jet quenching consists of three pieces: a kinetic term which induces a phase change in the time evolution, a diffusion term caused by transverse momentum (Glauber) exchanges between the high energy partons and the medium, and a splitting term accounting for parton radiation and recombination. We use $n$-particle states in momentum space as the basis of the physical Hilbert space and estimated the qubit cost. In this basis, the kinetic Hamiltonian becomes diagonal, which can be efficiently simulated on a quantum computer. Furthermore, the matrices of the diffusion and splitting parts are sparse. Therefore, one may be able to efficiently simulate multiple coherent radiations in a medium on a quantum computer and study the LPM effect therein. The diffusion term in the Hamiltonian depends on some classical background fields, of which the medium is a source. When constructing a quantum circuit for the Hamiltonian evolution, one needs to sample these classical background fields on a classical computer and then plug their values into the quantum circuit. This classical sampling scales as $\ml{O}(t V_k)$ where $t$ is the length of the time evolution and $V_k$ is the volume of the momentum space. Quantum trajectories with different sets of classical background fields in the simulation need to be averaged to give estimates of physical results. Then we applied this framework to study a toy model, by explicitly constructing a quantum circuit to simulate the time evolution. We also studied the gluon radiation process in a hot medium with and without the LPM effect. We observed the quantum decoherence effect in both the toy model and the gluon case that suppresses the total radiation probability, which is the essence of the LPM effect, despite a caveat caused by the small momentum lattice. For future physical applications, one should use larger and finer momentum grids for the simulation and investigate the effect of the zero mode and how to take the continuum and infinite volume limits. One should also verify the well-known result of the LPM effect in one splitting from quantum simulation and then study cases with multiple coherent splittings.

The framework developed here is general and it can be used to study jet quenching for various media that are either static or expanding, thin or thick, hot or cold. It can also be applied for cases where the classical background fields satisfy some non-Gaussian correlations. Since the framework automatically keeps track of quantum interference, it can be applied to study the LPM effect with more than two coherent splittings in a dynamically evolving medium, which is beyond the scope of state-of-the-art analyses. This framework of quantum simulation may help to deepen our understanding of jet quenching in nuclear environments in the near future with the advancement of quantum technology that provides more qubits of high fidelity, which is important for studies of jet production in current heavy ion collisions and in the forthcoming Electron-Ion Collider.

\acknowledgments
XY thanks Anthony Ciavarella, Thomas Mehen, Gerald Miller, Krishna Rajagopal, Martin Savage and Marc Illa Subina for useful discussions. The work of XY was supported by the U.S. Department of Energy, Office of Science, Office of Nuclear Physics, InQubator for Quantum Simulation (IQuS) under Award Number DOE (NP) Award DE-SC0020970 and grant DE-SC0011090.

\appendix
\section{Light-Front Hamiltonian of QCD}
\label{app:lfqcd}
Here we review the light-front Hamiltonian approach for QCD. Recent reviews can be found in Refs.~\cite{Brodsky:1997de,Bakker:2013cea}.

We start with the QCD Lagrangian density with one massive fermion field
\be
\ml{L} = \overline{\psi} (i\slashed{D} - m) \psi - \frac{1}{2}\rm{Tr}\big( F^{\mu\nu} F_{\mu\nu} \big)\,,
\ee
where $\slashed{D} = \gamma^\mu D_\mu$, $D_\mu = \partial_\mu - ig A_\mu$ and $F^{\mu\nu}=\frac{i}{g}[ D^\mu, D^\nu]$. Writing the color indexes out explicitly leads to 
\be
\ml{L} = \overline{\psi}_i (i\slashed{D}_{ij} - m \delta_{ij}) \psi_j - \frac{1}{4} F^{\mu\nu a} F_{\mu\nu}^a\,,
\ee
where $i,j,\cdots$ denote the fundamental color indexes and $a,b,\cdots$ represent the adjoint color indexes and we have used
$A^\mu = A^{\mu a}T^a$, $F^{\mu\nu a} = \partial^\mu A^{\nu a} - \partial^\nu A^{\mu a} + g f^{abc} A^{\mu b} A^{\nu c}$ and ${\rm Tr}(T^a T^b) = \frac{1}{2}\delta^{ab}$. Here we only raise and lower the Lorentz indexes but not the color indexes.

We will use the light-cone coordinates defined by
\be
x^\pm = x^0 \pm x^3\,,\quad\quad \gamma^\pm = \gamma^0 \pm \gamma^3\,,\quad\quad A^\pm = A^0 \pm A^3 \,,
\ee
where $x^+$ denotes the light-cone time while $x^-$ is the light-cone longitudinal coordinate. The metric is fixed as ($\mu=+,1,2,-$)
\be
g_{\mu\nu} = \begin{pmatrix}
0 & 0 & 0 & \frac{1}{2} \\
0 & -1 & 0 & 0 \\
0 & 0 & -1 & 0 \\
\frac{1}{2} & 0 & 0 & 0 \\
\end{pmatrix}\,, \quad\qquad 
g^{\mu\nu} = \begin{pmatrix}
0 & 0 & 0 & 2 \\
0 & -1 & 0 & 0 \\
0 & 0 & -1 & 0 \\
2 & 0 & 0 & 0 \\
\end{pmatrix} \,.
\ee
The inner product between two vectors is given by
\be
x\cdot y = x^\mu y_\mu = \frac{x^+y^- + x^- y^+}{2} + x_\perp \cdot y_\perp = \frac{x^+y^- + x^- y^+}{2} - {\bs x}_\perp \cdot {\bs y}_\perp \,,
\ee
where a bold symbol is used for Euclidean vectors to make them distinct from Minkowski vectors that are not bold. 
For the transverse components, we define the notation 
\be
x_\perp \cdot y_\perp = - {\bs x}_\perp \cdot {\bs y}_\perp = x_\perp^i y_{\perp i} = -x_{\perp i} y_{\perp i}  = -x_{\perp}^i y_{\perp}^i \,.
\ee
The momentum component conjugated to $x^+$ is the light-cone energy $p^-$ while the momentum component conjugated to $x^-$ is the longitudinal momentum $p^+$. From the on-shell condition $p^2 = m^2$, we find $p^- = ({\bs p}_\perp^2 + m^2) / p^+$. If $p^+$ is large, $p^-$ will be small. For $m\neq0$, $p^+>0$. If $m=0$, $p^+$ can be zero, which is the case for gluons. In this work, we focus on collinear radiation processes where both the mother and daughter partons have large $+$ momenta. Therefore, we neglect the effect of the zero mode, which should be investigated in future studies.

In the following, we will use light-cone gauge $A^+=0$ and derive the light-front Hamiltonian density defined by
\be
\ml{H} = \sum_{\phi=\psi,\,A^\mu} \Pi_\phi\, \dot{\phi} - \ml{L}(\phi, \dot{\phi})\,,
\ee
where the canonical momentum is given by
\be
\Pi_\phi = \frac{\partial \ml{L}(\phi, \dot{\phi})}{\partial \dot{\phi}} \,.
\ee

\subsection{Fermion Sector}
We will use the Dirac representation of the gamma matrices
\be
\gamma^0 = \begin{pmatrix}
1 & 0 \\
0 & -1 
\end{pmatrix}\,, \quad\quad \gamma^i = \begin{pmatrix}
0 & \sigma_i \\
-\sigma_i & 0 
\end{pmatrix}\,, \quad\quad \gamma^\pm = \begin{pmatrix}
1 & \pm \sigma_z \\
\mp \sigma_z & -1 
\end{pmatrix}\,.
\ee

We define two projection operators
\be
\Lambda^+ = \frac{1}{2}\gamma^0 \gamma^+ = \frac{1}{4}\gamma^-\gamma^+ \,, \quad\quad \Lambda^- = \frac{1}{2}\gamma^0 \gamma^- = \frac{1}{4}\gamma^+\gamma^- \,.
\ee
Some useful identities are $(\gamma^\pm)^\dagger = \gamma^\mp$, $(\gamma^+)^2 = (\gamma^-)^2 = 0$, $\gamma^+\gamma^-=2\gamma^+\gamma^0=2\gamma^0\gamma^-$, $\gamma^-\gamma^+=2\gamma^-\gamma^0=2\gamma^0\gamma^+$,  $\gamma^+\gamma^-\gamma^+ = 4\gamma^+$ and $\gamma^-\gamma^+\gamma^- = 4\gamma^-$, with which one can easily show $(\Lambda^\pm)^\dagger=\Lambda^\pm$, $\Lambda^\pm \Lambda^\pm = \Lambda^\pm$ and $\Lambda^\pm \Lambda^\mp = 0$. Using the projection operators, we can decompose the fermion field
\be
\psi = \psi_+ + \psi_- = \Lambda^+ \psi + \Lambda^- \psi \,,
\ee
where the two fields are defined by $\psi_+^\dagger = \psi^\dagger \Lambda^+$ and $\psi_-^\dagger = \psi^\dagger \Lambda^-$ respectively.

The equation of motion for the fermion field $(i\slashed{D}-m)\psi = 0$ can be written out explicitly as
\be
\frac{1}{2}( \gamma^+ D^- + \gamma^-\partial^+ ) ( \psi_+ + \psi_- ) +  (\slashed{D}_\perp+im ) ( \psi_+ + \psi_- ) = 0 \,,
\ee
where we have set $A^+=0$. Using $\gamma^+\Lambda^- = \gamma^- \Lambda^+ =0$ and multiplying both sides on the left by $\gamma^0$, we find
\be
\label{app:dirac1}
D^- \psi_+ + \partial^+ \psi_- + \gamma^0 (\slashed{D}_\perp+im ) ( \psi_+ + \psi_- ) = 0 \,.
\ee 
Using $\Lambda^\pm \gamma^0 \Lambda^\pm = 0$ and $\Lambda^\pm \gamma^0 \gamma_\perp \Lambda^\pm=0$, we can project Eq.~\eqref{app:dirac1} onto the two fermion field components $\psi_\pm$ and obtain
\begin{align}
\label{app:psi+}
D^- \psi_+ + \gamma^0 (\slashed{D}_\perp+im ) \psi_- &= 0 \,, \\
\partial^+ \psi_- + \gamma^0 (\slashed{D}_\perp+im ) \psi_+ &= 0 \,. \nn
\end{align}
The derivative $\partial^-$ is with respect to the light-cone time while the derivative $\partial^+$ is with respect to the longitudinal coordinate. So we can solve $\psi_-$ in terms of the $\psi_+$ at the same light-cone time
\be
\label{app:psi-}
\psi_- = -\frac{1}{\partial^+} \gamma^0 (\slashed{D}_\perp+im) \psi_+ \,.
\ee
In other words, the $\psi_-$ field is not dynamical. Plugging Eq.~\eqref{app:psi-} into Eq.~\eqref{app:psi+}, we find the equation of motion for the $\psi_+$ field is given by
\be
\partial^- \psi_+ - ig A^- \psi_+ - \gamma^0 (\slashed{D}_\perp+im ) \frac{1}{\partial^+} \gamma^0 (\slashed{D}_\perp+im ) \psi_+ = 0\,.
\ee
Since $\partial^- = \frac{\partial}{\partial^+}$ is the derivative with respect to the light-cone time, $\psi_+$ is a dynamical degree of freedom.

Using the identities shown above, we can write the fermionic part of the Lagrangian density in light-cone gauge as
\begin{align}
\ml{L}_f &= i\big( \psi^\dagger_+ D^- \psi_+ + \psi_-^\dagger \partial^+ \psi_-
+ \psi_-^\dagger \gamma^0 (\slashed{D}_\perp+im ) \psi_+ 
+ \psi_+^\dagger \gamma^0 (\slashed{D}_\perp+im ) \psi_-  \big) \\
&= i\big( \psi^\dagger_+ D^- \psi_+ 
- \psi_+^\dagger \gamma^0 (\slashed{D}_\perp+im ) \frac{1}{\partial^+} \gamma^0 (\slashed{D}_\perp+im ) \psi_+
\big) \,, \nn
\end{align}
where we have used Eq.~\eqref{app:psi-}. Then the fermionic part of the Hamiltonian density in light-cone gauge is given by
\begin{align}
\label{app:Hq}
\ml{H}_f &= -g \psi^\dagger_+ A^- \psi_+ 
+i \psi_+^\dagger \gamma^0 (\slashed{D}_\perp+im ) \frac{1}{\partial^+} \gamma^0 (\slashed{D}_\perp+im ) \psi_+ \,.
\end{align}
To quantize the theory canonically, we decompose the $\psi_+$ field as
\begin{align}
\psi_+^i(x^+=0,x_\perp, x^-) = \sum_{\sigma=\pm\frac{1}{2}} \int_{k^+>0} \frac{\diff k^+ \diff^2 k_\perp}{2(2\pi)^3 k^+} 
\Big( b^i(k,\sigma) u_+(k, \sigma) e^{-ik\cdot x} + d^{i\dagger}(k,\sigma) v_+(k,\sigma) e^{ik\cdot x} \Big)\,, 
\end{align}
where $i$ is the color index in the fundamental representation and the quark (antiquark) creation $b^{i\dagger}(d^{i\dagger})$ and annihilation $b^{i}(d^i)$ operators satisfy the anticommutation relations
\begin{align}
\big\{ b^i(k,\sigma), b^{j\dagger}(k',\sigma') \big\} = \big\{ d^i(k,\sigma), d^{j\dagger}(k',\sigma') \big\} = 2(2\pi)^3 k^+ \delta^{ij} \delta_{\sigma \sigma'} \delta^3(k-k') \,,
\end{align}
and all the other anticommutators vanish. Here $\delta^3(k) = \delta(k^+)\delta^2(k_\perp)$. Using $u_+ = \Lambda_+ u$ and $\bar{u}_+ = \bar{u} \Lambda_-$, one can easily show
\begin{align}
\sum_{\sigma=\pm\frac{1}{2}} u_+(k,\sigma) \bar{u}_+(k,\sigma) = \Lambda_+ \Big(\sum_{\sigma=\pm\frac{1}{2}} u(k,\sigma) \bar{u}(k,\sigma) \Big) \Lambda_- = \Lambda_+ \slashed{k} \Lambda_- = k^+ \Lambda_+ \gamma^0\,.
\end{align}
So we have
\begin{align}
\sum_{\sigma=\pm\frac{1}{2}} u_+(k,\sigma) u_+^\dagger(k,\sigma) = k^+ \Lambda_+ \,, \quad \qquad \sum_{\sigma=\pm\frac{1}{2}} v_+(k,\sigma) v_+^\dagger(k,\sigma) = k^+ \Lambda_+ \,.
\end{align}
With these we can show the quark field satisfies the following anticommutation relation
\be
\big\{ \psi_+^i(x), \psi_+^{j\dagger}(y) \big\}_{x^+=y^+=0} = \Lambda_+ \delta^{ij} \delta^3(x-y) \,,
\ee
where the delta function in space is defined as $\delta^3(x)=\delta(x^-)\delta^2(x_\perp)$. Furthermore, from $\bar{u}(k,\sigma) \gamma^\mu u(k, \sigma') = 2 p^\mu \delta_{\sigma\sigma'} = \bar{v}(k,\sigma) \gamma^\mu v(k, \sigma')$, we can show
\begin{align}
\label{eqn:u+du+}
u_+^\dagger(k,\sigma) u_+(k,\sigma') = \frac{1}{2}u^\dagger(k,\sigma) \gamma^0\gamma^+ u(k,\sigma') = k^+\delta_{\sigma\sigma'}\,,\quad\qquad v_+^\dagger(k,\sigma) v_+(k,\sigma') = k^+\delta_{\sigma\sigma'}\,.
\end{align}

The kinetic term in the fermionic part of the Hamiltonian can be worked out to give
\begin{align}
H_{f,\,{\rm kin}} & = \int \diff^3 x \Big( i\psi_+^\dagger \gamma^0 (\slashed{\partial}_\perp +im) \frac{1}{\partial^+} \gamma^0 (\slashed{\partial}_\perp +im) \psi_+ \Big) \\
&= \int \diff x^- \diff^2 x_\perp \Big( i\psi_+^\dagger  \frac{{\bs \partial}_\perp^2 - m^2}{\partial^+}  \psi_+ \Big) \nn\\
& = \sum_i \sum_{\sigma=\pm\frac{1}{2}} \sum_{\sigma'=\pm\frac{1}{2}} \int_{k^+>0} \frac{\diff k^+ \diff^2 k_\perp}{2(2\pi)^3 k^+} \frac{{\bs k}_\perp^2+m^2}{(k^+)^2} \Big( b^{i\dagger}(k,\sigma) b^{i}(k,\sigma') u_+^\dagger(k,\sigma) u_+(k,\sigma')
\nn\\
& \quad\quad\quad\quad\quad\quad\quad\quad\quad\quad\quad\quad\quad\quad\quad\quad\quad -d^{i}(k,\sigma) d^{i\dagger}(k,\sigma') v_+^\dagger(k,\sigma) v_+(k,\sigma') \Big) \nn\\
& = \sum_i \sum_{\sigma=\pm\frac{1}{2}} \int_{k^+>0} \frac{\diff k^+ \diff^2 k_\perp}{2(2\pi)^3 k^+} \frac{{\bs k}_\perp^2+m^2}{k^+} \Big( b^{i\dagger}(k,\sigma) b^{i}(k,\sigma)
+d^{i\dagger}(k,\sigma) d^{i}(k,\sigma) \Big) + {\rm const} \,, \nn
\end{align}
where we have used $\diff^3x = \diff x^- \diff^2 x_\perp$, $\partial_\perp^2 = - {\bs \partial}_\perp^2$,  Eq.~\eqref{eqn:u+du+} and 
\be
\label{app:int_k+0}
\int_{k_1^+>0} \diff k_1^+ \int_{k_2^+>0} \diff k_2^+ \, \delta(k_1^+ + k_2^+) = 0\,.
\ee

\subsection{Gauge Sector}
The gauge part of the Lagrangian density is given by
\be
\ml{L}_g = -\frac{1}{4}F^{\mu\nu a} F_{\mu\nu}^a + g\bar{\psi} \slashed{A} \psi\,,
\ee
where $F^{\mu\nu a} = \partial^\mu A^{\nu a} - \partial^\nu A^{\mu a} + g f^{abc} A^{\mu b} A^{\nu c}$. The equation of motion is determined from the Lagrangian equation
\be
\frac{\partial \ml{L}_g}{\partial A^{\nu a}} = \partial^\mu \frac{\partial \ml{L}_g}{ \partial ( \partial^\mu A^{\nu a} )} \,,
\ee
which leads to
\be
-g\bar{\psi}\gamma_\nu T^a \psi + g f^{abc} F_{\mu\nu}^b A^{\mu c} = \partial^\mu F_{\mu\nu}^a \,.
\ee
In light-cone gauge $A^+=0$ and for $\nu=+$ (we raise $\nu$ to an upper index), using
\be
F^{++a}=0\,,\quad\qquad F^{-+a} = -\partial^+ A^{-a}\,, \quad\qquad F^{i+a} = -\partial^+ A^{i a}_\perp \,,
\ee
we obtain
\be
- g\bar{\psi} \gamma^+ T^a \psi + g f^{abc} F_i^{\,+b} A^{ic}_\perp  = \frac{1}{2}\partial^+ F^{-+a} + \partial^i F_i^{\,+a} \,,
\ee
which is simplified to be
\be
\label{app:eom_A}
- 2g \psi_+^\dagger T^a \psi_+ + g f^{abc} ( \partial^+ A^{ib}_\perp ) A^{i c}_\perp = \frac{1}{2} \partial^+ ( -\partial^+ A^{-a} ) + \partial^i \partial^+ A^{i a}_\perp \,.
\ee
Since the derivative $\partial^+$ is with respect to the longitudinal coordinate, we can invert Eq.~(\ref{app:eom_A}) to obtain
\be
\label{app:A-}
A^{-a} = \frac{2}{\partial^+} \partial^i A^{ia}_\perp - \frac{2g}{\partial^{+2}} \Big(
f^{abc} ( \partial^+ A^{ib}_\perp ) A^{i c}_\perp - 2\psi_+^\dagger T^a \psi_+
\Big) \,.
\ee
As a result the $A^{-a}$ is determined by the transverse components and thus not a dynamical degree of freedom. Since we have chosen $A^+=0$ in light-cone gauge, only two field degrees of freedom are left in the gauge part, i.e., $A_\perp$. We choose the two gluon polarization vectors to be (the determination of the transverse plane replies on choosing the $+$ and $-$ directions of the spacetime)
\be
\varepsilon_\perp (\pm) = \frac{1}{\sqrt{2}}\big( 1, \pm i \big) \,,
\ee
which satisfies the completeness relation
\be
\sum_{\lambda = \pm} \varepsilon_\perp^\mu(\lambda) 
\varepsilon_\perp^{\nu*}(\lambda) = -g_\perp^{\mu\nu} \,, \quad\qquad \sum_{i=1,2} \varepsilon_\perp^i(\lambda_1) \varepsilon_\perp^{i*}(\lambda_2) = \delta_{\lambda_1\lambda_2} \,.
\ee
The four component polarization vector $\varepsilon^\mu$ can be chosen to be
\be
\varepsilon^+ = 0\,, \quad\qquad \varepsilon^- = -\frac{2 k_\perp \cdot \varepsilon_\perp}{k^+} \,.
\ee

The canonical momentum conjugated to the gauge field $A^{ia}_\perp$ is given by
\be
\Pi_{A^{ia}_\perp} = \frac{\partial \ml{L}}{\partial ( \partial^- A^{ia}_\perp ) } = -\frac{1}{2} \partial^+ A_{\perp i}^a = -\partial_- A_{\perp i}^a \,.
\ee
Then the gauge part of the Hamiltonian density in light-cone gauge is given by
\begin{align}
\label{app:Hg}
\ml{H}_g & = -\frac{1}{2} (\partial^+ A_{\perp i}^a) ( \partial^- A^{ia}_\perp ) - \ml{L}_g  \nn\\
& = -\frac{1}{8} ( \partial^+ A^{-a} )^2 + \frac{1}{2}( \partial^+ A^{i a}_\perp ) ( -\partial_i A^{-a} + g f^{abc} A^{-b} A_{\perp i}^c )
+ \frac{1}{4}F_\perp^{ija} F_{\perp ij}^a \,,
\end{align}
where the $A^{-a}$ component is fixed by Eq.~\eqref{app:A-}. In canonical quantization, the gauge field is decomposed as (we neglect the zero mode as mentioned earlier)
\begin{align}
A^{ib}_\perp(x^+=0, x_\perp, x^-) = \sum_{\lambda=\pm} \int_{k^+>0}\frac{\diff k^+ \diff^2 k_\perp}{2(2\pi)^3 k^+} \Big(
a^b(k,\lambda) \varepsilon_\perp^i(\lambda) e^{-ik\cdot x} + a^{b\dagger}(k,\lambda) \varepsilon_\perp^{i*}(\lambda) e^{ik\cdot x}
\Big) \,,
\end{align}
in which $i=1,2$ denotes the transverse coordinate components and the gluon creation and annihilation operators satisfy the commutation relation
\be
\big[ a^b(k, \lambda), a^{c\dagger}(k',\lambda') \big] = 2(2\pi)^3 k^+ \delta_{\lambda\lambda'} \delta^{bc} \delta^3(k-k') \,.
\ee
Then one can show the commutation relation for the gauge fields
\be
\big[ A^{ib}_\perp(x), \partial^+ A^{jc}_\perp(y) \big]_{x^+=y^+=0} = i\delta^{ij} \delta^{bc} \delta^3(x-y)\,.
\ee
The kinematic term in the gluon part of the Hamiltonian can be obtained by plugging Eq.~\eqref{app:A-} into Eq.~\eqref{app:Hg} and neglecting all interaction terms, which leads to
\begin{align}
H_{g,\,{\rm kin}} &= \int \diff x^- \diff^2 x_\perp \Big( -\frac{1}{2}\big(\partial^i A^{ia}_\perp\big)\big(\partial^j A^{ja}_\perp\big)
-\big(\partial^+ A^{ia}_\perp\big) \big( \partial_i \frac{1}{\partial^+} \partial^j A^{ja}_\perp\big)  \\
&\quad\quad\quad\quad\quad\quad\quad + \frac{1}{2}\big(\partial^i A^{ja}_\perp \big)\big(\partial_i A_{\perp j}^{\,a} \big)
- \frac{1}{2}\big(\partial^i A^{ja}_\perp \big)\big(\partial_j A_{\perp i}^{\,a} \big)
\Big) \nn \\
&= \frac{1}{2} \int \diff x^- \diff^2 x_\perp \big(\partial^i A^{ja}_\perp \big)\big(\partial^i A^{ja}_\perp \big)
 \nn \\
&= \sum_b\sum_{i=1,2}\sum_{\lambda_1=\pm} \sum_{\lambda_2=\pm} \int_{k^+>0} \frac{\diff k^+ \diff^2k_\perp }{2(2\pi)^3k^+} \frac{{\bs k}_\perp^2}{2k^+} \Big( a^b(k,\lambda_1) a^{b\dagger}(k,\lambda_2) \varepsilon_\perp^i(k,\lambda_1) \varepsilon_\perp^{i*}(k,\lambda_2) \nn\\
&\quad\quad\quad\quad\quad\quad\quad\quad\quad\quad\quad\quad\quad\quad\quad\quad\quad + a^{b\dagger}(k,\lambda_1) a^b(k,\lambda_2) \varepsilon_\perp^{i*}(k,\lambda_1) \varepsilon_\perp^i(k,\lambda_2)\Big) \nn\\
&= \sum_b \sum_{\lambda=\pm} \int_{k^+>0} \frac{\diff k^+ \diff^2k_\perp }{2(2\pi)^3k^+} \frac{{\bs k}_\perp^2}{k^+} a^{b\dagger}(k,\lambda) a^b(k,\lambda)
+ {\rm const} \,, \nn
\end{align}
where we have used Eq.~\eqref{app:int_k+0} again.

\subsection{Splitting}
\label{app:split}
The total Hamiltonian density is $\ml{H} = \ml{H}_q + \ml{H}_g$ where $\ml{H}_q$ and $\ml{H}_g$ are given by Eqs.~\eqref{app:Hq} and~\eqref{app:Hg} respectively. 
Now we organize the part of the Hamiltonian relevant for splitting in powers of $g$ and $\frac{1}{\partial^+}$. For the $i\psi_+^\dagger \gamma^0 (\slashed{D}_\perp +im) \frac{1}{\partial^+} \gamma^0 (\slashed{D}_\perp +im) \psi_+$, we find
\begin{align}
\ml{O}\Big(\frac{g}{\partial^+}\Big):& \qquad -g \psi_+^\dagger A_{\perp i}\gamma^i \frac{1}{\partial^+} (\partial_{\perp j} \gamma^j+ im) \psi_+ -g \psi_+^\dagger (\partial_{\perp i} \gamma^i - im) \frac{1}{\partial^+} \big( A_{\perp j}\gamma^j  \psi_+ \big) \\
&\quad =  -g\psi_+^\dagger A_{\perp i} \gamma^i \gamma^j \Big(\frac{\partial_{\perp j}}{\partial^+} \psi_+\Big) - g \Big( \frac{\partial_{\perp i}}{\partial^+} \psi_+^\dagger \Big) A_{\perp j} \gamma^i \gamma^j \psi_+ \nn\\
& \qquad\qquad\qquad\qquad\qquad\qquad\qquad -img \Big( \psi_+^\dagger \slashed{A}_\perp \frac{1}{\partial^+} \psi_+ - \psi_+^\dagger \frac{1}{\partial^+} \big(\slashed{A}_\perp \psi_+\big) \Big) \,, \nn\\
\ml{O}\Big( \frac{g^2}{\partial^+} \Big): &\qquad ig^2 \psi_+^\dagger \gamma^j\gamma^j A_{\perp i} \frac{1}{\partial^+} \big(A_{\perp j} \psi_+ \big) \,, \nn
\end{align}
where we follow a notation that derivatives inside parentheses act on everything on their right inside the same parentheses, while if there are no parentheses, derivatives act on everything on their right.
The term $-g\psi_+^\dagger A^- \psi_+$ with Eq.~\eqref{app:A-} leads to
\begin{align}
\ml{O}\Big( \frac{g}{\partial^+} \Big) : & \qquad \!\! -2g\, \psi_+^\dagger T^a \psi_+ \Big( \frac{\partial^i}{\partial^+} A^{ia}_\perp \Big)  \,,\\
\ml{O}\Big( \frac{g^2}{\partial^+} \Big) : & \qquad 2g^2 f^{abc} \, \psi_+^\dagger T^a \psi_+ \Big( \frac{1}{\partial^{+2}} \big(\partial^+ A_\perp^{ib} \big) A_\perp^{ic}\Big)  \,,\nn\\ 
\ml{O}\Big( \frac{g^2}{\partial^{+2}} \Big) : & \qquad \!\!-4g^2 \psi_+^\dagger T^a \psi_+ \Big( \frac{1}{\partial^{+2}} \psi_+^\dagger T^a \psi_+ \Big) \,. \nn
\end{align}
Next the term $-\frac{1}{8}(\partial^+ A^{-a})^2$ with Eq.~\eqref{app:A-} gives
\begin{align}
\ml{O}( g ) : & \qquad g f^{abc}\, \big(\partial^i A_\perp^{ia} \big) \Big( \frac{1}{\partial^+} \big( \partial^+ A_\perp^{jb}  \big) A_\perp^{jc} \Big) = - g f^{abc} \Big( \frac{\partial^i}{\partial^+} A_\perp^{ia} \Big)  \big( \partial^+ A_\perp^{jb}  \big) A_\perp^{jc} \,, \\
\ml{O}\Big( \frac{g}{\partial^+} \Big) : & \qquad \!\!- 2g\big( \partial^i A_\perp^{ia} \big)\Big( \frac{1}{\partial^+} \psi_+^\dagger T^a \psi_+ \Big) \,,\nn\\ 
\ml{O}( g^2 ) : & \qquad \!\! -\frac{g^2}{2} \Big( f^{abc}\frac{1}{\partial^+} \big(\partial^+ A_\perp^{ib} \big) A_\perp^{ic} \Big)^2 \,,\nn\\
\ml{O}\Big( \frac{g^2}{\partial^+} \Big) : & \qquad 2g^2 f^{abc}\Big( \frac{1}{\partial^+} \psi_+^\dagger T^a \psi_+ \Big) \Big( \frac{1}{\partial^+} \big( \partial^+ A_\perp^{ib} \big) A_\perp^{ic} \Big)\,,\nn\\
\ml{O}\Big( \frac{g^2}{\partial^{+2}} \Big) : & \qquad \!\! -2g^2 \Big( \frac{1}{\partial^+} \psi_+^\dagger T^a \psi_+ \Big)^2  \,.\nn
\end{align}
Furthermore, we find the term $-\frac{1}{2}(\partial^+ A_\perp^{ia}) (\partial_i A^{-a}) $ contributes as
\begin{align}
\ml{O}( g ) : & \qquad gf^{abc} \big( \partial_i A_\perp^{ia} \big) \Big( \frac{1}{\partial^+} \big( \partial^+ A_\perp^{jb} \big)A_\perp^{jc} \Big) = gf^{abc}\Big( \frac{\partial^i}{\partial^+} A_\perp^{ia}\Big) \big( \partial^+ A_\perp^{jb} \big)A_\perp^{jc} \,, \\
\ml{O}\Big( \frac{g}{\partial^+} \Big) : & \qquad \!\!- 2g\big( \partial_i A_\perp^{ia} \big) \Big( \frac{1}{\partial^+} \psi_+^\dagger T^a \psi_+ \Big) \,. \nn
\end{align}
Then we obtain the contribution from the $\frac{1}{2}gf^{abc} ( \partial^+ A_\perp^{ia}) A^{-b} A_{\perp i}^{c}$ term
\begin{align}
\ml{O}( g ) : & \qquad gf^{abc} \big( \partial^+ A_\perp^{ia} \big) \Big( \frac{\partial^j}{\partial^+} A_\perp^{jb} \Big) A_{\perp i}^{c} \,,\\
\ml{O}( g^2 ) : & \qquad \! \!-g^2f^{abc}f^{bde} \big( \partial^+ A_\perp^{ia} \big) \Big( \frac{1}{\partial^{+2}} \big(\partial^+ A_\perp^{jd} \big) A_\perp^{je} \Big) A_{\perp i}^{c} \,,\nn\\
\ml{O}\Big( \frac{g^2}{\partial^+} \Big) : & \qquad 2g^2 f^{abc} \big( \partial^+ A_\perp^{ia} \big) \Big( \frac{1}{\partial^{+2}} \psi_+^\dagger T^b \psi_+ \Big) A_{\perp i}^{c} \,. \nn
\end{align}
Finally, the term $-\frac{1}{4}F_\perp^{ija} F_{\perp ij}^{a}$ leads to
\begin{align}
\ml{O}( g ) : & \qquad \!\! -gf^{abc} \big( \partial^i A_\perp^{ja} \big) A_{\perp i}^b A_{\perp j}^c \,,\\
\ml{O}( g^2 ) : & \qquad \!\! -\frac{1}{4}g^2 f^{abc} f^{ade} A^{ib}_\perp A^{jc}_\perp A_{\perp i}^d A_{\perp j}^e \,. \nn
\end{align}

\section{Improved Decomposition into Tensor Products of Pauli Matrices}
\label{app:alter}
The generic method mentioned in section~\ref{sect:decomposition} to decompose a Hamiltonian matrix into tensor products of Pauli matrices involves calculating an exponential number of coefficients, which can be very expensive computationally as the size of the Hilbert space increases. Here we discuss a more efficient way of doing the decomposition. We take the toy model as an illustrative example.

\subsection{Kinetic Term}
We start with the 1-particle case and use the generic method mentioned in section~\ref{sect:decomposition} to decompose the 1-particle kinetic Hamiltonian Eq.~\eqref{equ:kin_1par}, which gives
\begin{align}
\label{eqn:app_Hkin1}
H_{\rm kin}^{(1)}(1,2) = \frac{(K_{\rm max}^\perp)^2}{K_{\rm max}^+} \Big( \frac{3}{4} \mathbb{1}_1\otimes \mathbb{1}_2 - \frac{3}{4} \mathbb{1}_1\otimes \sigma^z_2 + \frac{1}{4} \sigma^z_1\otimes \mathbb{1}_2 - \frac{1}{4} \sigma^z_1\otimes \sigma^z_2 \Big) \,,
\end{align}
where the argument $(1,2)$ on the left indicates the index of the qubits that $H_{\rm kin}^{(1)}$ acts on.
We will use Eq.~\eqref{eqn:app_Hkin1} as a building block to construct the full kinetic Hamiltonian for the Hilbert space made up of both the 1-particle and 2-particle states. First we notice that no matter whether the state is 1-particle or 2-particle, we always have $H_{\rm kin}^{(1)}(4,5)$, i.e., $H_{\rm kin}^{(1)}$ acting on the fourth and fifth qubits. Furthermore, if the state is a 2-particle state, we also have $H_{\rm kin}^{(1)}(2,3)$ with the first qubit being in ``1''. Therefore we can write
\begin{align}
\label{eqn:app_Hkin}
H_{\rm kin} = \mathbb{1}_1\otimes \mathbb{1}_2\otimes \mathbb{1}_3 \otimes H_{\rm kin}^{(1)}(4,5) + \frac{1}{2}\big(\mathbb{1}_1 - \sigma^z_1 \big) \otimes H_{\rm kin}^{(1)}(2,3) \otimes \mathbb{1}_4 \otimes \mathbb{1}_5 \,.
\end{align}
We have checked that Eq.~\eqref{eqn:app_Hkin} reproduces the kinetic Hamiltonian matrix elements for physical states. It differs from Eq.~\eqref{eqn:Hkin_toy} for unphysical states that are $|010q_4q_5\rangle$, $|001q_4q_5\rangle$ and $|011q_4q_5\rangle$, which is fine if the initial state of the time evolution is a physical state and the implementation of the other parts of the Hamiltonian does not connect physical states with unphysical ones.

In this way, we only need to apply the generic method of Pauli decomposition once for the 1-particle kinetic Hamiltonian, which is much cheaper computationally than applying the generic method to the full kinetic Hamiltonian. Once we have the decomposition of $H_{\rm kin}^{(1)}$, we can easily obtain the Pauli decomposition for the multi-particle Hamiltonian.

\subsection{Diffusion Term}
The strategy we employ is similar to the construction of the kinetic Hamiltonian discussed in the previous subsection, since the diffusion Hamiltonian does not change the number of particles in the state. From the 1-particle diffusion Hamiltonian~\eqref{eqn:Hdiff_single}, we obtain
\begin{align}
H_{\rm diff}^{(1)}(1,2) = g_d\bar{A}^-(K_{\rm max}^\perp)\, \mathbb{1}_1\otimes \sigma_2^x \,,
\end{align}
where we have neglected the term proportional to $\bar{A}^-(0)$ that only leads to a global phase change in the time evolution.

Generalizing to the case with both 1-particle and 2-particle states as in the previous section, we have
\begin{align}
\label{eqn:app_Hdiff}
H_{\rm diff} = \mathbb{1}_1\otimes \mathbb{1}_2\otimes \mathbb{1}_3 \otimes H_{\rm diff}^{(1)}(4,5) + \frac{1}{2}\big(\mathbb{1}_1 - \sigma^z_1 \big) \otimes H_{\rm diff}^{(1)}(2,3) \otimes \mathbb{1}_4 \otimes \mathbb{1}_5 \,.
\end{align}
We have checked that Eq.~\eqref{eqn:app_Hdiff} reproduces Eq.~\eqref{eqn:Hdiff_toy} for physical states and does not introduce transitions between physical and unphysical states. (The unphysical states are $|010q_4q_5\rangle$, $|001q_4q_5\rangle$ and $|011q_4q_5\rangle$.)

\subsection{Splitting Term}
\label{app:split_decom}
The splitting Hamiltonian involves both 1-particle and 2-particle states. So the generalization method used above for $H_{\rm kin}$ and $H_{\rm diff}$ only applies if we consider transitions between $n$-particle and $n+1$-particle states for $n>1$. To decompose Eq.~\eqref{eqn:Hsplit_toy} into tensor products of Pauli matrices, one may just apply the standard decomposition formula explained in section~\ref{sect:decomposition}.

Here we illustrate another way of decomposition. We take $\langle 10000 | H_{\rm split} | 00010 \rangle = g_s$ as an example. We can explicitly write out the operator for the change of each qubit and put them together as a tensor product
\begin{align}
\langle 10000 | H_{\rm split} | 00010 \rangle \to g_s \,\sigma_1^- \otimes \frac{\mathbb{1}_2+\sigma_2^z}{2} \otimes \frac{\mathbb{1}_3+\sigma_3^z}{2} \otimes \sigma_4^+ \otimes \frac{\mathbb{1}_5+\sigma_5^z}{2} \,,
\end{align}
where $\sigma_k^\pm = (\sigma_k^x \pm i \sigma_k^y)/2$. The general rule is as follows: if a qubit stays as 0 or 1, we use the operator $(\mathbb{1}+\sigma^z)/2$ or $(\mathbb{1}-\sigma^z)/2$; if a qubit turns to 0 from 1, we use $\sigma^+$; if a qubit changes from 0 to 1, we use $\sigma^-$. We have checked that by using this way of decomposition, we can reproduce Eq.~\eqref{eqn:Hsplit_toy} exactly.

\section{Some Decomposition Formulas for the Gluon Radiation Case}
\label{app:qcd_decom}
\subsection{Kinetic Term}
To decompose the kinetic Hamiltonian of the 15-qubit system, we first consider the kinetic Hamiltonian for 1-particle states, which are represented by 7 qubits in Eq.~\eqref{eqn:1gluon}. The first three qubits encode the momentum and thus will have nontrivial operators act on them in the kinetic Hamiltonian. The color and spin degrees of freedom are degenerate in the kinetic Hamiltonian. Taking $(k_x^2+k_y^2)/k^+$ as the kinetic energy and using the discretization in Eq.~\eqref{eqn:1gluon_k}, we find the decomposition of the kinetic Hamiltonian for the three qubits is given by
\begin{align}
H_{\rm kin}^{(1)}(1,2,3) =& \frac{(K_{\rm max}^\perp)^2}{K_{\rm max}^+} \Big( \frac{3}{2}
\mathbb{1}_1\otimes\mathbb{1}_2\otimes\mathbb{1}_3 - \frac{3}{4}
\mathbb{1}_1\otimes\mathbb{1}_2\otimes\sigma^z_3 - \frac{3}{4}
\mathbb{1}_1\otimes\sigma^z_2\otimes\mathbb{1}_3 \\
&\qquad\quad + \frac{1}{2}
\sigma^z_1\otimes\mathbb{1}_2\otimes\mathbb{1}_3 - \frac{1}{4}
\sigma^z_1\otimes\mathbb{1}_2\otimes\sigma^z_3
 - \frac{1}{4}
\sigma^z_1\otimes\sigma^z_2\otimes\mathbb{1}_3
\Big) \,,\nn
\end{align}
where the argument $(1,2,3)$ on the left hand side indicates the indices of the qubits that the operators act on. Including the color and spin degrees of freedom that are represented by the fourth to the seventh qubit, we have
\begin{align}
H_{\rm kin}^{(1)}(1,2,3) \otimes\mathbb{1}_4 \otimes\mathbb{1}_5 \otimes\mathbb{1}_6 \otimes\mathbb{1}_7 \,,
\end{align}
for the kinetic Hamiltonian of 1-particle states. 
Using Eq.~\eqref{eqn:app_Hkin} we can easily generalize this for states with both 1-particle and 2-particle states
\begin{align}
H_{\rm kin}^{(1)}(9,10,11) + \frac{1}{2}\big(\mathbb{1}_1 - \sigma_1^z\big) \otimes H_{\rm kin}^{(1)}(2,3,4) \,,
\end{align}
where we have omitted identities.
The generalization for multi-particle states can be similarly done.

\subsection{SU(3) Structure Constants}
In this subsection we discuss the decomposition for the SU(3) structure constants $f^{abc}$ since they appear in the diffusion and splitting parts of the Hamiltonian. In the gluon diffusion part of the Hamiltonian~\eqref{eqn:Hdiff_dis1}, the color structure constant appears as 
\be
\label{eqn:color_diff_app}
\langle a_1 |H_{\rm diff} | a_2 \rangle = f^{a_2ba_1} - f^{a_1ba_2} \,,
\ee
where we omitted other terms that are factorized from the color part. Since the classical background field $\bar{A}^{-b}$ that is contracted with Eq.~\eqref{eqn:color_diff_app} is random, we need to decompose Eq.~\eqref{eqn:color_diff_app} into tensor products of Pauli matrices acting on three qubits representing the color degree of freedom, for each $b\in\{1,2,\cdots,8\}$. Using the generic decomposition method, we find $f^{a_2ba_1} - f^{a_1ba_2}$ can be decomposed as
\begin{align}
b=1:\quad &  \frac{i}{4} \Big( \sigma_3^y + 2\sigma^x_2\otimes \sigma^y_3 - 2\sigma^y_2\otimes \sigma^x_3 + \sigma_2^z\otimes\sigma_3^y + \sigma_1^x\otimes \sigma_3^y -\sigma_1^y\otimes \sigma_3^x - \sigma_1^x\otimes \sigma_2^z\otimes \sigma_3^y
 \nn\\
&+ \sigma_1^y\otimes \sigma_2^z\otimes \sigma_3^x - \sigma_1^z\otimes \sigma_3^y + 2 \sigma_1^z\otimes \sigma_2^x\otimes \sigma_3^y
- 2\sigma_1^z\otimes \sigma_2^y\otimes \sigma_3^x - \sigma_1^z\otimes \sigma_2^z\otimes \sigma_3^y \Big) \nn\\
b=2:\quad &  \frac{i}{4} \Big( \sigma_2^y+\sigma_2^y\otimes \sigma_3^z + \sigma_1^x\otimes \sigma_2^y - \sigma_1^y\otimes \sigma_2^x - \sigma_1^x\otimes \sigma_2^y\otimes \sigma_3^z  + \sigma_1^y\otimes \sigma_2^x\otimes \sigma_3^z \nn\\
&+  3\sigma_1^z\otimes \sigma_2^y + 3\sigma_1^z\otimes \sigma_2^y \otimes \sigma_3^z 
\Big) \nn\\
b=3:\quad &  \frac{i}{4} \Big( -2\sigma_3^y - \sigma^x_2\otimes \sigma^y_3 + \sigma^y_2\otimes \sigma^x_3 -2 \sigma_2^z\otimes\sigma_3^y + \sigma_1^x\otimes \sigma_2^x\otimes\sigma_3^y + \sigma_1^x\otimes \sigma_2^y\otimes \sigma_3^x
 \nn\\
&- \sigma_1^y\otimes \sigma_2^x\otimes \sigma_3^x + \sigma_1^y\otimes \sigma_2^y\otimes \sigma_3^y -2 \sigma_1^z\otimes \sigma_3^y + \sigma_1^z\otimes \sigma_2^x\otimes \sigma_3^y
- \sigma_1^z\otimes \sigma_2^y\otimes \sigma_3^x \nn\\
& - 2 \sigma_1^z\otimes \sigma_2^z\otimes \sigma_3^y \Big) \nn\\
b=4: \quad &  \frac{i}{4} \Big( -\sqrt{3} \sigma_2^x\otimes\sigma_3^y - \sqrt{3}\sigma_2^y\otimes\sigma_3^x + \sigma_1^y - \sigma_1^y\otimes\sigma_3^z + 2\sigma_1^y\otimes \sigma_2^x + 2\sigma_1^y\otimes \sigma_2^x \otimes \sigma_3^z \nn\\
& + \sigma_1^y\otimes\sigma_2^z - \sigma_1^y\otimes\sigma_2^z\otimes\sigma_3^z + \sqrt{3}\sigma_1^z\otimes \sigma_2^x \otimes \sigma_3^y + \sqrt{3}\sigma_1^z\otimes \sigma_2^y \otimes \sigma_3^x
\Big) \nn\\
b=5:\quad &  \frac{i}{4} \Big( -\sigma_3^y + \sigma^z_2\otimes \sigma^y_3 - \sigma^x_1\otimes \sigma^y_3 - \sigma_1^y\otimes \sigma_3^x - \sigma^x_1\otimes \sigma_2^x\otimes\sigma_3^y + \sigma_1^x\otimes \sigma_2^y\otimes\sigma_3^x
 \nn\\
& -\sigma_1^x\otimes\sigma_2^z\otimes\sigma_3^y + \sqrt{3}\sigma_1^y - \sigma_1^y\otimes\sigma_3^x - \sqrt{3} \sigma_1^y\otimes \sigma_3^z
+ \sigma_1^y\otimes \sigma_2^x\otimes \sigma_3^x + \sigma_1^y\otimes \sigma_2^y\otimes \sigma_3^y
 \nn\\
& - \sqrt{3} \sigma_1^y\otimes \sigma_2^z - \sigma_1^y\otimes \sigma_2^z\otimes \sigma_3^x +\sqrt{3} \sigma_1^y\otimes \sigma_2^z\otimes \sigma_3^z - \sigma_1^z\otimes \sigma_3^y + \sigma_1^z\otimes \sigma_2^z\otimes \sigma_3^y \Big) \nn\\
b=6: \quad &  \frac{i}{4} \Big( -\sqrt{3} \sigma_3^y - \sigma_2^y + \sigma_2^y\otimes\sigma_3^z + \sqrt{3} \sigma_2^z\otimes\sigma_3^y + 2 \sigma_1^y\otimes\sigma_2^z + 2\sigma_1^y\otimes \sigma_2^z \otimes \sigma_3^z  \nn\\
& + \sqrt{3} \sigma_1^z\otimes \sigma_3^y - \sigma_1^z\otimes\sigma_2^y + \sigma_1^z\otimes\sigma_2^y\otimes\sigma_3^z - \sqrt{3} \sigma_1^z\otimes\sigma_2^z\otimes\sigma_3^y
\Big) \nn\\
b=7:\quad &  \frac{i}{4} \Big( -\sigma_2^x\otimes \sigma^y_3 + \sqrt{3} \sigma^y_2 - \sigma^y_2\otimes \sigma^x_3 - \sqrt{3} \sigma_2^y\otimes \sigma_3^z + \sigma^x_1\otimes \sigma_3^y + \sigma_1^x\otimes \sigma_2^x\otimes\sigma_3^y
 \nn\\
& -\sigma_1^x\otimes\sigma_2^y\otimes\sigma_3^x + \sigma_1^x\otimes\sigma_2^z\otimes\sigma_3^y - \sigma_1^y\otimes\sigma_3^x + \sigma_1^y\otimes\sigma_2^x\otimes \sigma_3^x
+ \sigma_1^y\otimes \sigma_2^y\otimes \sigma_3^y 
 \nn\\
&- \sigma_1^y\otimes \sigma_2^z\otimes \sigma_3^x - \sigma_1^z\otimes \sigma_2^x \otimes \sigma_3^y - \sqrt{3} \sigma_1^z\otimes \sigma_2^y - \sigma_1^z\otimes \sigma_2^y\otimes \sigma_3^x +\sqrt{3} \sigma_1^z\otimes \sigma_2^y\otimes \sigma_3^z \Big) \nn\\
b=8:\quad &  \frac{i\sqrt{3}}{4} \Big( \sigma_2^x\otimes \sigma_3^y - \sigma_2^y\otimes \sigma_3^x + \sigma_1^x\otimes \sigma_2^x \otimes \sigma_3^y + \sigma_1^x\otimes \sigma_2^y \otimes \sigma_3^x - \sigma_1^y\otimes\sigma_2^x\otimes\sigma_3^x \nn\\
& + \sigma_1^y\otimes\sigma_2^y\otimes\sigma_3^y - \sigma_1^z\otimes \sigma_2^x \otimes \sigma_3^y + \sigma_1^z\otimes \sigma_2^y \otimes \sigma_3^x
\Big) \,,
\end{align}
where the qubits 1, 2 and 3 describe the color index $a_2$.

Next we consider the color structure constant in the gluon splitting Hamiltonian~\eqref{eqn:Hsplit_dis1}, which appears as
\be
\langle a_2, a_3 | H_{\rm split} | a_1 \rangle = f^{a_1a_2a_3} \,.
\ee
This matrix involves 6 qubits since it is a transition between 1-particle and 2-particle states. In terms of qubits representing the color degrees of freedom, we have
\be
\langle \underbrace{q_1q_2q_3}_\text{color $a_2$} \overbrace{q_4 q_5q_6}^\text{color $a_3$} | H_{\rm split} | \underbrace{000}_\text{unoccupied} \overbrace{q_4 q_5q_6}^\text{color $a_1$} \rangle \,.
\ee
Its decomposition into tensor products of Pauli matrices can be worked out by using the generic method explained in section~\ref{sect:decomposition}. We will not write the decomposition out explicitly since it is very lengthy. This 6-qubit representation of the color structure constant serves as a building block for the full splitting Hamiltonian.

After obtaining the qubit representation for the color part of the Hamiltonian, we can take its tensor product with the part describing the momentum and spin changes to obtain the complete Pauli matrix representation of $H_{\rm diff}$ and $H_{\rm split}$. The qubit representation of the momentum and spin parts can be worked out by using the generic method explained in section~\ref{sect:decomposition} or the method introduced in appendix~\ref{app:split_decom}.

\bibliography{main.bib}

\providecommand{\href}[2]{#2}\begingroup\raggedright\begin{thebibliography}{100}

\bibitem{Butterworth:2008iy}
J.~M. Butterworth, A.~R. Davison, M.~Rubin and G.~P. Salam, \emph{{Jet
  substructure as a new Higgs search channel at the LHC}},
  \href{http://dx.doi.org/10.1103/PhysRevLett.100.242001}{\emph{Phys.Rev.Lett.}
  {\bf 100} (2008) 242001}, [\href{http://arxiv.org/abs/0802.2470}{{\tt
  0802.2470}}].

\bibitem{Ellis:2009su}
S.~D. Ellis, C.~K. Vermilion and J.~R. Walsh, \emph{{Techniques for improved
  heavy particle searches with jet substructure}},
  \href{http://dx.doi.org/10.1103/PhysRevD.80.051501}{\emph{Phys.Rev.} {\bf
  D80} (2009) 051501}, [\href{http://arxiv.org/abs/0903.5081}{{\tt
  0903.5081}}].

\bibitem{Stewart:2010tn}
I.~W. Stewart, F.~J. Tackmann and W.~J. Waalewijn, \emph{{N-Jettiness: An
  Inclusive Event Shape to Veto Jets}},
  \href{http://dx.doi.org/10.1103/PhysRevLett.105.092002}{\emph{Phys. Rev.
  Lett.} {\bf 105} (2010) 092002}, [\href{http://arxiv.org/abs/1004.2489}{{\tt
  1004.2489}}].

\bibitem{Ellis:2010rwa}
S.~D. Ellis, C.~K. Vermilion, J.~R. Walsh, A.~Hornig and C.~Lee, \emph{{Jet
  Shapes and Jet Algorithms in SCET}},
  \href{http://dx.doi.org/10.1007/JHEP11(2010)101}{\emph{JHEP} {\bf 1011}
  (2010) 101}, [\href{http://arxiv.org/abs/1001.0014}{{\tt 1001.0014}}].

\bibitem{Abdesselam:2010pt}
A.~Abdesselam et~al., \emph{{Boosted objects: A Probe of beyond the Standard
  Model physics}},
  \href{http://dx.doi.org/10.1140/epjc/s10052-011-1661-y}{\emph{Eur. Phys. J.}
  {\bf C71} (2011) 1661}, [\href{http://arxiv.org/abs/1012.5412}{{\tt
  1012.5412}}].

\bibitem{Altheimer:2012mn}
A.~Altheimer et~al., \emph{{Jet Substructure at the Tevatron and LHC: New
  results, new tools, new benchmarks}},
  \href{http://dx.doi.org/10.1088/0954-3899/39/6/063001}{\emph{J. Phys.} {\bf
  G39} (2012) 063001}, [\href{http://arxiv.org/abs/1201.0008}{{\tt
  1201.0008}}].

\bibitem{Larkoski:2013eya}
A.~J. Larkoski, G.~P. Salam and J.~Thaler, \emph{{Energy Correlation Functions
  for Jet Substructure}},
  \href{http://dx.doi.org/10.1007/JHEP06(2013)108}{\emph{JHEP} {\bf 06} (2013)
  108}, [\href{http://arxiv.org/abs/1305.0007}{{\tt 1305.0007}}].

\bibitem{Altheimer:2013yza}
A.~Altheimer et~al., \emph{{Boosted objects and jet substructure at the LHC.
  Report of BOOST2012, held at IFIC Valencia, 23rd-27th of July 2012}},
  \href{http://dx.doi.org/10.1140/epjc/s10052-014-2792-8}{\emph{Eur. Phys. J.}
  {\bf C74} (2014) 2792}, [\href{http://arxiv.org/abs/1311.2708}{{\tt
  1311.2708}}].

\bibitem{Dasgupta:2013ihk}
M.~Dasgupta, A.~Fregoso, S.~Marzani and G.~P. Salam, \emph{{Towards an
  understanding of jet substructure}},
  \href{http://dx.doi.org/10.1007/JHEP09(2013)029}{\emph{JHEP} {\bf 09} (2013)
  029}, [\href{http://arxiv.org/abs/1307.0007}{{\tt 1307.0007}}].

\bibitem{Larkoski:2014wba}
A.~J. Larkoski, S.~Marzani, G.~Soyez and J.~Thaler, \emph{{Soft Drop}},
  \href{http://dx.doi.org/10.1007/JHEP05(2014)146}{\emph{JHEP} {\bf 05} (2014)
  146}, [\href{http://arxiv.org/abs/1402.2657}{{\tt 1402.2657}}].

\bibitem{Adams:2015hiv}
D.~Adams et~al., \emph{{Towards an Understanding of the Correlations in Jet
  Substructure}},
  \href{http://dx.doi.org/10.1140/epjc/s10052-015-3587-2}{\emph{Eur. Phys. J.}
  {\bf C75} (2015) 409}, [\href{http://arxiv.org/abs/1504.00679}{{\tt
  1504.00679}}].

\bibitem{Chien:2015cka}
Y.-T. Chien, A.~Hornig and C.~Lee, \emph{{Soft-collinear mode for jet cross
  sections in soft collinear effective theory}},
  \href{http://dx.doi.org/10.1103/PhysRevD.93.014033}{\emph{Phys. Rev.} {\bf
  D93} (2016) 014033}, [\href{http://arxiv.org/abs/1509.04287}{{\tt
  1509.04287}}].

\bibitem{Larkoski:2015kga}
A.~J. Larkoski, I.~Moult and D.~Neill, \emph{{Analytic Boosted Boson
  Discrimination}},
  \href{http://dx.doi.org/10.1007/JHEP05(2016)117}{\emph{JHEP} {\bf 05} (2016)
  117}, [\href{http://arxiv.org/abs/1507.03018}{{\tt 1507.03018}}].

\bibitem{Moult:2016cvt}
I.~Moult, L.~Necib and J.~Thaler, \emph{{New Angles on Energy Correlation
  Functions}}, \href{http://dx.doi.org/10.1007/JHEP12(2016)153}{\emph{JHEP}
  {\bf 12} (2016) 153}, [\href{http://arxiv.org/abs/1609.07483}{{\tt
  1609.07483}}].

\bibitem{Frye:2016okc}
C.~Frye, A.~J. Larkoski, M.~D. Schwartz and K.~Yan, \emph{{Precision physics
  with pile-up insensitive observables}},
  \href{http://arxiv.org/abs/1603.06375}{{\tt 1603.06375}}.

\bibitem{Frye:2016aiz}
C.~Frye, A.~J. Larkoski, M.~D. Schwartz and K.~Yan, \emph{{Factorization for
  groomed jet substructure beyond the next-to-leading logarithm}},
  \href{http://dx.doi.org/10.1007/JHEP07(2016)064}{\emph{JHEP} {\bf 07} (2016)
  064}, [\href{http://arxiv.org/abs/1603.09338}{{\tt 1603.09338}}].

\bibitem{Kang:2016mcy}
Z.-B. Kang, F.~Ringer and I.~Vitev, \emph{{The semi-inclusive jet function in
  SCET and small radius resummation for inclusive jet production}},
  \href{http://arxiv.org/abs/1606.06732}{{\tt 1606.06732}}.

\bibitem{Kang:2016ehg}
Z.-B. Kang, F.~Ringer and I.~Vitev, \emph{{Jet substructure using
  semi-inclusive jet functions in SCET}}, {\emph{JHEP} {\bf 11} (2016) 155},
  [\href{http://arxiv.org/abs/1606.07063}{{\tt 1606.07063}}].

\bibitem{Kolodrubetz:2016dzb}
D.~W. Kolodrubetz, P.~Pietrulewicz, I.~W. Stewart, F.~J. Tackmann and W.~J.
  Waalewijn, \emph{{Factorization for Jet Radius Logarithms in Jet Mass Spectra
  at the LHC}}, \href{http://dx.doi.org/10.1007/JHEP12(2016)054}{\emph{JHEP}
  {\bf 12} (2016) 054}, [\href{http://arxiv.org/abs/1605.08038}{{\tt
  1605.08038}}].

\bibitem{Moult:2016fqy}
I.~Moult, L.~Rothen, I.~W. Stewart, F.~J. Tackmann and H.~X. Zhu,
  \emph{{Subleading Power Corrections for N-Jettiness Subtractions}},
  {\emph{Phys. Rev. D} {\bf 95} (2017) 074023},
  [\href{http://arxiv.org/abs/1612.00450}{{\tt 1612.00450}}].

\bibitem{Chien:2016led}
Y.-T. Chien and I.~Vitev, \emph{{Probing the Hardest Branching within Jets in
  Heavy-Ion Collisions}},
  \href{http://dx.doi.org/10.1103/PhysRevLett.119.112301}{\emph{Phys. Rev.
  Lett.} {\bf 119} (2017) 112301}, [\href{http://arxiv.org/abs/1608.07283}{{\tt
  1608.07283}}].

\bibitem{Moult:2017jsg}
I.~Moult, L.~Rothen, I.~W. Stewart, F.~J. Tackmann and H.~X. Zhu, \emph{{N
  -jettiness subtractions for $gg\to H$ at subleading power}}, {\emph{Phys.
  Rev. D} {\bf 97} (2018) 014013}, [\href{http://arxiv.org/abs/1710.03227}{{\tt
  1710.03227}}].

\bibitem{Moult:2017okx}
I.~Moult, B.~Nachman and D.~Neill, \emph{{Convolved Substructure: Analytically
  Decorrelating Jet Substructure Observables}}, {\emph{JHEP} {\bf 05} (2018)
  002}, [\href{http://arxiv.org/abs/1710.06859}{{\tt 1710.06859}}].

\bibitem{Larkoski:2017jix}
A.~J. Larkoski, I.~Moult and B.~Nachman, \emph{{Jet Substructure at the Large
  Hadron Collider: A Review of Recent Advances in Theory and Machine
  Learning}},  \href{http://arxiv.org/abs/1709.04464}{{\tt 1709.04464}}.

\bibitem{Kang:2018jwa}
Z.-B. Kang, K.~Lee, X.~Liu and F.~Ringer, \emph{{The groomed and ungroomed jet
  mass distribution for inclusive jet production at the LHC}},
  \href{http://dx.doi.org/10.1007/JHEP10(2018)137}{\emph{JHEP} {\bf 10} (2018)
  137}, [\href{http://arxiv.org/abs/1803.03645}{{\tt 1803.03645}}].

\bibitem{Ebert:2018lzn}
M.~A. Ebert, I.~Moult, I.~W. Stewart, F.~J. Tackmann, G.~Vita and H.~X. Zhu,
  \emph{{Power Corrections for N-Jettiness Subtractions at ${\cal
  O}(\alpha_s)$}}, {\emph{JHEP} {\bf 12} (2018) 084},
  [\href{http://arxiv.org/abs/1807.10764}{{\tt 1807.10764}}].

\bibitem{Moult:2018jjd}
I.~Moult, I.~W. Stewart, G.~Vita and H.~X. Zhu, \emph{{First Subleading Power
  Resummation for Event Shapes}}, {\emph{JHEP} {\bf 08} (2018) 013},
  [\href{http://arxiv.org/abs/1804.04665}{{\tt 1804.04665}}].

\bibitem{Chien:2018lmv}
Y.-T. Chien, D.~Kang, K.~Lee and Y.~Makris, \emph{{Subtracted Cumulants:
  Mitigating Large Background in Jet Substructure}}, {\emph{Phys. Rev. D} {\bf
  100} (2019) 074030}, [\href{http://arxiv.org/abs/1812.06977}{{\tt
  1812.06977}}].

\bibitem{Kang:2018vgn}
Z.-B. Kang, K.~Lee, X.~Liu and F.~Ringer, \emph{{Soft drop groomed jet
  angularities at the LHC}}, {\emph{Phys. Lett. B} {\bf 793} (2019) 41--47},
  [\href{http://arxiv.org/abs/1811.06983}{{\tt 1811.06983}}].

\bibitem{Dasgupta:2018nvj}
M.~Dasgupta, F.~A. Dreyer, K.~Hamilton, P.~F. Monni and G.~P. Salam,
  \emph{{Logarithmic accuracy of parton showers: a fixed-order study}},
  \href{http://dx.doi.org/10.1007/JHEP09(2018)033}{\emph{JHEP} {\bf 09} (2018)
  033}, [\href{http://arxiv.org/abs/1805.09327}{{\tt 1805.09327}}].

\bibitem{Asquith:2018igt}
R.~Kogler et~al., \emph{{Jet Substructure at the Large Hadron Collider:
  Experimental Review}}, {\emph{Rev. Mod. Phys.} {\bf 91} (2019) 045003},
  [\href{http://arxiv.org/abs/1803.06991}{{\tt 1803.06991}}].

\bibitem{Marzani:2019hun}
S.~Marzani, G.~Soyez and M.~Spannowsky, \emph{{Looking inside jets: an
  introduction to jet substructure and boosted-object phenomenology}},
  vol.~958.
\newblock Springer, 2019.

\bibitem{Hoang:2019ceu}
A.~H. Hoang, S.~Mantry, A.~Pathak and I.~W. Stewart, \emph{{Nonperturbative
  Corrections to Soft Drop Jet Mass}},
  \href{http://arxiv.org/abs/1906.11843}{{\tt 1906.11843}}.

\bibitem{Kang:2019prh}
Z.-B. Kang, K.~Lee, X.~Liu, D.~Neill and F.~Ringer, \emph{{The soft drop
  groomed jet radius at NLL}}, {\emph{JHEP} {\bf 02} (2020) 054},
  [\href{http://arxiv.org/abs/1908.01783}{{\tt 1908.01783}}].

\bibitem{Chien:2019gyf}
Y.-T. Chien, D.~Y. Shao and B.~Wu, \emph{{Resummation of Boson-Jet Correlation
  at Hadron Colliders}}, {\emph{JHEP} {\bf 11} (2019) 025},
  [\href{http://arxiv.org/abs/1905.01335}{{\tt 1905.01335}}].

\bibitem{Chien:2019osu}
Y.-T. Chien and I.~W. Stewart, \emph{{Collinear Drop}}, {\emph{JHEP} {\bf 06}
  (2020) 064}, [\href{http://arxiv.org/abs/1907.11107}{{\tt 1907.11107}}].

\bibitem{Stewart:2022ari}
I.~W. Stewart and X.~Yao, \emph{{Pure Quark and Gluon Observables in Collinear
  Drop}},  \href{http://arxiv.org/abs/2203.14980}{{\tt 2203.14980}}.

\bibitem{Vaidya:2020cyi}
V.~Vaidya and X.~Yao, \emph{{Transverse momentum broadening of a jet in
  quark-gluon plasma: an open quantum system EFT}},
  \href{http://dx.doi.org/10.1007/JHEP10(2020)024}{\emph{JHEP} {\bf 10} (2020)
  024}, [\href{http://arxiv.org/abs/2004.11403}{{\tt 2004.11403}}].

\bibitem{Vaidya:2020lih}
V.~Vaidya, \emph{{Effective Field Theory for jet substructure in heavy ion
  collisions}}, \href{http://dx.doi.org/10.1007/JHEP11(2021)064}{\emph{JHEP}
  {\bf 11} (2021) 064}, [\href{http://arxiv.org/abs/2010.00028}{{\tt
  2010.00028}}].

\bibitem{CasalderreySolana:2004qm}
J.~Casalderrey-Solana, E.~Shuryak and D.~Teaney, \emph{{Conical flow induced by
  quenched QCD jets}},
  \href{http://dx.doi.org/10.1088/1742-6596/27/1/003}{\emph{J. Phys. Conf.
  Ser.} {\bf 27} (2005) 22--31},
  [\href{http://arxiv.org/abs/hep-ph/0411315}{{\tt hep-ph/0411315}}].

\bibitem{Ruppert:2005uz}
J.~Ruppert and B.~Müller, \emph{{Waking the colored plasma}},
  \href{http://dx.doi.org/10.1016/j.physletb.2005.04.075}{\emph{Phys. Lett. B}
  {\bf 618} (2005) 123--130}, [\href{http://arxiv.org/abs/hep-ph/0503158}{{\tt
  hep-ph/0503158}}].

\bibitem{Chaudhuri:2005vc}
A.~Chaudhuri and U.~Heinz, \emph{{Effect of jet quenching on the hydrodynamical
  evolution of QGP}},
  \href{http://dx.doi.org/10.1103/PhysRevLett.97.062301}{\emph{Phys. Rev.
  Lett.} {\bf 97} (2006) 062301},
  [\href{http://arxiv.org/abs/nucl-th/0503028}{{\tt nucl-th/0503028}}].

\bibitem{CasalderreySolana:2006sq}
J.~Casalderrey-Solana, E.~Shuryak and D.~Teaney, \emph{{Hydrodynamic flow from
  fast particles}},  \href{http://arxiv.org/abs/hep-ph/0602183}{{\tt
  hep-ph/0602183}}.

\bibitem{Chesler:2007an}
P.~M. Chesler and L.~G. Yaffe, \emph{{The Wake of a quark moving through a
  strongly-coupled plasma}},
  \href{http://dx.doi.org/10.1103/PhysRevLett.99.152001}{\emph{Phys. Rev.
  Lett.} {\bf 99} (2007) 152001}, [\href{http://arxiv.org/abs/0706.0368}{{\tt
  0706.0368}}].

\bibitem{Gubser:2007ga}
S.~S. Gubser, S.~S. Pufu and A.~Yarom, \emph{{Sonic booms and diffusion wakes
  generated by a heavy quark in thermal AdS/CFT}},
  \href{http://dx.doi.org/10.1103/PhysRevLett.100.012301}{\emph{Phys. Rev.
  Lett.} {\bf 100} (2008) 012301}, [\href{http://arxiv.org/abs/0706.4307}{{\tt
  0706.4307}}].

\bibitem{Chesler:2007sv}
P.~M. Chesler and L.~G. Yaffe, \emph{{The Stress-energy tensor of a quark
  moving through a strongly-coupled N=4 supersymmetric Yang-Mills plasma:
  Comparing hydrodynamics and AdS/CFT}},
  \href{http://dx.doi.org/10.1103/PhysRevD.78.045013}{\emph{Phys. Rev. D} {\bf
  78} (2008) 045013}, [\href{http://arxiv.org/abs/0712.0050}{{\tt 0712.0050}}].

\bibitem{Chesler:2008wd}
P.~M. Chesler, K.~Jensen and A.~Karch, \emph{{Jets in strongly-coupled N = 4
  super Yang-Mills theory}},
  \href{http://dx.doi.org/10.1103/PhysRevD.79.025021}{\emph{Phys. Rev. D} {\bf
  79} (2009) 025021}, [\href{http://arxiv.org/abs/0804.3110}{{\tt 0804.3110}}].

\bibitem{Chesler:2008uy}
P.~M. Chesler, K.~Jensen, A.~Karch and L.~G. Yaffe, \emph{{Light quark energy
  loss in strongly-coupled N = 4 supersymmetric Yang-Mills plasma}},
  \href{http://dx.doi.org/10.1103/PhysRevD.79.125015}{\emph{Phys. Rev. D} {\bf
  79} (2009) 125015}, [\href{http://arxiv.org/abs/0810.1985}{{\tt 0810.1985}}].

\bibitem{Neufeld:2008fi}
R.~Neufeld, B.~Müller and J.~Ruppert, \emph{{Sonic Mach Cones Induced by Fast
  Partons in a Perturbative Quark-Gluon Plasma}},
  \href{http://dx.doi.org/10.1103/PhysRevC.78.041901}{\emph{Phys. Rev. C} {\bf
  78} (2008) 041901}, [\href{http://arxiv.org/abs/0802.2254}{{\tt 0802.2254}}].

\bibitem{Neufeld:2008dx}
R.~Neufeld, \emph{{Mach cones in the quark-gluon plasma: Viscosity, speed of
  sound, and effects of finite source structure}},
  \href{http://dx.doi.org/10.1103/PhysRevC.79.054909}{\emph{Phys. Rev. C} {\bf
  79} (2009) 054909}, [\href{http://arxiv.org/abs/0807.2996}{{\tt 0807.2996}}].

\bibitem{Qin:2009uh}
G.-Y. Qin, A.~Majumder, H.~Song and U.~Heinz, \emph{{Energy and momentum
  deposited into a QCD medium by a jet shower}},
  \href{http://dx.doi.org/10.1103/PhysRevLett.103.152303}{\emph{Phys. Rev.
  Lett.} {\bf 103} (2009) 152303}, [\href{http://arxiv.org/abs/0903.2255}{{\tt
  0903.2255}}].

\bibitem{Neufeld:2009ep}
R.~Neufeld and B.~Müller, \emph{{The sound produced by a fast parton in the
  quark-gluon plasma is a 'crescendo'}},
  \href{http://dx.doi.org/10.1103/PhysRevLett.103.042301}{\emph{Phys. Rev.
  Lett.} {\bf 103} (2009) 042301}, [\href{http://arxiv.org/abs/0902.2950}{{\tt
  0902.2950}}].

\bibitem{Gubser:2009sn}
S.~S. Gubser, S.~S. Pufu, F.~D. Rocha and A.~Yarom, \emph{{Energy loss in a
  strongly coupled thermal medium and the gauge-string duality}},
  \href{http://arxiv.org/abs/0902.4041}{{\tt 0902.4041}}.

\bibitem{Chesler:2011nc}
P.~M. Chesler, Y.-Y. Ho and K.~Rajagopal, \emph{{Shining a Gluon Beam Through
  Quark-Gluon Plasma}},
  \href{http://dx.doi.org/10.1103/PhysRevD.85.126006}{\emph{Phys. Rev. D} {\bf
  85} (2012) 126006}, [\href{http://arxiv.org/abs/1111.1691}{{\tt 1111.1691}}].

\bibitem{Betz:2010qh}
B.~Betz, J.~Noronha, G.~Torrieri, M.~Gyulassy and D.~H. Rischke,
  \emph{{Universal Flow-Driven Conical Emission in Ultrarelativistic Heavy-Ion
  Collisions}},
  \href{http://dx.doi.org/10.1103/PhysRevLett.105.222301}{\emph{Phys. Rev.
  Lett.} {\bf 105} (2010) 222301}, [\href{http://arxiv.org/abs/1005.5461}{{\tt
  1005.5461}}].

\bibitem{Ayala:2012bv}
A.~Ayala, I.~Dominguez and M.~E. Tejeda-Yeomans, \emph{{Head shock vs Mach
  cone: Azimuthal correlations from $2 \to 3$ parton processes in relativistic
  heavy-ion collisions}},
  \href{http://dx.doi.org/10.1103/PhysRevC.88.025203}{\emph{Phys. Rev. C} {\bf
  88} (2013) 025203}, [\href{http://arxiv.org/abs/1212.1127}{{\tt 1212.1127}}].

\bibitem{Ayala:2014sua}
A.~Ayala, J.~D. Casta\~no Yepes, I.~Dominguez and M.~E. Tejeda-Yeomans,
  \emph{{Impact of the energy-loss spatial profile and shear-viscosity to
  entropy-density ratio for the Mach cone versus head-shock signals produced by
  a fast-moving parton in a quark-gluon plasma}},
  \href{http://dx.doi.org/10.1103/PhysRevC.92.024910}{\emph{Phys. Rev. C} {\bf
  92} (2015) 024910}, [\href{http://arxiv.org/abs/1412.5879}{{\tt 1412.5879}}].

\bibitem{Floerchinger:2014yqa}
S.~Floerchinger and K.~C. Zapp, \emph{{Hydrodynamics and Jets in Dialogue}},
  \href{http://dx.doi.org/10.1140/epjc/s10052-014-3189-4}{\emph{Eur. Phys. J.
  C} {\bf 74} (2014) 3189}, [\href{http://arxiv.org/abs/1407.1782}{{\tt
  1407.1782}}].

\bibitem{Tachibana:2014lja}
Y.~Tachibana and T.~Hirano, \emph{{Momentum transport away from a jet in an
  expanding nuclear medium}},
  \href{http://dx.doi.org/10.1103/PhysRevC.90.021902}{\emph{Phys. Rev. C} {\bf
  90} (2014) 021902}, [\href{http://arxiv.org/abs/1402.6469}{{\tt 1402.6469}}].

\bibitem{Yan:2017rku}
L.~Yan, S.~Jeon and C.~Gale, \emph{{Jet-medium interaction and conformal
  relativistic fluid dynamics}},
  \href{http://dx.doi.org/10.1103/PhysRevC.97.034914}{\emph{Phys. Rev. C} {\bf
  97} (2018) 034914}, [\href{http://arxiv.org/abs/1707.09519}{{\tt
  1707.09519}}].

\bibitem{Chen:2017zte}
W.~Chen, S.~Cao, T.~Luo, L.-G. Pang and X.-N. Wang, \emph{{Effects of
  jet-induced medium excitation in $\gamma$-hadron correlation in A+A
  collisions}},
  \href{http://dx.doi.org/10.1016/j.physletb.2017.12.015}{\emph{Phys. Lett. B}
  {\bf 777} (2018) 86--90}, [\href{http://arxiv.org/abs/1704.03648}{{\tt
  1704.03648}}].

\bibitem{Tachibana:2020mtb}
Y.~Tachibana, C.~Shen and A.~Majumder, \emph{{Bulk medium evolution has
  considerable effects on jet observables!}},
  \href{http://arxiv.org/abs/2001.08321}{{\tt 2001.08321}}.

\bibitem{Casalderrey-Solana:2020rsj}
J.~Casalderrey-Solana, J.~G. Milhano, D.~Pablos, K.~Rajagopal and X.~Yao,
  \emph{{Jet Wake from Linearized Hydrodynamics}},
  \href{http://dx.doi.org/10.1007/JHEP05(2021)230}{\emph{JHEP} {\bf 05} (2021)
  230}, [\href{http://arxiv.org/abs/2010.01140}{{\tt 2010.01140}}].

\bibitem{Brewer:2021hmh}
J.~Brewer, Q.~Brodsky and K.~Rajagopal, \emph{{Disentangling jet modification
  in jet simulations and in Z+jet data}},
  \href{http://dx.doi.org/10.1007/JHEP02(2022)175}{\emph{JHEP} {\bf 02} (2022)
  175}, [\href{http://arxiv.org/abs/2110.13159}{{\tt 2110.13159}}].

\bibitem{Chesler:2014jva}
P.~M. Chesler and K.~Rajagopal, \emph{{Jet quenching in strongly coupled
  plasma}}, \href{http://dx.doi.org/10.1103/PhysRevD.90.025033}{\emph{Phys.
  Rev. D} {\bf 90} (2014) 025033}, [\href{http://arxiv.org/abs/1402.6756}{{\tt
  1402.6756}}].

\bibitem{Chesler:2015nqz}
P.~M. Chesler and K.~Rajagopal, \emph{{On the Evolution of Jet Energy and
  Opening Angle in Strongly Coupled Plasma}},
  \href{http://dx.doi.org/10.1007/JHEP05(2016)098}{\emph{JHEP} {\bf 05} (2016)
  098}, [\href{http://arxiv.org/abs/1511.07567}{{\tt 1511.07567}}].

\bibitem{Casalderrey-Solana:2014bpa}
J.~Casalderrey-Solana, D.~C. Gulhan, J.~G. Milhano, D.~Pablos and K.~Rajagopal,
  \emph{{A Hybrid Strong/Weak Coupling Approach to Jet Quenching}},
  \href{http://dx.doi.org/10.1007/JHEP10(2014)019}{\emph{JHEP} {\bf 10} (2014)
  19}, [\href{http://arxiv.org/abs/1405.3864}{{\tt 1405.3864}}].

\bibitem{Casalderrey-Solana:2015vaa}
J.~Casalderrey-Solana, D.~C. Gulhan, J.~G. Milhano, D.~Pablos and K.~Rajagopal,
  \emph{{Predictions for Boson-Jet Observables and Fragmentation Function
  Ratios from a Hybrid Strong/Weak Coupling Model for Jet Quenching}},
  \href{http://arxiv.org/abs/1508.00815}{{\tt 1508.00815}}.

\bibitem{Casalderrey-Solana:2016jvj}
J.~Casalderrey-Solana, D.~Gulhan, G.~Milhano, D.~Pablos and K.~Rajagopal,
  \emph{{Angular Structure of Jet Quenching Within a Hybrid Strong/Weak
  Coupling Model}},
  \href{http://dx.doi.org/10.1007/JHEP03(2017)135}{\emph{JHEP} {\bf 03} (2017)
  135}, [\href{http://arxiv.org/abs/1609.05842}{{\tt 1609.05842}}].

\bibitem{Hulcher:2017cpt}
Z.~Hulcher, D.~Pablos and K.~Rajagopal, \emph{{Resolution Effects in the Hybrid
  Strong/Weak Coupling Model}},
  \href{http://dx.doi.org/10.1007/JHEP03(2018)010}{\emph{JHEP} {\bf 03} (2018)
  010}, [\href{http://arxiv.org/abs/1707.05245}{{\tt 1707.05245}}].

\bibitem{Casalderrey-Solana:2018wrw}
J.~Casalderrey-Solana, Z.~Hulcher, G.~Milhano, D.~Pablos and K.~Rajagopal,
  \emph{{Simultaneous description of hadron and jet suppression in heavy-ion
  collisions}}, \href{http://dx.doi.org/10.1103/PhysRevC.99.051901}{\emph{Phys.
  Rev. C} {\bf 99} (2019) 051901}, [\href{http://arxiv.org/abs/1808.07386}{{\tt
  1808.07386}}].

\bibitem{Casalderrey-Solana:2019ubu}
J.~Casalderrey-Solana, G.~Milhano, D.~Pablos and K.~Rajagopal,
  \emph{{Modification of Jet Substructure in Heavy Ion Collisions as a Probe of
  the Resolution Length of Quark-Gluon Plasma}},
  \href{http://dx.doi.org/10.1007/JHEP01(2020)044}{\emph{JHEP} {\bf 01} (2020)
  044}, [\href{http://arxiv.org/abs/1907.11248}{{\tt 1907.11248}}].

\bibitem{Gyulassy:1993hr}
M.~Gyulassy and X.-n. Wang, \emph{{Multiple collisions and induced gluon
  Bremsstrahlung in QCD}},
  \href{http://dx.doi.org/10.1016/0550-3213(94)90079-5}{\emph{Nucl. Phys.} {\bf
  B420} (1994) 583--614}, [\href{http://arxiv.org/abs/nucl-th/9306003}{{\tt
  nucl-th/9306003}}].

\bibitem{Wang:1994fx}
X.-N. Wang, M.~Gyulassy and M.~Plumer, \emph{{The LPM effect in QCD and
  radiative energy loss in a quark gluon plasma}},
  \href{http://dx.doi.org/10.1103/PhysRevD.51.3436}{\emph{Phys. Rev.} {\bf D51}
  (1995) 3436--3446}, [\href{http://arxiv.org/abs/hep-ph/9408344}{{\tt
  hep-ph/9408344}}].

\bibitem{Baier:1994bd}
R.~Baier, Y.~L. Dokshitzer, S.~Peigne and D.~Schiff, \emph{{Induced gluon
  radiation in a QCD medium}},
  \href{http://dx.doi.org/10.1016/0370-2693(94)01617-L}{\emph{Phys. Lett. B}
  {\bf 345} (1995) 277--286}, [\href{http://arxiv.org/abs/hep-ph/9411409}{{\tt
  hep-ph/9411409}}].

\bibitem{Baier:1996kr}
R.~Baier, Y.~L. Dokshitzer, A.~H. Mueller, S.~Peigne and D.~Schiff,
  \emph{{Radiative energy loss of high-energy quarks and gluons in a finite
  volume quark - gluon plasma}},
  \href{http://dx.doi.org/10.1016/S0550-3213(96)00553-6}{\emph{Nucl. Phys.}
  {\bf B483} (1997) 291--320}, [\href{http://arxiv.org/abs/hep-ph/9607355}{{\tt
  hep-ph/9607355}}].

\bibitem{Zakharov:1996fv}
B.~G. Zakharov, \emph{{Fully quantum treatment of the Landau-Pomeranchuk-Migdal
  effect in QED and QCD}}, \href{http://dx.doi.org/10.1134/1.567126}{\emph{JETP
  Lett.} {\bf 63} (1996) 952--957},
  [\href{http://arxiv.org/abs/hep-ph/9607440}{{\tt hep-ph/9607440}}].

\bibitem{Baier:1996sk}
R.~Baier, Y.~L. Dokshitzer, A.~H. Mueller, S.~Peigne and D.~Schiff,
  \emph{{Radiative energy loss and p(T) broadening of high-energy partons in
  nuclei}}, \href{http://dx.doi.org/10.1016/S0550-3213(96)00581-0}{\emph{Nucl.
  Phys. B} {\bf 484} (1997) 265--282},
  [\href{http://arxiv.org/abs/hep-ph/9608322}{{\tt hep-ph/9608322}}].

\bibitem{Gyulassy:1999zd}
M.~Gyulassy, P.~Levai and I.~Vitev, \emph{{Jet quenching in thin quark gluon
  plasmas. 1. Formalism}},
  \href{http://dx.doi.org/10.1016/S0550-3213(99)00713-0}{\emph{Nucl. Phys. B}
  {\bf 571} (2000) 197--233}, [\href{http://arxiv.org/abs/hep-ph/9907461}{{\tt
  hep-ph/9907461}}].

\bibitem{Gyulassy:2000fs}
M.~Gyulassy, P.~Levai and I.~Vitev, \emph{{NonAbelian energy loss at finite
  opacity}},
  \href{http://dx.doi.org/10.1103/PhysRevLett.85.5535}{\emph{Phys.Rev.Lett.}
  {\bf 85} (2000) 5535--5538},
  [\href{http://arxiv.org/abs/nucl-th/0005032}{{\tt nucl-th/0005032}}].

\bibitem{Wiedemann:2000za}
U.~A. Wiedemann, \emph{{Gluon radiation off hard quarks in a nuclear
  environment: Opacity expansion}},
  \href{http://dx.doi.org/10.1016/S0550-3213(00)00457-0}{\emph{Nucl. Phys.}
  {\bf B588} (2000) 303--344}, [\href{http://arxiv.org/abs/hep-ph/0005129}{{\tt
  hep-ph/0005129}}].

\bibitem{Arnold:2002ja}
P.~B. Arnold, G.~D. Moore and L.~G. Yaffe, \emph{{Photon and gluon emission in
  relativistic plasmas}},
  \href{http://dx.doi.org/10.1088/1126-6708/2002/06/030}{\emph{JHEP} {\bf 06}
  (2002) 030}, [\href{http://arxiv.org/abs/hep-ph/0204343}{{\tt
  hep-ph/0204343}}].

\bibitem{CasalderreySolana:2011rz}
J.~Casalderrey-Solana and E.~Iancu, \emph{{Interference effects in
  medium-induced gluon radiation}},
  \href{http://dx.doi.org/10.1007/JHEP08(2011)015}{\emph{JHEP} {\bf 08} (2011)
  015}, [\href{http://arxiv.org/abs/1105.1760}{{\tt 1105.1760}}].

\bibitem{MehtarTani:2011tz}
Y.~Mehtar-Tani, C.~Salgado and K.~Tywoniuk, \emph{{Jets in QCD Media: From
  Color Coherence to Decoherence}},
  \href{http://dx.doi.org/10.1016/j.physletb.2011.12.042}{\emph{Phys. Lett. B}
  {\bf 707} (2012) 156--159}, [\href{http://arxiv.org/abs/1102.4317}{{\tt
  1102.4317}}].

\bibitem{Ovanesyan:2011xy}
G.~Ovanesyan and I.~Vitev, \emph{{An effective theory for jet propagation in
  dense QCD matter: jet broadening and medium-induced bremsstrahlung}},
  \href{http://dx.doi.org/10.1007/JHEP06(2011)080}{\emph{JHEP} {\bf 1106}
  (2011) 080}, [\href{http://arxiv.org/abs/1103.1074}{{\tt 1103.1074}}].

\bibitem{MehtarTani:2011gf}
Y.~Mehtar-Tani, C.~A. Salgado and K.~Tywoniuk, \emph{{The radiation pattern of
  a QCD antenna in a dilute medium}},
  \href{http://dx.doi.org/10.1007/JHEP04(2012)064}{\emph{JHEP} {\bf 04} (2012)
  064}, [\href{http://arxiv.org/abs/1112.5031}{{\tt 1112.5031}}].

\bibitem{MehtarTani:2012cy}
Y.~Mehtar-Tani, C.~A. Salgado and K.~Tywoniuk, \emph{{The Radiation pattern of
  a QCD antenna in a dense medium}},
  \href{http://dx.doi.org/10.1007/JHEP10(2012)197}{\emph{JHEP} {\bf 10} (2012)
  197}, [\href{http://arxiv.org/abs/1205.5739}{{\tt 1205.5739}}].

\bibitem{Blaizot:2012fh}
J.-P. Blaizot, F.~Dominguez, E.~Iancu and Y.~Mehtar-Tani, \emph{{Medium-induced
  gluon branching}},
  \href{http://dx.doi.org/10.1007/JHEP01(2013)143}{\emph{JHEP} {\bf 01} (2013)
  143}, [\href{http://arxiv.org/abs/1209.4585}{{\tt 1209.4585}}].

\bibitem{Blaizot:2013hx}
J.-P. Blaizot, E.~Iancu and Y.~Mehtar-Tani, \emph{{Medium-induced QCD cascade:
  democratic branching and wave turbulence}},
  \href{http://dx.doi.org/10.1103/PhysRevLett.111.052001}{\emph{Phys. Rev.
  Lett.} {\bf 111} (2013) 052001}, [\href{http://arxiv.org/abs/1301.6102}{{\tt
  1301.6102}}].

\bibitem{Blaizot:2013vha}
J.-P. Blaizot, F.~Dominguez, E.~Iancu and Y.~Mehtar-Tani, \emph{{Probabilistic
  picture for medium-induced jet evolution}},
  \href{http://dx.doi.org/10.1007/JHEP06(2014)075}{\emph{JHEP} {\bf 06} (2014)
  075}, [\href{http://arxiv.org/abs/1311.5823}{{\tt 1311.5823}}].

\bibitem{Ghiglieri:2015ala}
J.~Ghiglieri, G.~D. Moore and D.~Teaney, \emph{{Jet-Medium Interactions at NLO
  in a Weakly-Coupled Quark-Gluon Plasma}},
  \href{http://dx.doi.org/10.1007/JHEP03(2016)095}{\emph{JHEP} {\bf 03} (2016)
  095}, [\href{http://arxiv.org/abs/1509.07773}{{\tt 1509.07773}}].

\bibitem{Salgado:2003gb}
C.~A. Salgado and U.~A. Wiedemann, \emph{{Calculating quenching weights}},
  \href{http://dx.doi.org/10.1103/PhysRevD.68.014008}{\emph{Phys. Rev. D} {\bf
  68} (2003) 014008}, [\href{http://arxiv.org/abs/hep-ph/0302184}{{\tt
  hep-ph/0302184}}].

\bibitem{Adhya:2019qse}
S.~P. Adhya, C.~A. Salgado, M.~Spousta and K.~Tywoniuk, \emph{{Medium-induced
  cascade in expanding media}},
  \href{http://dx.doi.org/10.1007/JHEP07(2020)150}{\emph{JHEP} {\bf 07} (2020)
  150}, [\href{http://arxiv.org/abs/1911.12193}{{\tt 1911.12193}}].

\bibitem{Mehtar-Tani:2019ygg}
Y.~Mehtar-Tani and K.~Tywoniuk, \emph{{Improved opacity expansion for
  medium-induced parton splitting}},
  \href{http://dx.doi.org/10.1007/JHEP06(2020)187}{\emph{JHEP} {\bf 06} (2020)
  187}, [\href{http://arxiv.org/abs/1910.02032}{{\tt 1910.02032}}].

\bibitem{Barata:2021wuf}
J.~a. Barata, Y.~Mehtar-Tani, A.~Soto-Ontoso and K.~Tywoniuk,
  \emph{{Medium-induced radiative kernel with the Improved Opacity Expansion}},
  \href{http://dx.doi.org/10.1007/JHEP09(2021)153}{\emph{JHEP} {\bf 09} (2021)
  153}, [\href{http://arxiv.org/abs/2106.07402}{{\tt 2106.07402}}].

\bibitem{Arnold:2020uzm}
P.~Arnold, T.~Gorda and S.~Iqbal, \emph{{The LPM effect in sequential
  bremsstrahlung: nearly complete results for QCD}},
  \href{http://dx.doi.org/10.1007/JHEP11(2020)053}{\emph{JHEP} {\bf 11} (2020)
  053}, [\href{http://arxiv.org/abs/2007.15018}{{\tt 2007.15018}}].

\bibitem{Arnold:2021pin}
P.~Arnold, T.~Gorda and S.~Iqbal, \emph{{The LPM effect in sequential
  bremsstrahlung: analytic results for sub-leading (single) logarithms}},
  \href{http://dx.doi.org/10.1007/JHEP04(2022)085}{\emph{JHEP} {\bf 04} (2022)
  085}, [\href{http://arxiv.org/abs/2112.05161}{{\tt 2112.05161}}].

\bibitem{Arnold:2022epx}
P.~Arnold and O.~Elgedawy, \emph{{The LPM Effect in sequential bremsstrahlung:
  $1/N_c^2$ corrections}},  \href{http://arxiv.org/abs/2202.04662}{{\tt
  2202.04662}}.

\bibitem{feynman1986quantum}
R.~P. Feynman, \emph{Quantum mechanical computers}, {\emph{Foundations of
  physics} {\bf 16} (1986) 507--531}.

\bibitem{Devoret2013}
M.~Devoret and R.~Schoelkopf, \emph{Superconducting circuits for quantum
  information: An outlook},
  \href{http://dx.doi.org/10.1126/science.1231930}{\emph{Science (New York,
  N.Y.)} {\bf 339} (03, 2013) 1169--74}.

\bibitem{annurev-conmatphys-031119-050605}
M.~Kjaergaard, M.~E. Schwartz, J.~Braumüller, P.~Krantz, J.~I.-J. Wang,
  S.~Gustavsson et~al., \emph{Superconducting qubits: Current state of play},
  \href{http://dx.doi.org/10.1146/annurev-conmatphys-031119-050605}{\emph{Annual
  Review of Condensed Matter Physics} {\bf 11} (2020) 369--395}.

\bibitem{doi:10.1063/1.5088164}
C.~D. Bruzewicz, J.~Chiaverini, R.~McConnell and J.~M. Sage, \emph{Trapped-ion
  quantum computing: Progress and challenges},
  \href{http://dx.doi.org/10.1063/1.5088164}{\emph{Applied Physics Reviews}
  {\bf 6} (2019) 021314}.

\bibitem{google_supremacy}
F.~Arute, K.~Arya, R.~Babbush, D.~Bacon, J.~C. Bardin, R.~Barends et~al.,
  \emph{Quantum supremacy using a programmable superconducting processor},
  \href{http://dx.doi.org/10.1038/s41586-019-1666-5}{\emph{Nature} {\bf 574}
  (2019) 505--510}.

\bibitem{Lamm:2018siq}
H.~Lamm and S.~Lawrence, \emph{{Simulation of Nonequilibrium Dynamics on a
  Quantum Computer}},
  \href{http://dx.doi.org/10.1103/PhysRevLett.121.170501}{\emph{Phys. Rev.
  Lett.} {\bf 121} (2018) 170501}, [\href{http://arxiv.org/abs/1806.06649}{{\tt
  1806.06649}}].

\bibitem{Bauer:2019qxa}
C.~W. Bauer, W.~A. de~Jong, B.~Nachman and D.~Provasoli, \emph{{Quantum
  Algorithm for High Energy Physics Simulations}},
  \href{http://dx.doi.org/10.1103/PhysRevLett.126.062001}{\emph{Phys. Rev.
  Lett.} {\bf 126} (2021) 062001}, [\href{http://arxiv.org/abs/1904.03196}{{\tt
  1904.03196}}].

\bibitem{Mueller:2019qqj}
N.~Mueller, A.~Tarasov and R.~Venugopalan, \emph{{Deeply inelastic scattering
  structure functions on a hybrid quantum computer}},
  \href{http://dx.doi.org/10.1103/PhysRevD.102.016007}{\emph{Phys. Rev. D} {\bf
  102} (2020) 016007}, [\href{http://arxiv.org/abs/1908.07051}{{\tt
  1908.07051}}].

\bibitem{Wei:2019rqy}
A.~Y. Wei, P.~Naik, A.~W. Harrow and J.~Thaler, \emph{{Quantum Algorithms for
  Jet Clustering}},
  \href{http://dx.doi.org/10.1103/PhysRevD.101.094015}{\emph{Phys. Rev. D} {\bf
  101} (2020) 094015}, [\href{http://arxiv.org/abs/1908.08949}{{\tt
  1908.08949}}].

\bibitem{Smith2019}
A.~Smith, M.~S. Kim, F.~Pollmann and J.~Knolle, \emph{Simulating quantum
  many-body dynamics on a current digital quantum computer},
  \href{http://dx.doi.org/10.1038/s41534-019-0217-0}{\emph{npj Quantum
  Information} {\bf 5} (11, 2019) 106}.

\bibitem{Barata:2020jtq}
J.~a. Barata, N.~Mueller, A.~Tarasov and R.~Venugopalan, \emph{{Single-particle
  digitization strategy for quantum computation of a $\phi^4$ scalar field
  theory}}, \href{http://dx.doi.org/10.1103/PhysRevA.103.042410}{\emph{Phys.
  Rev. A} {\bf 103} (2021) 042410},
  [\href{http://arxiv.org/abs/2012.00020}{{\tt 2012.00020}}].

\bibitem{Liu:2020eoa}
J.~Liu and Y.~Xin, \emph{{Quantum simulation of quantum field theories as
  quantum chemistry}},
  \href{http://dx.doi.org/10.1007/JHEP12(2020)011}{\emph{JHEP} {\bf 12} (2020)
  011}, [\href{http://arxiv.org/abs/2004.13234}{{\tt 2004.13234}}].

\bibitem{Liu:2020wtr}
J.~Liu and Y.-Z. Li, \emph{{On Quantum Simulation Of Cosmic Inflation}},
  \href{http://dx.doi.org/10.1103/PhysRevD.104.086013}{\emph{Phys. Rev. D} {\bf
  104} (2021) 086013}, [\href{http://arxiv.org/abs/2009.10921}{{\tt
  2009.10921}}].

\bibitem{Buser:2020cvn}
A.~J. Buser, H.~Gharibyan, M.~Hanada, M.~Honda and J.~Liu, \emph{{Quantum
  simulation of gauge theory via orbifold lattice}},
  \href{http://dx.doi.org/10.1007/JHEP09(2021)034}{\emph{JHEP} {\bf 09} (2021)
  034}, [\href{http://arxiv.org/abs/2011.06576}{{\tt 2011.06576}}].

\bibitem{Kan:2021nyu}
A.~Kan, L.~Funcke, S.~K\"uhn, L.~Dellantonio, J.~Zhang, J.~F. Haase et~al.,
  \emph{{Investigating a (3+1)D topological \ensuremath{\theta}-term in the
  Hamiltonian formulation of lattice gauge theories for quantum and classical
  simulations}},
  \href{http://dx.doi.org/10.1103/PhysRevD.104.034504}{\emph{Phys. Rev. D} {\bf
  104} (2021) 034504}, [\href{http://arxiv.org/abs/2105.06019}{{\tt
  2105.06019}}].

\bibitem{Martyn:2021eaf}
J.~M. Martyn, Z.~M. Rossi, A.~K. Tan and I.~L. Chuang, \emph{{Grand Unification
  of Quantum Algorithms}},
  \href{http://dx.doi.org/10.1103/PRXQuantum.2.040203}{\emph{PRX Quantum} {\bf
  2} (2021) 040203}, [\href{http://arxiv.org/abs/2105.02859}{{\tt
  2105.02859}}].

\bibitem{Klco:2021lap}
N.~Klco, A.~Roggero and M.~J. Savage, \emph{{Standard Model Physics and the
  Digital Quantum Revolution: Thoughts about the Interface}},
  \href{http://arxiv.org/abs/2107.04769}{{\tt 2107.04769}}.

\bibitem{Bauer:2021gup}
C.~W. Bauer, M.~Freytsis and B.~Nachman, \emph{{Simulating Collider Physics on
  Quantum Computers Using Effective Field Theories}},
  \href{http://dx.doi.org/10.1103/PhysRevLett.127.212001}{\emph{Phys. Rev.
  Lett.} {\bf 127} (2021) 212001}, [\href{http://arxiv.org/abs/2102.05044}{{\tt
  2102.05044}}].

\bibitem{Czajka:2021yll}
A.~M. Czajka, Z.-B. Kang, H.~Ma and F.~Zhao, \emph{{Quantum Simulation of
  Chiral Phase Transitions}},  \href{http://arxiv.org/abs/2112.03944}{{\tt
  2112.03944}}.

\bibitem{Ciavarella:2022zhe}
A.~Ciavarella, N.~Klco and M.~J. Savage, \emph{{Some Conceptual Aspects of
  Operator Design for Quantum Simulations of Non-Abelian Lattice Gauge
  Theories}},  3, 2022.
\newblock \href{http://arxiv.org/abs/2203.11988}{{\tt 2203.11988}}.

\bibitem{Bauer:2022hpo}
C.~W. Bauer et~al., \emph{{Quantum Simulation for High Energy Physics}},
  \href{http://arxiv.org/abs/2204.03381}{{\tt 2204.03381}}.

\bibitem{Jordan:2011ci}
S.~P. Jordan, K.~S.~M. Lee and J.~Preskill, \emph{{Quantum Computation of
  Scattering in Scalar Quantum Field Theories}}, {\emph{Quant. Inf. Comput.}
  {\bf 14} (2014) 1014--1080}, [\href{http://arxiv.org/abs/1112.4833}{{\tt
  1112.4833}}].

\bibitem{Jordan:2012xnu}
S.~P. Jordan, K.~S.~M. Lee and J.~Preskill, \emph{{Quantum Algorithms for
  Quantum Field Theories}},
  \href{http://dx.doi.org/10.1126/science.1217069}{\emph{Science} {\bf 336}
  (2012) 1130--1133}, [\href{http://arxiv.org/abs/1111.3633}{{\tt 1111.3633}}].

\bibitem{Jordan:2017lea}
S.~P. Jordan, H.~Krovi, K.~S.~M. Lee and J.~Preskill, \emph{{BQP-completeness
  of Scattering in Scalar Quantum Field Theory}},
  \href{http://dx.doi.org/10.22331/q-2018-01-08-44}{\emph{Quantum} {\bf 2}
  (2018) 44}, [\href{http://arxiv.org/abs/1703.00454}{{\tt 1703.00454}}].

\bibitem{Klco:2018zqz}
N.~Klco and M.~J. Savage, \emph{{Digitization of scalar fields for quantum
  computing}}, \href{http://dx.doi.org/10.1103/PhysRevA.99.052335}{\emph{Phys.
  Rev. A} {\bf 99} (2019) 052335}, [\href{http://arxiv.org/abs/1808.10378}{{\tt
  1808.10378}}].

\bibitem{Jordan:2014tma}
S.~P. Jordan, K.~S.~M. Lee and J.~Preskill, \emph{{Quantum Algorithms for
  Fermionic Quantum Field Theories}},
  \href{http://arxiv.org/abs/1404.7115}{{\tt 1404.7115}}.

\bibitem{hauke2013quantum}
P.~Hauke, D.~Marcos, M.~Dalmonte and P.~Zoller, \emph{Quantum simulation of a
  lattice schwinger model in a chain of trapped ions}, {\emph{Physical Review
  X} {\bf 3} (2013) 041018}.

\bibitem{kuhn2014quantum}
S.~K{\"u}hn, J.~I. Cirac and M.-C. Ba{\~n}uls, \emph{Quantum simulation of the
  schwinger model: A study of feasibility}, {\emph{Physical Review A} {\bf 90}
  (2014) 042305}.

\bibitem{Klco:2018kyo}
N.~Klco, E.~F. Dumitrescu, A.~J. McCaskey, T.~D. Morris, R.~C. Pooser, M.~Sanz
  et~al., \emph{{Quantum-classical computation of Schwinger model dynamics
  using quantum computers}},
  \href{http://dx.doi.org/10.1103/PhysRevA.98.032331}{\emph{Phys. Rev. A} {\bf
  98} (2018) 032331}, [\href{http://arxiv.org/abs/1803.03326}{{\tt
  1803.03326}}].

\bibitem{Zache:2018cqq}
T.~V. Zache, N.~Mueller, J.~T. Schneider, F.~Jendrzejewski, J.~Berges and
  P.~Hauke, \emph{{Dynamical Topological Transitions in the Massive Schwinger
  Model with a $\theta$ Term}},
  \href{http://dx.doi.org/10.1103/PhysRevLett.122.050403}{\emph{Phys. Rev.
  Lett.} {\bf 122} (2019) 050403}, [\href{http://arxiv.org/abs/1808.07885}{{\tt
  1808.07885}}].

\bibitem{Klco:2019evd}
N.~Klco, J.~R. Stryker and M.~J. Savage, \emph{{SU(2) non-Abelian gauge field
  theory in one dimension on digital quantum computers}},
  \href{http://dx.doi.org/10.1103/PhysRevD.101.074512}{\emph{Phys. Rev. D} {\bf
  101} (2020) 074512}, [\href{http://arxiv.org/abs/1908.06935}{{\tt
  1908.06935}}].

\bibitem{Chakraborty:2020uhf}
B.~Chakraborty, M.~Honda, T.~Izubuchi, Y.~Kikuchi and A.~Tomiya,
  \emph{{Classically Emulated Digital Quantum Simulation of the Schwinger Model
  with Topological Term via Adiabatic State Preparation}},
  \href{http://arxiv.org/abs/2001.00485}{{\tt 2001.00485}}.

\bibitem{Nguyen:2021hyk}
N.~H. Nguyen, M.~C. Tran, Y.~Zhu, A.~M. Green, C.~H. Alderete, Z.~Davoudi
  et~al., \emph{{Digital Quantum Simulation of the Schwinger Model and Symmetry
  Protection with Trapped Ions}},
  \href{http://dx.doi.org/10.1103/PRXQuantum.3.020324}{\emph{PRX Quantum} {\bf
  3} (2022) 020324}, [\href{http://arxiv.org/abs/2112.14262}{{\tt
  2112.14262}}].

\bibitem{deJong:2021wsd}
W.~A. de~Jong, K.~Lee, J.~Mulligan, M.~P\l{}osko\'n, F.~Ringer and X.~Yao,
  \emph{{Quantum simulation of non-equilibrium dynamics and thermalization in
  the Schwinger model}},  \href{http://arxiv.org/abs/2106.08394}{{\tt
  2106.08394}}.

\bibitem{Ciavarella:2021nmj}
A.~Ciavarella, N.~Klco and M.~J. Savage, \emph{{Trailhead for quantum
  simulation of SU(3) Yang-Mills lattice gauge theory in the local multiplet
  basis}}, \href{http://dx.doi.org/10.1103/PhysRevD.103.094501}{\emph{Phys.
  Rev. D} {\bf 103} (2021) 094501},
  [\href{http://arxiv.org/abs/2101.10227}{{\tt 2101.10227}}].

\bibitem{Gonzalez-Cuadra:2022hxt}
D.~Gonz\'alez-Cuadra, T.~V. Zache, J.~Carrasco, B.~Kraus and P.~Zoller,
  \emph{{Hardware efficient quantum simulation of non-abelian gauge theories
  with qudits on Rydberg platforms}},
  \href{http://arxiv.org/abs/2203.15541}{{\tt 2203.15541}}.

\bibitem{DeJong:2020riy}
W.~A. De~Jong, M.~Metcalf, J.~Mulligan, M.~P\l{}osko\'n, F.~Ringer and X.~Yao,
  \emph{{Quantum simulation of open quantum systems in heavy-ion collisions}},
  \href{http://dx.doi.org/10.1103/PhysRevD.104.L051501}{\emph{Phys. Rev. D}
  {\bf 104} (2021) 051501}, [\href{http://arxiv.org/abs/2010.03571}{{\tt
  2010.03571}}].

\bibitem{Barata:2021yri}
J.~a. Barata and C.~A. Salgado, \emph{{A quantum strategy to compute the jet
  quenching parameter $\hat{q}$}},
  \href{http://dx.doi.org/10.1140/epjc/s10052-021-09674-9}{\emph{Eur. Phys. J.
  C} {\bf 81} (2021) 862}, [\href{http://arxiv.org/abs/2104.04661}{{\tt
  2104.04661}}].

\bibitem{Qian:2021jxp}
W.~Qian, R.~Basili, S.~Pal, G.~Luecke and J.~P. Vary, \emph{{Solving hadron
  structures using the basis light-front quantization approach on quantum
  computers}},  \href{http://arxiv.org/abs/2112.01927}{{\tt 2112.01927}}.

\bibitem{preskill2018quantum}
J.~Preskill, \emph{Quantum computing in the nisq era and beyond},
  {\emph{Quantum} {\bf 2} (2018) 79}.

\bibitem{He:2020udd}
A.~He, B.~Nachman, W.~A. de~Jong and C.~W. Bauer, \emph{{Zero-noise
  extrapolation for quantum-gate error mitigation with identity insertions}},
  \href{http://dx.doi.org/10.1103/PhysRevA.102.012426}{\emph{Phys. Rev. A} {\bf
  102} (2020) 012426}, [\href{http://arxiv.org/abs/2003.04941}{{\tt
  2003.04941}}].

\bibitem{Pascuzzi:2021mhw}
V.~R. Pascuzzi, A.~He, C.~W. Bauer, W.~A. de~Jong and B.~Nachman,
  \emph{{Computationally Efficient Zero Noise Extrapolation for Quantum Gate
  Error Mitigation}},
  \href{http://dx.doi.org/10.1103/PhysRevA.105.042406}{\emph{Phys. Rev. A} {\bf
  105} (2022) 042406}, [\href{http://arxiv.org/abs/2110.13338}{{\tt
  2110.13338}}].

\bibitem{Brodsky:1997de}
S.~J. Brodsky, H.-C. Pauli and S.~S. Pinsky, \emph{{Quantum chromodynamics and
  other field theories on the light cone}},
  \href{http://dx.doi.org/10.1016/S0370-1573(97)00089-6}{\emph{Phys. Rept.}
  {\bf 301} (1998) 299--486}, [\href{http://arxiv.org/abs/hep-ph/9705477}{{\tt
  hep-ph/9705477}}].

\bibitem{Bakker:2013cea}
B.~L.~G. Bakker et~al., \emph{{Light-Front Quantum Chromodynamics: A framework
  for the analysis of hadron physics}},
  \href{http://dx.doi.org/10.1016/j.nuclphysbps.2014.05.004}{\emph{Nucl. Phys.
  B Proc. Suppl.} {\bf 251-252} (2014) 165--174},
  [\href{http://arxiv.org/abs/1309.6333}{{\tt 1309.6333}}].

\bibitem{Li:2020uhl}
M.~Li, X.~Zhao, P.~Maris, G.~Chen, Y.~Li, K.~Tuchin et~al.,
  \emph{{Ultrarelativistic quark-nucleus scattering in a light-front
  Hamiltonian approach}},
  \href{http://dx.doi.org/10.1103/PhysRevD.101.076016}{\emph{Phys. Rev. D} {\bf
  101} (2020) 076016}, [\href{http://arxiv.org/abs/2002.09757}{{\tt
  2002.09757}}].

\bibitem{Li:2021zaw}
M.~Li, T.~Lappi and X.~Zhao, \emph{{Scattering and gluon emission in a color
  field: A light-front Hamiltonian approach}},
  \href{http://dx.doi.org/10.1103/PhysRevD.104.056014}{\emph{Phys. Rev. D} {\bf
  104} (2021) 056014}, [\href{http://arxiv.org/abs/2107.02225}{{\tt
  2107.02225}}].

\bibitem{Bender:1992yd}
C.~M. Bender, S.~Pinsky and B.~Van~de Sande, \emph{{Spontaneous symmetry
  breaking of Phi**4 in (1+1)-dimensions in light front field theory}},
  \href{http://dx.doi.org/10.1103/PhysRevD.48.816}{\emph{Phys. Rev. D} {\bf 48}
  (1993) 816--821}, [\href{http://arxiv.org/abs/hep-th/9212009}{{\tt
  hep-th/9212009}}].

\bibitem{Ji:2020baz}
X.~Ji, \emph{{Fundamental Properties of the Proton in Light-Front Zero Modes}},
  \href{http://dx.doi.org/10.1016/j.nuclphysb.2020.115181}{\emph{Nucl. Phys.}
  {\bf B} (2020) 115181}, [\href{http://arxiv.org/abs/2003.04478}{{\tt
  2003.04478}}].

\bibitem{Gelis:2005pt}
F.~Gelis and Y.~Mehtar-Tani, \emph{{Gluon propagation inside a high-energy
  nucleus}}, \href{http://dx.doi.org/10.1103/PhysRevD.73.034019}{\emph{Phys.
  Rev. D} {\bf 73} (2006) 034019},
  [\href{http://arxiv.org/abs/hep-ph/0512079}{{\tt hep-ph/0512079}}].

\bibitem{farhi2000quantum}
E.~Farhi, J.~Goldstone, S.~Gutmann and M.~Sipser, \emph{Quantum computation by
  adiabatic evolution}, {\emph{arXiv preprint quant-ph/0001106} (2000) }.

\bibitem{Alexandru:2019dmv}
A.~Alexandru, P.~F. Bedaque and S.~Lawrence, \emph{{Quantum algorithms for
  disordered physics}},
  \href{http://dx.doi.org/10.1103/PhysRevA.101.032325}{\emph{Phys. Rev. A} {\bf
  101} (2020) 032325}, [\href{http://arxiv.org/abs/1911.11117}{{\tt
  1911.11117}}].

\bibitem{low2017optimal}
G.~H. Low and I.~L. Chuang, \emph{Optimal hamiltonian simulation by quantum
  signal processing}, {\emph{Physical review letters} {\bf 118} (2017) 010501}.

\bibitem{martyn2021efficient}
J.~M. Martyn, Y.~Liu, Z.~E. Chin and I.~L. Chuang, \emph{Efficient
  fully-coherent hamiltonian simulation}, {\emph{arXiv preprint
  arXiv:2110.11327} (2021) }.

\end{thebibliography}\endgroup
\end{document}